\documentclass[conference]{IEEEtran}
\IEEEoverridecommandlockouts

\usepackage{cite}
\def\BibTeX{{\rm B\kern-.05em{\sc i\kern-.025em b}\kern-.08em
    T\kern-.1667em\lower.7ex\hbox{E}\kern-.125emX}}

\usepackage{tikz}
\usepackage{amsmath}
\usepackage[hyphens]{url}
\usepackage{makecell}
\usepackage{subfigure}
\usepackage{soul}
\usepackage{enumitem}
\usepackage{graphicx}
\usepackage{textcomp}
\usepackage{xcolor}
\usepackage{colortbl}
\usepackage{booktabs}
\usepackage{multirow}
\usepackage[hidelinks]{hyperref}

\newcommand*\circled[1]{\tikz[baseline=(char.base)]{
            \node[shape=circle,fill,draw=black,color=black,text=white,inner sep=0.05pt](char){#1};}}

\newcommand\narrow[1]{\scalebox{.94}[1.0]{#1}}
\newcommand\vnarrow[1]{\scalebox{.85}[1.0]{#1}}

\let\textttinner\texttt
\renewcommand\textasciitilde{\raisebox{0.5ex}{\texttildelow}}
\renewcommand{\texttt}[1]{\scalebox{.85}[1.0]{\textttinner{#1}}}

\newcommand{\worse}{\cellcolor{red!10}}
\newcommand{\better}{\cellcolor{blue!10}}

\newcommand{\SystemName}{DisDP}

\title{\SystemName: Disaggregating Compute, Network, and Storage for Model-Sharded Data-Parallel Training}

\author{
    \IEEEauthorblockN{Mo Sun, Zihan Yang, Changyue Liao, Yingtao Li, Jie Zhang, Kaiqi Chen, Fei Wu and Zeke Wang}
    \IEEEauthorblockA{Zhejiang University, China}
    \IEEEauthorblockA{\{sunmo,zihanyang,changyueliao,Li\_Yingtao,carlzhang4,chiaki\_cage,wufei,wangzeke\}@zju.edu.cn}
}

\begin{document}

\maketitle
\thispagestyle{plain}
\pagestyle{plain}

\begin{abstract}
Model-sharded data parallelism (MSDP), e.g., ZeRO, evenly shards the model states across all GPUs, and thus has been widely adopted by LLM pre-training, such as Llama and DeepSeek, due to its low GPU memory capacity requirement. However, MSDP introduces severe overhead from additional network communication collectives (i.e., \narrow{\texttt{AllGather}} and \narrow{\texttt{ReduceScatter}}). Although the collectives themselves only occupy fewer than 10\% of GPU SMs, their execution time increases by 41\% due to the serial execution of aggregated CPU/GPU-managed compute (i.e., GEMM), network (i.e., NCCL), and storage (i.e., optimizer states).
To this end, we present \SystemName, a fully disaggregated distributed data-parallel architecture that first fully disaggregates compute, network, and storage for MSDP, 
such that GPUs only focus on the computing part, and thus the GPU utilization is maximized. The key idea is 1) fully offloading collectives to SmartNICs and SmartSwitch to avoid interference between GEMM kernels and collective kernels, and 2) fully offloading storage to a SmartSwitch-enhanced parameter server that allows a single PS to serve massive workers with linear scalability. 
\SystemName{} on 8 distributed GPUs outperforms the state-of-the-art training systems by 3.98$\times$ when training on a 175B model, validating the efficiency of disaggregation. 
\end{abstract}


\section{Introduction}


Large language models (LLMs) have made advances in application domains such as natural language processing~\cite{bert, t5}, programming~\cite{vega_cgo2025}, and computer vision~\cite{vit, swin-transformer, dalle}. Along with the advances of LLM are their fast-growing model sizes, from 100M-scale~\cite{bert} to 100B-scale~\cite{gpt-3, llama3, mt-nlg, palm}. Training a large-scale model requires using many GPUs with efficient parallelism strategies~\cite{kim2019parallax, hashemi2016performance, watcharapichat2016ako, inceptionn_micro2018, plink, bml, fold3d, submodel_ccgrid2024, ultima, mist, wlbllm, vela_asplos2025, mist_eurosys2025, pccl_iccd2024, baft_fcs2025, chimera_isca2025, traci_isca2025, disttrain_sigcomm25, bytescale_sigcomm25}.
Among these strategies, model replicated data parallelism~(MRDP or DP)~\cite{dp, allreduce, projectadam, ps, psnips, tensorflow, mxnet, pytorch-ddp, multitree_allreduce_isca2021, phub, psplus, herring, geeps, byteps, zenops, optimus, topoopt, megascale, lsmgnn_arxiv2024} replicates the model states across GPU workers, and each GPU needs to accommodate the entire model state, including all parameters and auxiliary optimizer states. So, an NVIDIA H100 GPU cluster (each GPU with 80 GB of device memory) even fails to train a 7B model (with 112 GB gradients and model states), regardless of the number of GPUs in the cluster. 

\begin{figure*}[t]
    \begin{center}
        \subfigure[\label{fig_overview_zero} ZeRO-Infinity: Full aggregation so suffering from interference between collectives and computation on GPU.]{
            \includegraphics[width=0.249\linewidth]{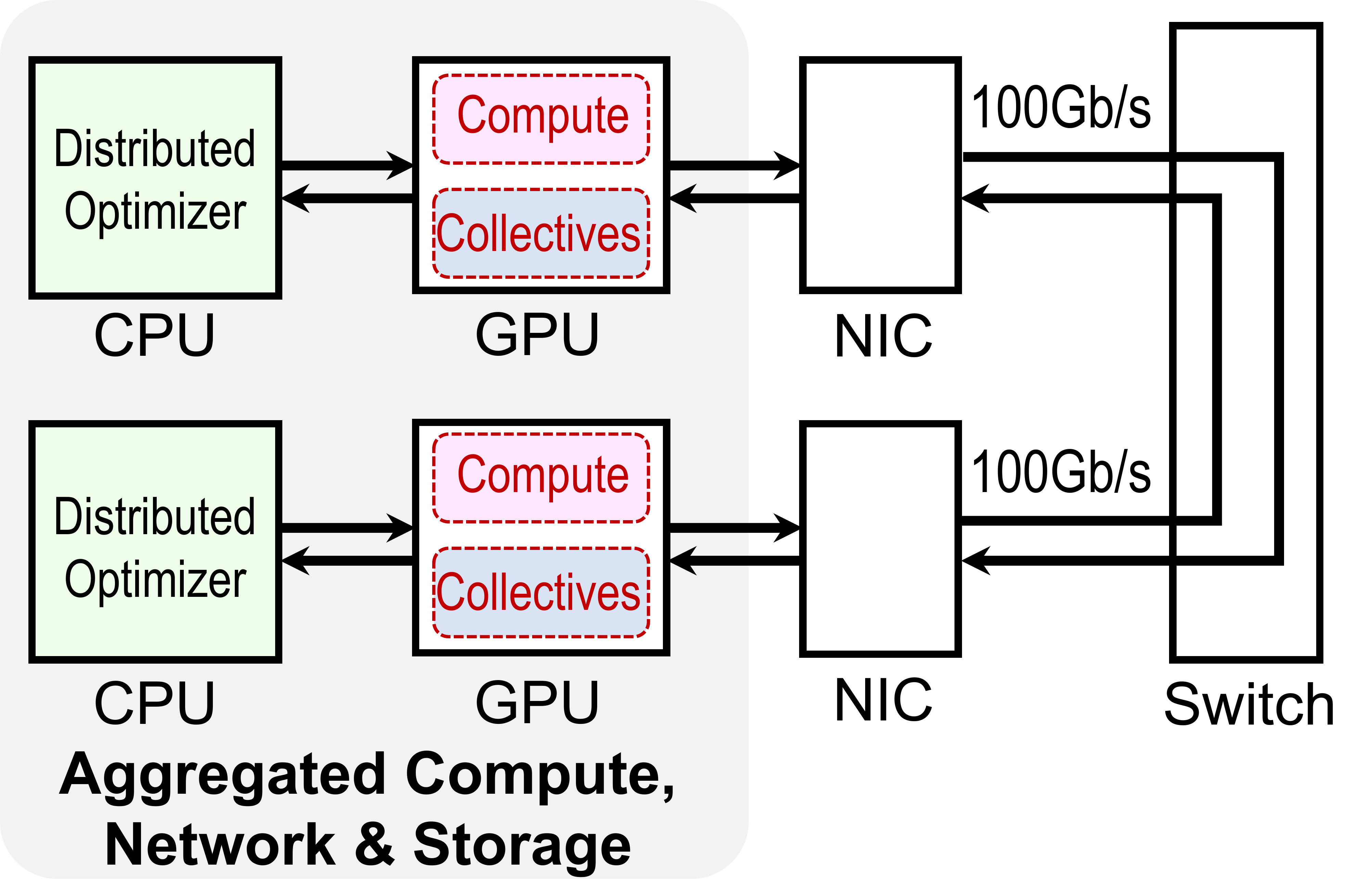}
        }
        \hfill
        \subfigure[\label{fig_overview_switchml} SwitchML: Partial disaggregation due to only offloading in-network aggregation logic to SmartSwitch.]{
            \includegraphics[width=0.267\linewidth]{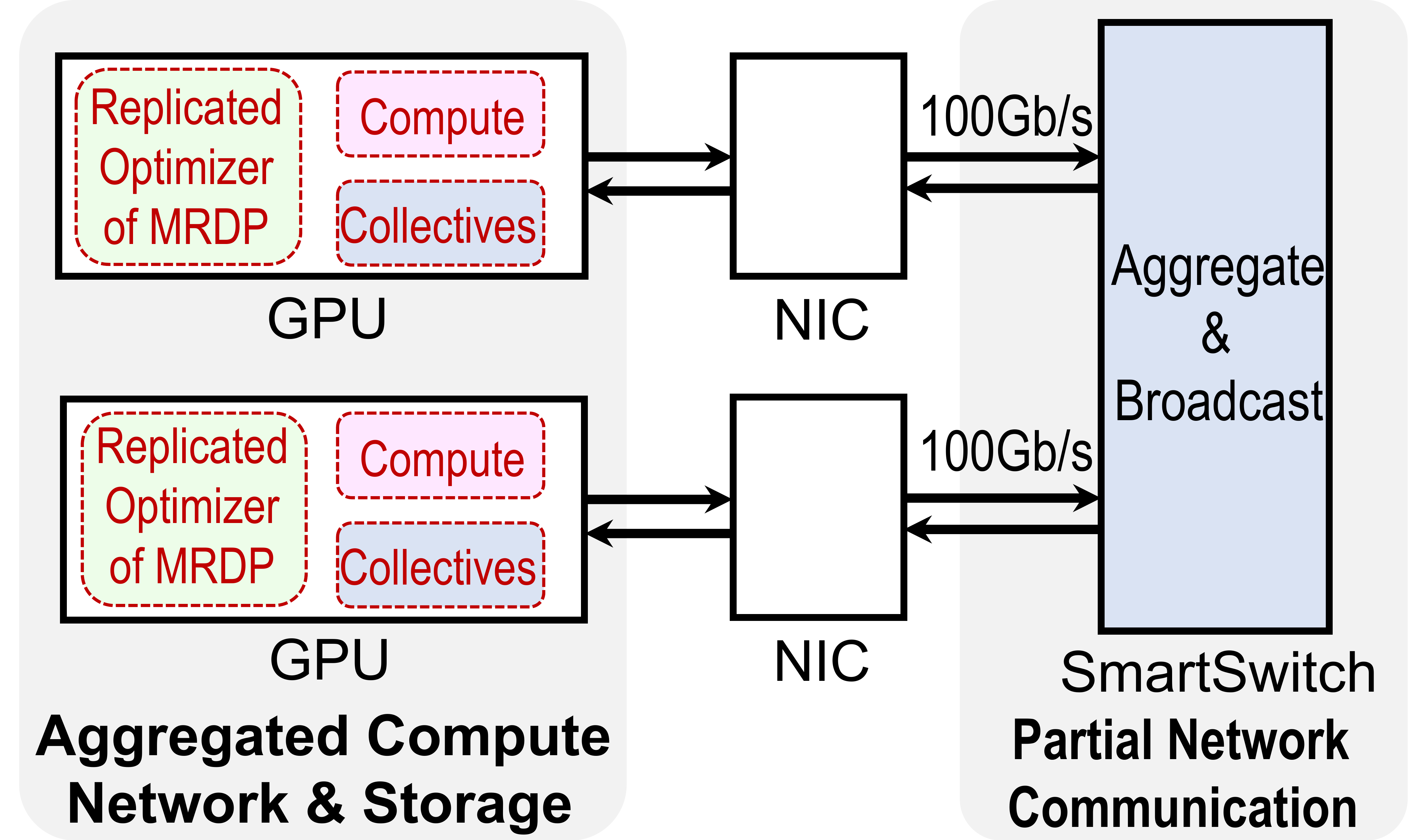}
        }
        \hfill
        \subfigure[\label{fig_overview_ours} \SystemName{}: Fully disaggregates compute, network (i.e., collectives), and storage (i.e., optimizer) such that GPUs only focus on the computing part to maximize GPU utilization.]{
            \includegraphics[width=0.424\linewidth]{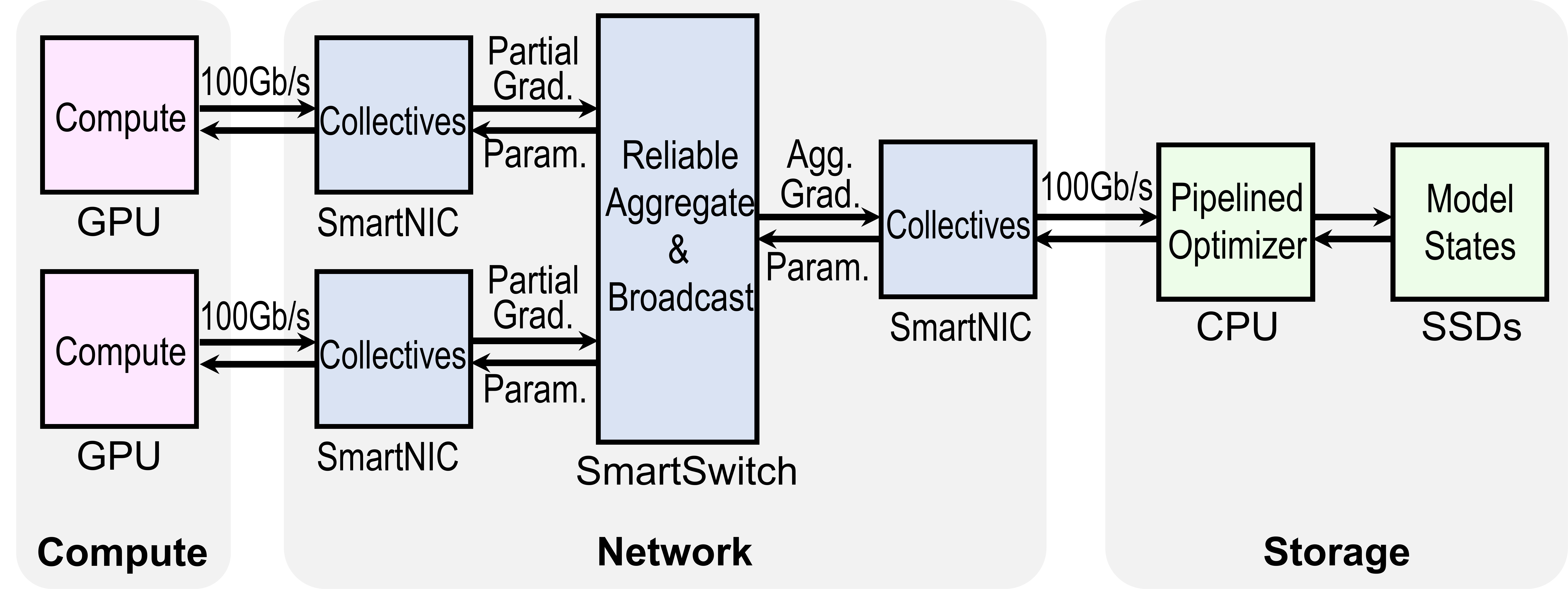}
        }
    \end{center}
    \vspace{-2ex}
    \caption{Comparison of different DP architectures. Both ZeRO-Infinity and SwitchML aggregate compute, network, and storage on worker CPU/GPU, leading to severe interference, while \SystemName{} fully disaggregates compute, network, and storage for high GPU utilization. }
    \label{fig_comparison}
    \vspace{-1ex}
\end{figure*}

\begin{table*}[t]
    \footnotesize
    \centering
    \vspace{-1ex}
    \caption{Comparison between DP architectures. \SystemName{} disaggregates compute, network and storage, so GPUs achieve maximum utilization.}
    \label{table_system_comparison}
    \vspace{-2ex}
    \begin{tabular}{
        l
        >{\centering\arraybackslash}p{2.1cm}
        >{\centering\arraybackslash}p{4.1cm}
        >{\centering\arraybackslash}p{5.8cm}
    }
        \specialrule{.1em}{.05em}{.05em} 
        \textbf{System}  & \makecell{\textbf{\narrow{Interference-Free}} \\ \textbf{Network}}  & \makecell{\textbf{Traffic-Optimal} \\ \textbf{Collective Topology}} & \makecell{\textbf{Low-Overhead} \\ \textbf{Optimizer Execution}} \\ \specialrule{.1em}{.05em}{.05em} 
        \makecell[l]{\textbf{Model-Sharded DP} \\ \textbf{(\S\ref{subsec_motivation_msdp})}} & \worse\makecell{No \\ (GPU Contention)} & \worse\makecell{No \\ (Peer-Based AG/RS)} & \better\makecell{Yes (But colocated CPU optimizer could interfere \\ with GPU due to intra-node PCIe contention)} \\ \hline
        \makecell[l]{\textbf{\narrow{SmartSwitch-Enhanced}} \\ \textbf{MSDP (\S\ref{subsec_motivation_atp})}} & \worse\makecell{No \\ (GPU Contention)} & \better \makecell{Yes} & \better\makecell{Yes (But colocated CPU optimizer could interfere \\ with GPU due to intra-node PCIe contention)} \\ \hline
        \makecell[l]{\textbf{Model-in-Server DP} \\ \textbf{(\S\ref{subsec_motivation_ps})}} & \worse\makecell{No \\ (GPU Contention)} & \better\makecell{Yes \\ (But only for non-colocated PS)} & \worse\makecell{No (Requiring many additional machines \\ as PSs for aggregation/broadcast)} \\ \hline
        \textbf{\SystemName{} (Ours)} & \better Yes & \better Yes & \better\makecell{Yes (Requiring only 1 additional machine \\ as scalable PS)} \\ \specialrule{.1em}{.05em}{.05em}   
    \end{tabular}
    \vspace{-2ex}
\end{table*}

\noindent\textbf{Model-Sharded DP (MSDP).}
\
To address the limitation of trainable model size, memory-efficient MSDP such as ZeRO~\cite{xu2020automatic, zero, fsdp, zero++, mics, amsp, comet_arxiv2022, zero-offload, zero-infinity, patrickstar, colossal-ai, colossal-auto, sentinel, dynn-offload} and PyTorch FSDP~\cite{fsdp} are proposed to evenly shard the model states across all GPU workers, as shown in Figure~\ref{fig_overview_zero}.\footnote{In this paper, we refer to a worker as a GPU paired with a NIC.} With enough GPUs, MSDP allows training a huge model, and thus has been adopted by DeepSeek~\cite{deepseek-v3} and Llama~\cite{llama3_isca2025} to pre-train 100B-scale models.\footnote{Though DeepSeek and Llama use 3D parallelism, including DP, TP, and PP, MSDP is important, especially in the pertaining phase that takes 80\%\textasciitilde 95\% of the total training steps~\cite{longcontext_naaclhlt2024, llama3}. For example, Llama 3 applies 128-degree MSDP, while employing 8-degree TP and 16-degree PP~\cite{llama3_isca2025}.} Therefore, the optimization of MSDP itself is important.
However, MSDP comes at the cost of extra collective operations, i.e., \texttt{AllGather}, to fetch on-demand parameters within an iteration. 
We observe that although network communication itself consumes relatively low GPU resources, its aggregated computation and collectives on GPUs lead to severe interference. In particular, with GPUs' non-preemptive scheduling policy, current-round compute-intensive GEMM kernels would occupy all GPU SMs and block next-round communication kernels from launching, even without dependencies. 
In our experiment, ZeRO-Infinity~\cite{zero-infinity} on 8$\times$ 1-GPU machines under a 100Gbps network only achieves 15\% Model FLOPS Utilization (MFU) when training a 175B model. 

\noindent\textbf{Smart\-Switch-Enhanced MSDP.}
\
Prior works~\cite{atp, a2tp} adopt a SmartSwitch~\cite{tofino, programmableswitch} in MRDP to optimize \texttt{AllReduce} collectives by aggregating gradients in the SmartSwitch, thus reducing required network traffic and communication time, as shown in Figure~\ref{fig_overview_switchml}.
However, introducing SmartSwitch further to MSDP can barely reduce the required communication time, as both \texttt{AllGather} (AG) and \texttt{ReduceScatter} (RS) collectives used by MSDP can not be ideally optimized. For AG, SmartSwitch can only reduce the worker sending traffic, while the receiving traffic remains the same and thus bound the overall communication time. Similarly, SmartSwitch can only reduce the worker receiving traffic in RS, while the sending traffic remains the same and thus bound the overall communication time. 
Existing collective libraries like NCCL cannot efficiently overlap AG and RS due to network bandwidth contention on NIC, thus developers can not run AG and RS concurrently to combine the benefits of SmartSwitch-enabled AG and RS to fully exploit duplex network bandwidth, thus reducing communication time. 

\noindent\textbf{Model-in-Server DP (MiSDP).}
Prior works~\cite{ps, psnips, phub} adopt parameter servers (PSs) to minimize network traffic. Each worker only needs to \texttt{push} its gradients to PS and \texttt{pull} parameters from PS, while MiSDP needs PS to aggregate gradients to update parameters. As such, MiSDP reduces one-third of the collective traffic compared to MSDP. 
However, MiSDP on the one hand requires a large number of additional CPU machines (PSs) to provide adequate CPU compute power and memory bandwidth for heavy gradient aggregation and Adam optimizer. In our experiments, 16 GPU workers that run colocated PS processes require at least 13 additional CPU machines as non-colocated PSs to achieve line rate (Figure \ref{fig_motivation_ps}). On the other hand, MiSDP would still suffer from interference between GPU-centric network communication and computation, which leads to low MFU during training.
 

\noindent\textbf{Our Design: \SystemName{}.}
\
Inspired by DeepSeek~\cite{deepseek-v3, deepseekv3_isca2025} that argues to use a dedicated network processor for collectives that originally run on GPUs, we propose \SystemName{}, a \underline{F}ully \underline{D}isaggregated \underline{D}istributed \underline{D}ata \underline{P}arallelism that first disaggregates compute, network (i.e., collectives), and storage (i.e., optimizer) for MSDP to maximize the GPU's utilization in Figure~\ref{fig_overview_ours}. 
The key idea of \SystemName{} is twofold. First, it offloads collectives to SmartNICs to avoid interference between GEMM kernels and collective kernels (network disaggregation). Second, it offloads the optimizer to a scalable PS where a single PS can serve any number of workers (storage disaggregation). As such, GPUs focus on computation for which GPUs are originally designed, and thus maximizing their utilization, as Table~\ref{table_system_comparison} concludes.
To do so, \SystemName{} consists of three innovations: 


\setlist{leftmargin=1em}
\begin{itemize}
    \vspace{-\topsep}
    \setlength{\itemsep}{0pt}
    \setlength{\parsep}{0pt}
    \setlength{\parskip}{0pt}

    \item\textbf{SmartNIC-Managed Interference-Free Collectives.} We propose the first SmartNIC-managed collectives that disaggregate GEMM kernels and collective kernels to completely avoid interference between them. 

    \item\textbf{SmartSwitch-Assisted Many-to-One Reliable Protocol.} To address the incast problem introduced by serving many workers with a few PSs, \SystemName{} uses a SmartSwitch to aggregate gradients from workers to the PS and broadcast parameters reversely. However, each worker has to maintain a reliable connection with the PS, and the number of extra reliability-related packets increases with the number of workers. As the number of workers increases, reliability-related packets soon exhaust the PS network IOPS and degrade the overall network performance. 
    To this end, we propose the SmartSwitch-assisted reliable protocol, which leverages SmartSwitch to reduce the required reliability-related packets. 

    \item\textbf{Step-Centric Optimizer Pipelining. }Unlike traditional MiSDP that features a large number of PSs, \SystemName{} requires only a single PS to perform optimizer computation for massive workers. The challenge is that traditional layer-centric pipelining would soon exhaust all available CPU threads on the PS, and thus fails to consume line-rate aggregated gradients. Therefore, we propose a step-centric pipelining on the PS to maximize CPU thread usage, so as to consume the aggregated gradients at line rate. 
    \vspace{-\topsep}
\end{itemize}

We prototype the SmartNICs with Xilinx U50 FPGA boards. We evaluate \SystemName{} on eight 1$\times$A100 GPU machines and 1 CPU machine under a 100Gbps network. 
Experiment results show that \SystemName{} with 100Gb/s interference-free in-network collectives achieves 
1.17$\times$ higher throughput than ZeRO-Offload on the DGX machines (intra-machine GPUs are fully connected with 600GB/s NVLink), while keeping only 60\% capex cost of the DGX machines. 

\section{Background and Motivation}
\label{sec_motivation}

\subsection{Issues of Model-Sharded DP (MSDP)}
\label{subsec_motivation_msdp}

\begin{figure}[t]
    \begin{center}
        \subfigure[\label{fig_motivation_zero_interference} Execution time breakdown in a training iteration.]{
            \includegraphics[width=0.412\linewidth]{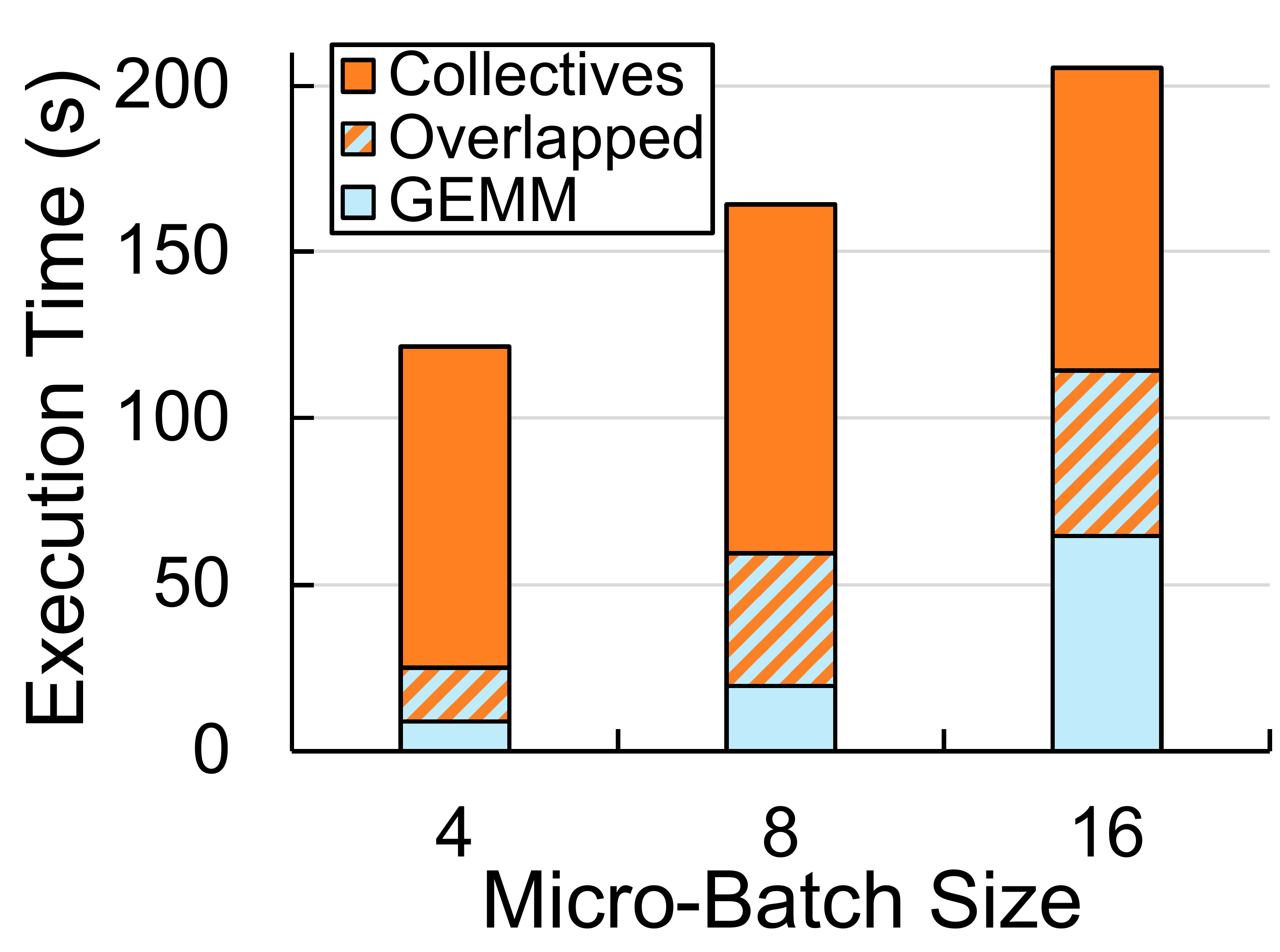}
        }
        \hspace{2mm}
        \subfigure[\label{fig_motivation_interference} Impact of concurrent GEMM to NCCL \texttt{AllReduce}.]{
            \includegraphics[width=0.388\linewidth]{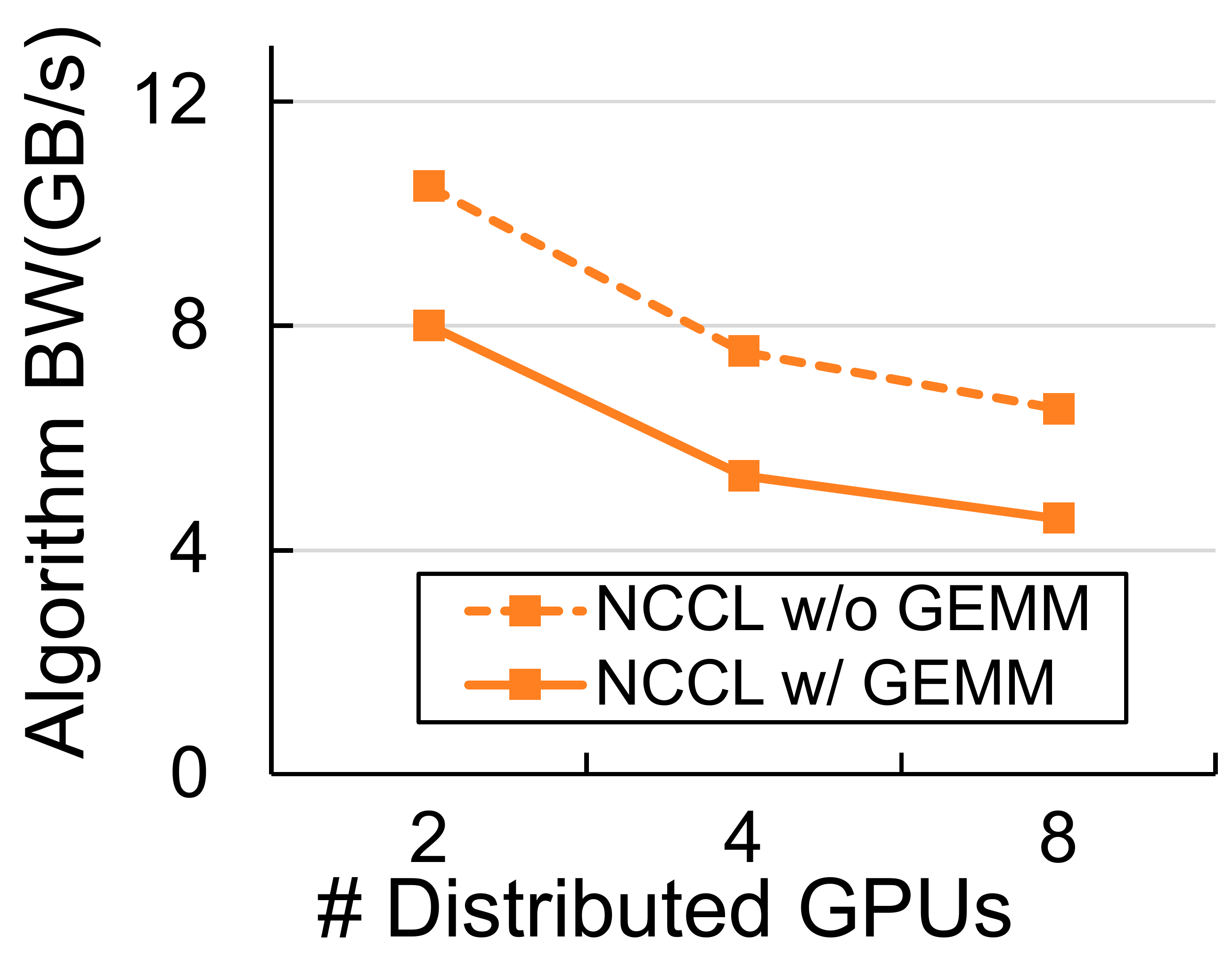}
        }
    \end{center}
    \vspace{-3ex}
    \caption{Issues of MSDP (ZeRO-Infinity). }
    \label{fig_motivation_zero}
\end{figure}

MSDP relies on two optimizations for memory-efficient training: 1)~it evenly shards model states and optimizer execution across all workers to aggregate the memory of multiple GPUs, and 2)~it offloads the optimizer to CPU memory or SSDs to further reduce the GPU memory capacity requirement. 
Compared to MRDP which performs one \texttt{AllReduce} to exchange gradients for each model layer, MSDP requires one RS and two AG collectives for each layer during forward and backward propagation, thus causing heavy collective traffic, e.g., the collectives take 1.25$\times$\textasciitilde 4.5$\times$ more time over GPU computation (Figure~\ref{fig_motivation_zero_interference}). 
Current MSDP systems employ collective libraries such as NCCL~\cite{nccl} to execute RS and AG on GPUs. These systems try to pipeline collectives and model computation (i.e., GEMM), assuming that the two processes utilize distinct hardware resources, thereby enabling their overlapping. 
However, MSDP's aggregated computation and collectives on GPU cause severe interference, mostly serial execution, thus leading to low GPU utilization, e.g., 15\% MFU on eight 1-GPU machines (Figure~\ref{fig_exp_thpt_175b}).

\begin{figure}[t]
    \begin{center}
        \subfigure[\label{fig_motivation_pattern_msdp}MSDP (w/ and w/o SmartSwitch): 1) A GEMM kernel that occupies all SMs blocks subsequent collective kernels, and 2) AG and RS kernels are on the same GPU stream, thus cannot run concurrently.]{
            \includegraphics[width=0.9\linewidth]{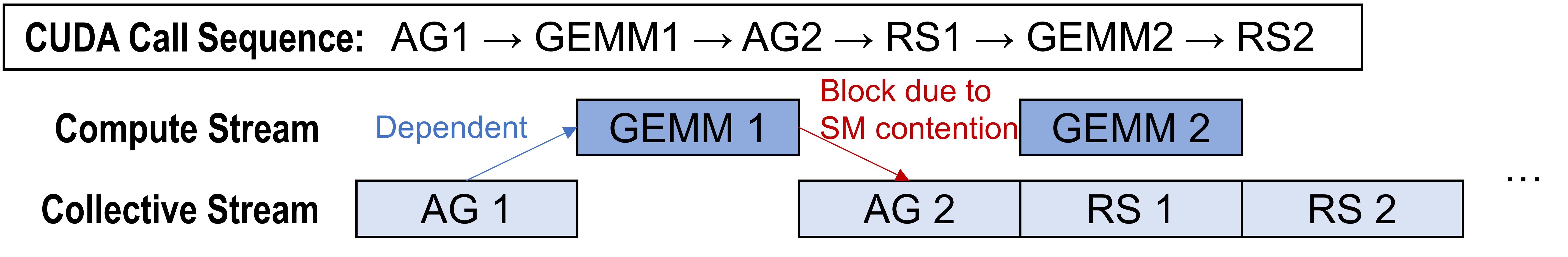}
        }
        \vspace{-1ex}
        \\
        \subfigure[\label{fig_motivation_pattern_mindp}\SystemName{}: 1) SmartNIC-managed collectives do not interfere with GEMM, and 2) contention-free primitives enable concurrent \texttt{push} (for partial gradients) and \texttt{pull} (for on-demand new parameters) operations.]{ 
            \includegraphics[width=0.9\linewidth]{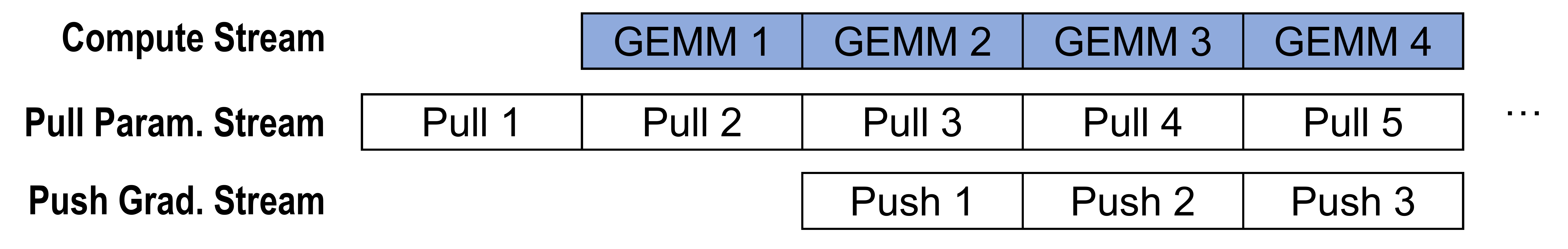}
        }
        \vspace{-2ex}
    \end{center}
    \caption{\label{fig_motivation_pattern} Computing and collective pattern during the backward stage: MSDP vs. \SystemName{}. A darker color indicates more SMs being used, and white indicates no SMs being used.}
    \vspace{-2ex}
\end{figure}

Even if we enabled to run GEMM and collective kernels concurrently, it still suffers from severe interference from the following two main issues: 

\noindent\textbf{1, GPU Computing Unit Contention.}
\
GEMM and collective kernels compete for GPU computing units, i.e., streaming multiprocessors (SMs). Collective kernels only require a small portion of SMs (a few percent to 10\%)~\cite{resccl_sigcomm25}, and intuitively, we think it is easy to overlap the GEMM and collective kernels to achieve a high GPU SM utilization. However, due to GPUs' non-preemptive scheduling policy, a compute-intensive GEMM kernel that occupies all SMs will block subsequent collectives from launching~\cite{olmedo2020dissecting}, even though a collective kernel has the highest priority. Figure~\ref{fig_motivation_pattern_msdp} shows an example where ``GEMM 1'' blocks ``AG 2'' from launching. After ``GEMM 1'' finishes and ``AG 2'' starts, ``GEMM 2'' cannot be launched because it depends on the completion of ``AG 2''. Consequently, a collective kernel cannot fully overlap with GEMM kernels. 

To illustrate the impact of this interference, we break down the execution time for training OPT-175B on 8$\times$ 1-GPU machines under a 100Gbps network using ZeRO-Infinity, as shown in Figure~\ref{fig_motivation_zero_interference}. 
Up to 65\% of the collective execution time cannot overlap with GEMM at the batch size of 16, incuring 41\% more execution time than ideal full overlapping.

\begin{figure}[t]
    \vspace{-4ex}
    \begin{center}
        \subfigure[Impact of concurrent DMA to GEMM execution time.]{
            \includegraphics[width=0.42\linewidth]{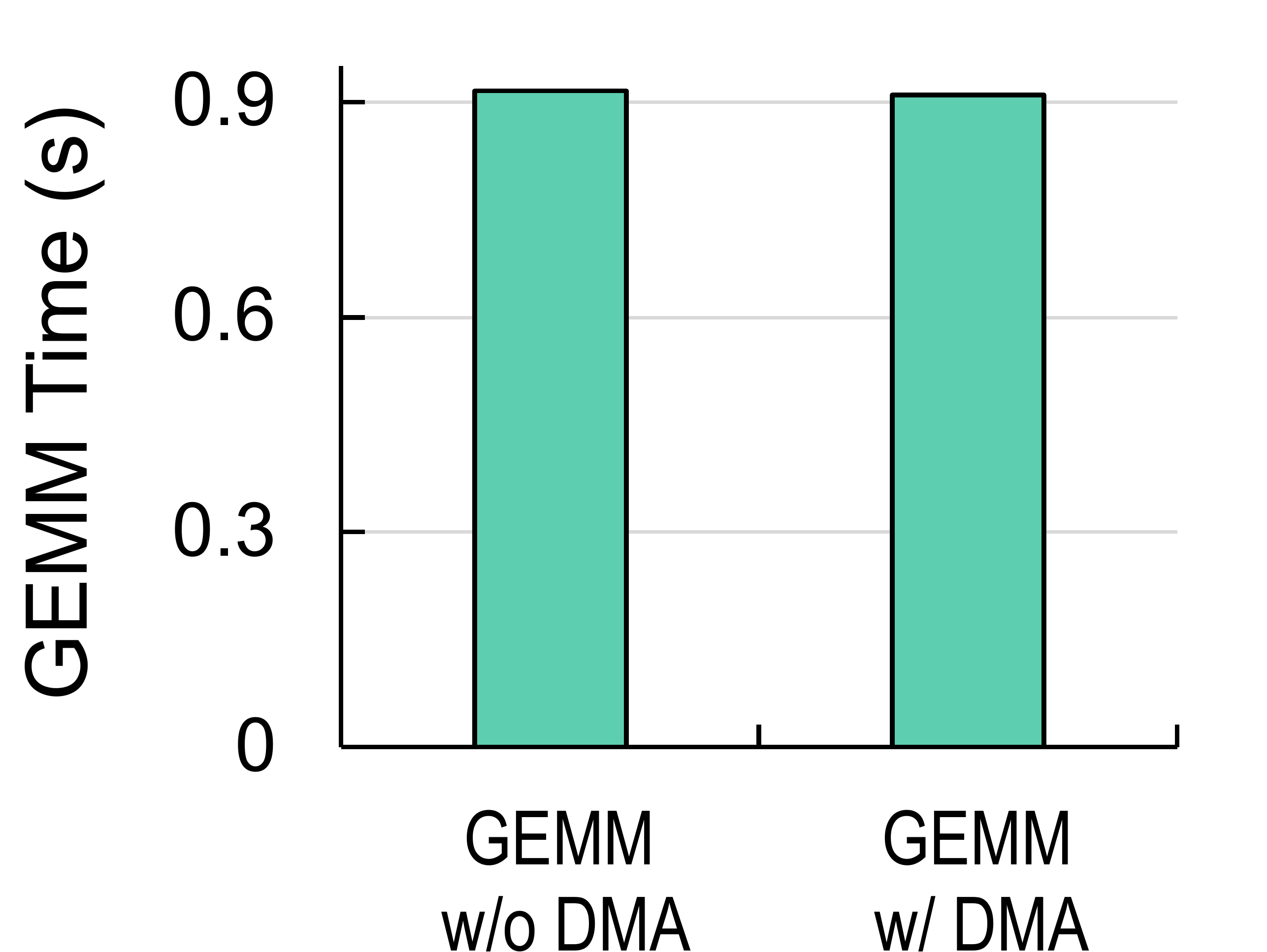}
        }
        \hspace{2mm}
        \subfigure[Impact of concurrent GEMM to GPU DMA bandwidth.]{
            \includegraphics[width=0.42\linewidth]{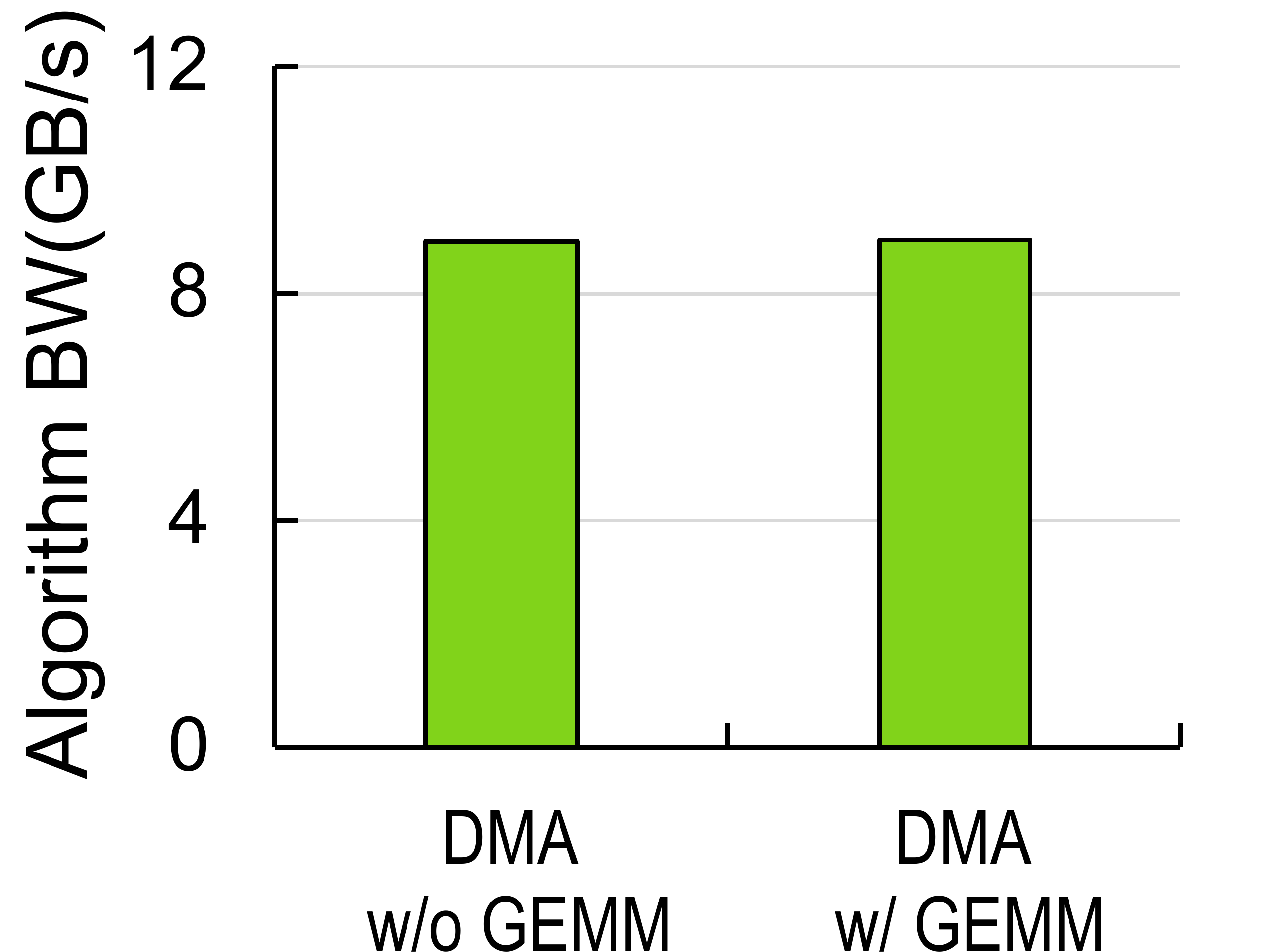}
        }
    \end{center}
    \vspace{-2ex}
    \caption{There is barely interference between GPU GEMM kernel and an external device accessing GPU memory. }
    \label{fig_motivation_interference_concurrent}
    \vspace{-2ex}
\end{figure}

\noindent\textbf{2, GPU Memory Bandwidth Contention. }GEMM and collective kernels compete for memory bandwidth and L2 cache, leading to performance degradation of the collective kernels~\cite{ace, olmedo2020dissecting}. Typically, we use algorithm bandwidth to characterize the collective performance, which is defined as total data size $S$ in a collective operation divided by collective execution time $t$, i.e., algorithm bandwidth $BW_{\rm alg}=\frac{S}{t}$. To show the algorithm bandwidth drop due to contention, we run independent GEMM and NCCL \texttt{AllReduce} kernels on distributed A100 GPUs under 100Gbps network and a typical configuration when training the OPT-175B model, and use CUDA multi-process service (MPS) to statically isolate SMs to avoid the influence of competing SMs.\footnote{Even though MPS is able to partition SMs, we can not adopt it to eliminate the contention for GPU computing units in actual training because MPS only allow partitioning SMs statically, and thus is incompatible with important features like dynamic parallelism and CUDA graph~\cite{mps} that LLM training frameworks heavily rely on. }
Figure~\ref{fig_motivation_interference} shows that the GEMM kernel causes a 30\% algorithm bandwidth drop to the concurrent NCCL kernel. 

\noindent{\bf Finding.} Although the concurrently running GEMM and collectives kernels incur interference regarding SMs and GPU memory subsystem, we observe that there is barely interference when the GEMM kernel is running and the external device is reading/writing GPU memory through the PCIe. To illustrate this, we 1) measure the GEMM execution time with and without a PCIe device (e.g., an FPGA-based NIC) performing DMA requests to read/write GPU memory concurrently; and 2) measure the read/write speed of PCIe devices accessing GPU memory through DMA with/without concurrently running GEMM kernels. 
Figure~\ref{fig_motivation_interference_concurrent} indicates there is barely interference between a GEMM kernel and a DMA task of an external PCIe device. Such a finding motivates us to offload the entire collective stack on SmartNIC (i.e., network and compute disaggregation) to avoid interference.

\subsection{Issues of Smart\-Switch-Enhanced MSDP}
\label{subsec_motivation_atp}

To relieve the issue of MSDP, a possible way is to implement in-network computing primitives in SmartSwitch~\cite{tofino, inc_hotnets19, in_network_computing_eurosys19, programmableswitch, p4v_sigcomm2018, yue_isscc2024, parserhawk_sigcomm25} to reduce the collective traffic via in-switch aggregation and broadcasting. 
However, current SmartSwitch can only provide \texttt{AllReduce} collective needed by MRDP, while MSDP relies on AG and RS. Even if we enabled SmartSwitch to run AG and RS, we still identify that the na\"ive solution has the following two issues.

\noindent\textbf{1, Unchanged Collective Time for Separate AG and RS.}
\
Considering $N$ workers performing an AG on a total data size $S$. Without a SmartSwitch, each worker sends its $\frac{S}{N}$ data to every other worker and receives different $\frac{S}{N}$ data from every other worker~\cite{allreduce}. As such, each worker has to send (or receive) $\frac{(N-1)S}{N}$ network traffic. 

If we enabled SmartSwitch to support AG and RS, 
for AG, SmartSwitch can help each worker to broadcast its $\frac{S}{N}$ to every other worker, thus each worker only has to send $\frac{S}{N}$ network traffic, while the receiving traffic remains at $\frac{(N-1)S}{N}$. 
Similarly, for RS, SmartSwitch allows each worker to receive $\frac{S}{N}$ traffic, while the sending traffic remains at $\frac{(N-1)S}{N}$. 
Therefore, a SmartSwitch can not reduce the network communication time for separate AG and RS because the communication time is decided by the maximum of the sending traffic and the receiving traffic. 

\textbf{2, Compute-Collective Interference.}
\
If we enabled SmartSwitch to support AG and RS, they still rely on GPU-centric NCCL for data chunk management, as the existing SHARP-based in-switch aggregation stack~\cite{nccl_sharp} does. Consequently, they still suffer from computation-collective interference that is identified in Subsection~\ref{subsec_motivation_msdp}.



\subsection{Issues of Model-in-Server DP (MiSDP)}
\label{subsec_motivation_ps}

To relieve the issue of MSDP, another possible way is to adopt the model-in-server DP (MiSDP) architecture~\cite{ps, psnips, phub, byteps} that offloads the model states onto PSs. 
Each worker performs 1) a \texttt{push} operation to send its gradients to PSs, which can aggregate gradients from workers and update the corresponding optimizer states, and 2) two \texttt{pull} operations (one in forward stage and one in backward stage) to receive the latest parameters from PSs. Compared to two AG and one RS of MSDP for a layer ($\frac{3(N-1)S}{N}$ in each direction), MiSDP ($2S$ receive and $S$ send) reduces the network traffic by at most one-third. 
However, existing MiSDP systems push the entire model gradients to PS after GPU finishes computation and then pulls the entire model parameters, so that each GPU has to accommodate the full model parameters. Therefore, they fail to train 100B-scale model that cannot fit in GPU memory. Even if MiSDP were enabled to train large models, we identify that MiSDP solutions still suffer from two issues. 

\noindent\textbf{1, Computation-Collective Interference.}
\
To enable large-model training, MiSDP systems need to use GPU-centric network stacks such as IBGDA (InfiniBand with GPUDirect Async) to enable GPU to directly interact with NICs for higher network bandwidth. Therefore, they would still suffer from interference between GPU computation kernels and collective kernels, as demonstrated in Subsection~\ref{subsec_motivation_msdp}.

\begin{figure}[t]
    \vspace{-1ex}
    \centering
    \begin{minipage}{\linewidth}
        \centering
        \includegraphics[width=0.43\linewidth]{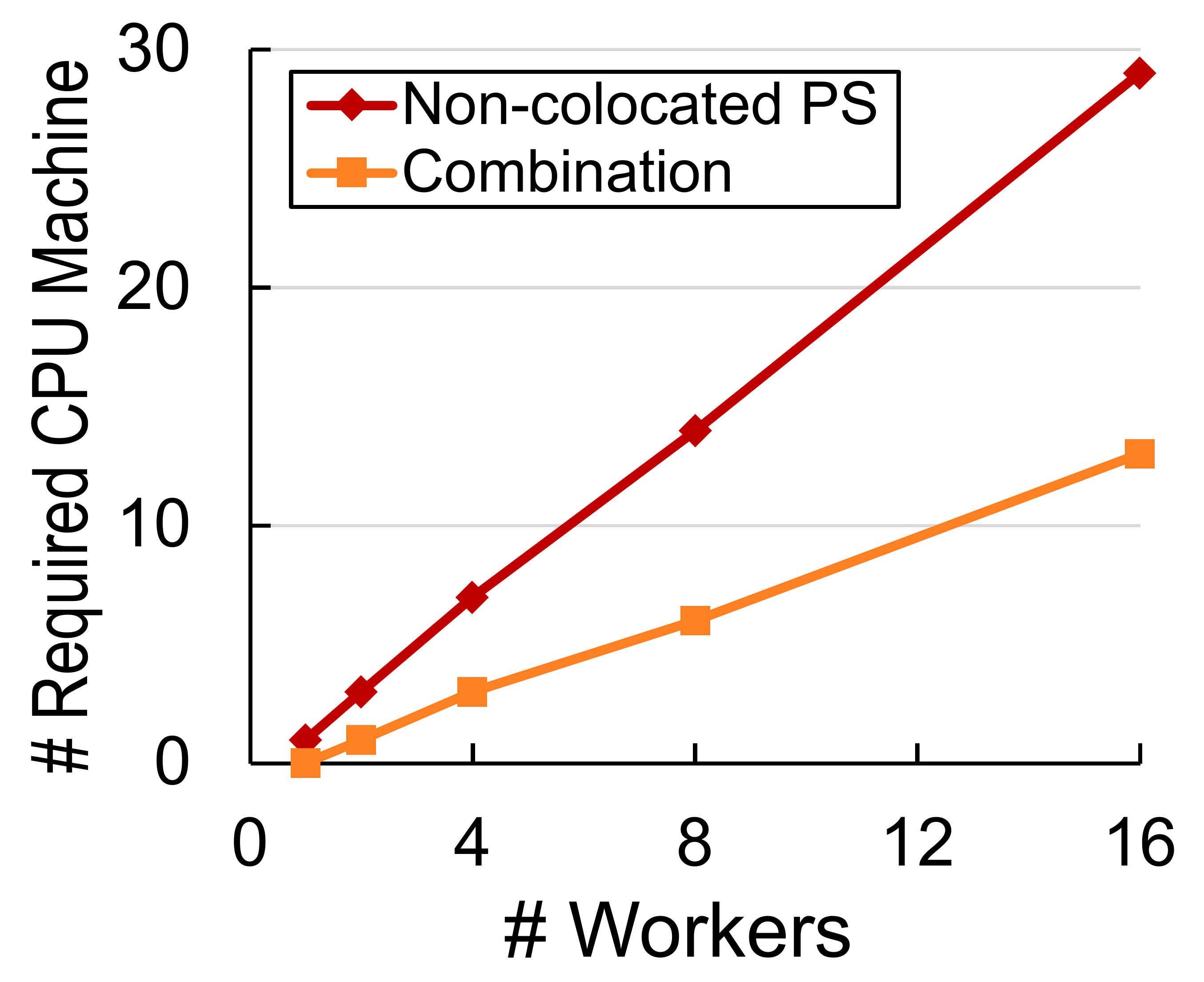}
        \vspace{-2ex}
        \caption{\label{fig_motivation_ps}Overhead of MiSDP: Many extra CPU machines.}
        \vspace{-3ex}
    \end{minipage} 
\end{figure}

\noindent\textbf{2, Escalating Hardware Requirement to Aggregate Partial Gradients.}
\
MiSDP requires PS to provide adequate CPU compute power, memory bandwidth, and network bandwidth to perform optimizer, gradient aggregation, and parameter broadcasts. Therefore, a growing number of workers would require many additional CPU machines as PSs to consume their 100Gbps-per-worker partial gradients and produce 100Gbps-per-worker parameters, which is not always acceptable due to their monetary and space costs. To illustrate the machine requirement, we simulate the minimal machine requirement to saturate 100Gbps with different numbers of 1-GPU worker machines, where each GPU machine and each CPU machine has 1$\times$100Gbps NIC. Each server runs a multi-threaded SIMD-optimized loop-unrolled Adam optimizer proposed by ZeRO-Offload~\cite{zero-offload}. We simulate on both non-colocated PS (PS processes only run on additional CPU machines) and a combination of colocated PS (PS processes run on GPU machines' spare CPUs) and non-colocated PS. Figure~\ref{fig_motivation_ps} shows that a 16-worker cluster requires 29 additional CPU machines in non-colocated PS configuration and 13 machines in a combined colocated and non-colocated PS configuration to achieve line-rate throughput. 

\noindent\textbf{Finding.}
\
We observe from Figure~\ref{fig_motivation_ps} that the scalability inefficiency of PS comes from more pressure on the memory subsystem of the server due to consuming more partial gradients from an increasing number of workers. This motivates us to offload gradient aggregation and parameter broadcasts to SmartSwitch, which produces reliable aggregated gradients to enable only one CPU machine (storage disaggregation) to accommodate 100B Adam states.\footnote{ATP~\cite{atp} only offloads in-network aggregation primitive to SmartSwitch to 1) provide an \texttt{AllReduce} primitive, to aggregate gradients to support small model training, rather than LLM; and 2) updates its optimizer states on GPU workers. Thus, ATP suffers from interference from its partial disaggregation. } Meanwhile, a SmartNIC would disaggregate the network and compute to eliminate the interference issue. This motivates us to use SmartNIC-SmartSwitch co-optimization to fully disaggregate compute, network, and storage, so as to benefit from reduced network traffic of MiSDP while addressing scalability and interference issues, thus achieving high MFU on scalable FSDP training.

\begin{figure}[t]
    \vspace{-1ex}
    \begin{center}
        \includegraphics[width=\linewidth]{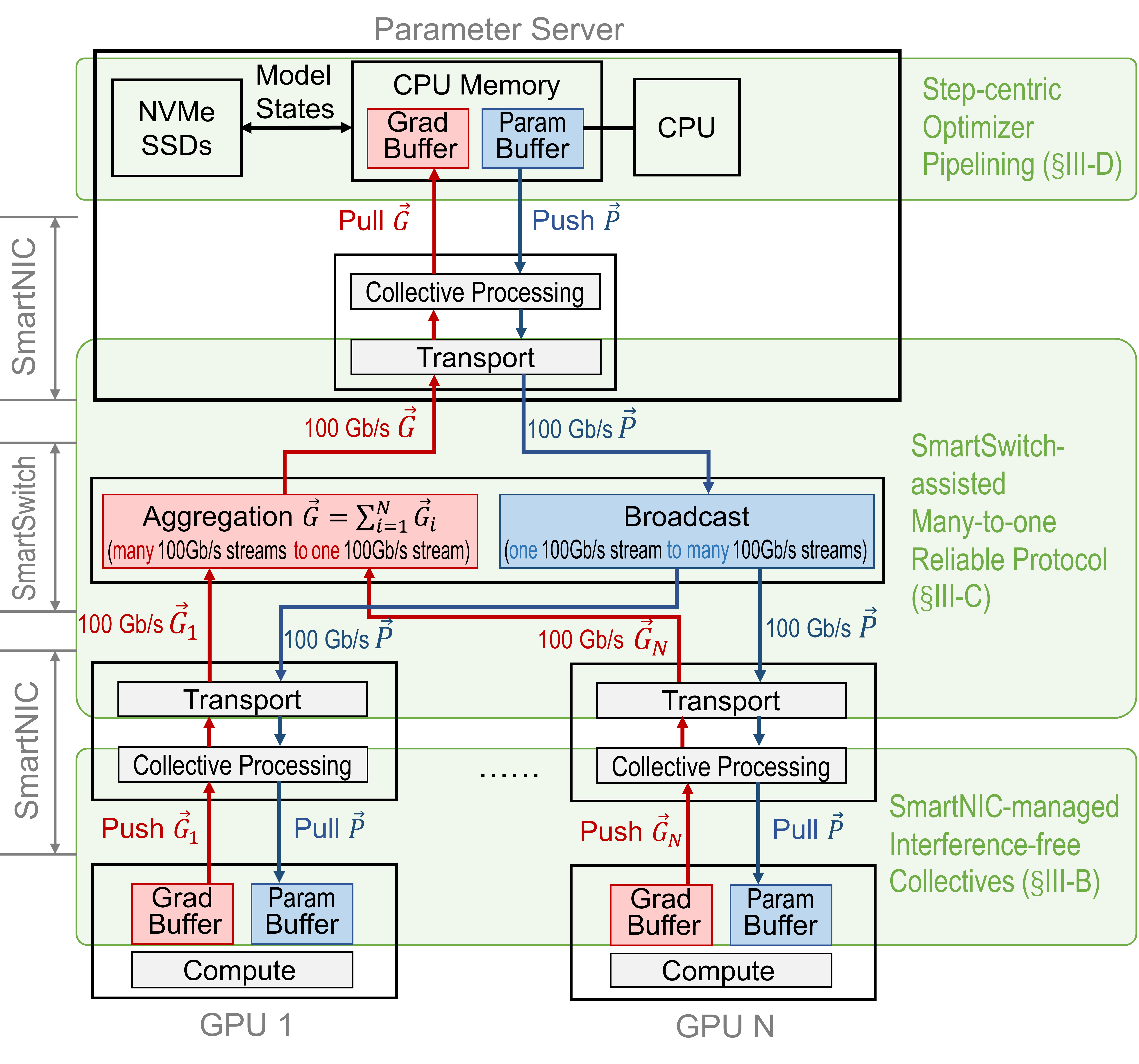}
    \end{center}
    \vspace{-3ex}
    \caption{\label{fig_system_overview} System overview of \SystemName{}. \SystemName{} offloads collectives and the optimizer to dedicated hardware, such that GPUs only focus on the computing part to maximize their utilization.}
    \vspace{-2ex}
\end{figure}

\section{Design and Implementation of \SystemName}
\label{sec_design}
\subsection{\SystemName{} Overview}

Inspired by the DeepSeek~\cite{deepseek-v3, deepseek_infra_sc24, deepseekv3_isca2025} that argues to use a dedicated network processor for collectives, we propose \SystemName{} that first fully disaggregates compute, network (i.e., collectives), and storage (i.e., optimizer) to maximize the GPU's utilization in Figure~\ref{fig_overview_ours}.
The key idea of \SystemName{} is twofold. First, it offloads collectives to SmartNICs and SmartSwitch to avoid interference between GEMM kernels and collective kernels (network disaggregation). Second, it offloads the optimizer to a scalable PS that can serve any number of workers (storage disaggregation). \SystemName{} leverages SmartSwitch to 1) aggregate line-rate partial gradients from GPU workers to the PS that maintains 100B optimizer states, and 2) broadcast the on-demand parameters reversely. As such, GPUs focus on computation for which GPUs are originally designed, and thus maximize their utilization. 
However, building \SystemName{} is not trivial, mainly due to three main challenges.

\noindent{\bf C1: Offloading collectives to SoC-based SmartNICs cannot saturate network line rate.} A possible solution to avoid SM contention is to offload collectives to SoC-based SmartNICs like BlueField~\cite{bluefield}. However, an SoC-based SmartNIC fails to process line-rate packets due to its internal switch link and Arm memory bandwidth bottlenecks.


\noindent{\bf C2: Existing reliable protocols would soon exhaust the asymmetrical PS network IOPS.} In \SystemName{}, the single PS has to maintain reliable connections with dozens of workers. Although the gradients/parameter packets can be aggregated/broadcast by the SmartSwitch, other packets to maintain reliable connections would exhaust the limited network IOPS of the PS, resulting in network throughput degradation.

\noindent{\bf C3: The optimizer computing would soon exhaust the PS computing power.} Optimizer on the PS needs to perform Adam on the line-rate aggregated gradients and serve line-rate parameters to GPU workers. \SystemName{} exploits SmartSwitch to reduce the number of parameter servers, which results in computing power contention on the PS. Traditional layer-centric pipelining of the optimizer would suffer from the limited CPU threads and lead to degraded serving performance. 


\noindent\textbf{Overall Architecture.}
\SystemName{} consists of a series of hardware-software co-designs, as shown in Figure~\ref{fig_system_overview}. The hardware part mainly consists of three components: 1) a per-GPU FPGA-based SmartNIC that provides an interference-free collective library for worker GPUs; 2) a SmartSwitch that aggregates gradients, broadcasts parameters, and maintains many-to-one reliable connections; 3) a single non-colocated PS that performs Adam on line-rate aggregated gradients and serves line-rate parameters. 


\subsection{SmartNIC-Managed Interference-Free Collectives}
\label{subsec_network}
To address \textbf{C1}, we propose SmartNIC-managed collectives that perform the entire collectives on FPGA-based SmartNICs, while GPUs only need to perform the model's forward and backward computation. Each SmartNIC is paired with a GPU, as shown in Figure~\ref{fig_system_overview}. 

\begin{table}[t]
    \centering\footnotesize
    \caption{Core software APIs of \SystemName{}.}
    \label{table_nic_api}
    \vspace{-2ex}
    \resizebox{\columnwidth}{!}{
        \begin{tabular}{ll}
            \specialrule{.1em}{.05em}{.05em} 
            \makecell[c]{\textbf{API}} & \makecell[c]{\textbf{Description}} \\ 
            \specialrule{.1em}{.05em}{.05em} 
            \narrow{\texttt{handle\_t push(void* buf, size\_t len)}} & \makecell[l]{Issue a contention-free push \\ request to the SmartNIC.} \\ \hline
            \narrow{\texttt{handle\_t pull(void* buf, size\_t len)}} & \makecell[l]{Issue a contention-free pull \\ request to the SmartNIC.} \\ \hline
            \narrow{\texttt{void wait(handle\_t request)}} & \makecell[l]{Block CPU/GPU execution until \\  completion of a specific request.}\\
            \specialrule{.1em}{.05em}{.05em} 
        \end{tabular}
    }
    \vspace{-2ex}
\end{table}

\noindent\textbf{Software API.}
\
\SystemName{} provides a two-sided software API for both workers and the PS, so as to enable the PS to prefetch parameters, thus reducing the latency of workers pulling parameters from the PS. Table~\ref{table_nic_api} lists the core APIs. During initialization, workers and the PS's CPU invoke the \texttt{register\_buf} function to enable direct data movement between the application buffer and SmartNICs. 
At runtime, \SystemName{} provides \textit{contention-free} \texttt{push} and \texttt{pull} to workers and the PS to enable concurrent \texttt{push} and \texttt{pull} calls to fully exploit the network bandwidth, as shown in Figure~\ref{fig_motivation_pattern_mindp}. 
Each worker invokes a \texttt{push} request to the SmartNIC to send partial gradients of a whole model layer from GPU memory, meanwhile, the PS invokes a \texttt{pull} request to the SmartNIC to receive gradients aggregated by Smart\-Switch to CPU memory. 
Similarly, the PS issues a \texttt{push} request to send parameters of a whole layer from CPU memory, and each worker issues a \texttt{pull} request to receive parameters broadcast by Smart\-Switch to GPU memory. Both \texttt{push} and \texttt{pull} return a handle, so workers and the PS can wait until the completion of the request via the \texttt{wait} function with the handle.


\subsubsection{Na\"ive Solution: SoC-Based SmartNIC-Managed Collective Library}
\label{subsec_design_bluefield}
To avoid interference, a straightforward solution is to offload collectives to SoC-based SmartNICs~\cite{suresh2023novel, graham2024optimizing, oodsmpi_sc25, khalilov2024network} (e.g., Nvidia BlueField-3~\cite{bluefield, intel_ipu_hcs2025}), which allows C/C++ software-programming. However, existing off-the-shelf SmartNICs suffer from two severe issues, which are also reported by prior work~\cite{smartns_eurosys26}.

\begin{figure}[t]
    \centering
    \begin{minipage}{0.459\linewidth}
        \begin{center}
            \includegraphics[width=\linewidth]{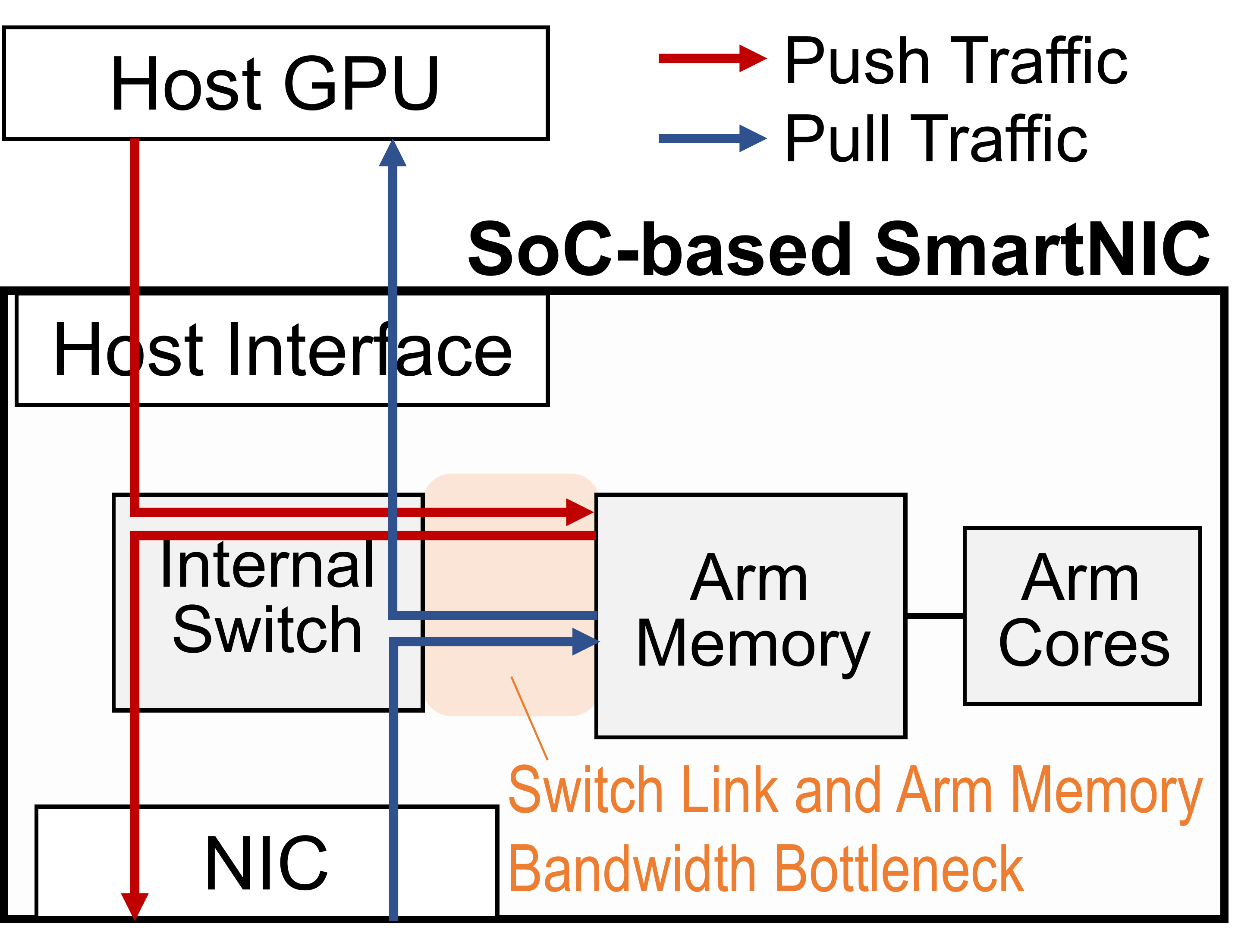}
        \end{center}
        \vspace{-2ex}
        \caption{\label{fig_soc_smartnic}\texttt{Push/pull} traffic insider a worker-side SoC-based SmartNIC.}
        \vspace{-4ex}
    \end{minipage}
    \hfill
    \begin{minipage}{0.484\linewidth}
        \begin{center}
            \includegraphics[width=\linewidth]{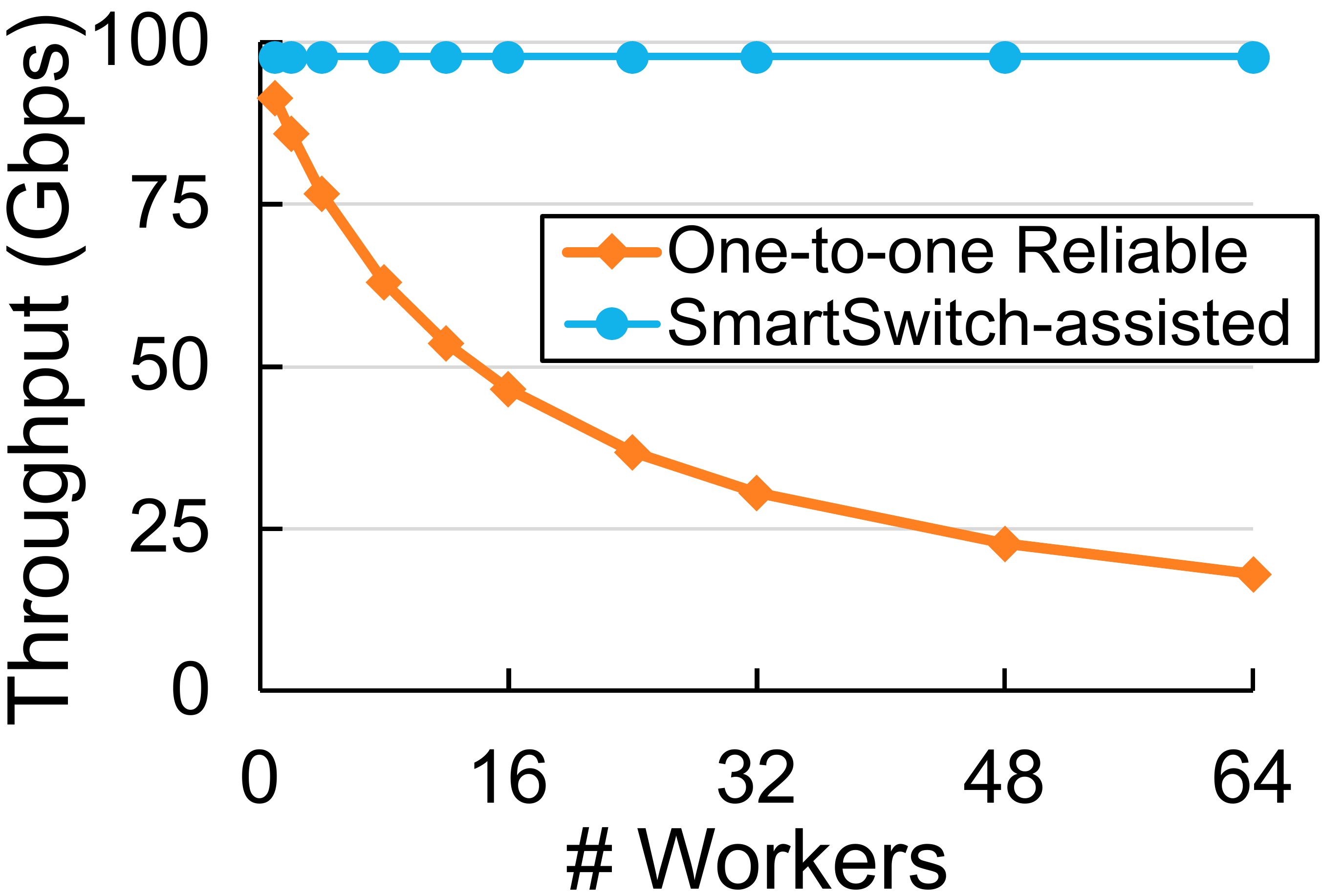}
        \end{center}
        \vspace{-1ex}
        \caption{\label{fig_ack_overhead}Throughput under different reliable protocols.}
        \vspace{-4ex}
    \end{minipage}
\end{figure}

\noindent\textbf{Issue 1: Off-path SmartNIC's Internal Switch Link Contention.}
\
Off-path NICs dominate the data center NIC market due to their ease of integration and full operating system support. In an off-path SmartNIC, the host interface, NIC, and Arm are connected by an internal switch, as Figure~\ref{fig_soc_smartnic} shows. The Arm core processes traffic between host and network in a lookaside manner. In the case of model training, both \texttt{push} and \texttt{pull} traffic first goes from the internal switch to Arm for processing, then from Arm back to the internal switch. However, this would result in severe Arm-Switch PCIe link contention when serving bi-directional network traffic at line-rate. Arm-switch bandwidth is usually only slightly higher than network bandwidth, where \texttt{push} and \texttt{pull} throughput would nearly be halved due to the contention. For example, BlueField-2's per-direction network bandwidth is 200Gbps, concurrent \texttt{push/pull} require 400Gbps per-direction Arm-Switch bandwidth to fully saturate the network bandwidth, while the actual per-direction Arm-Switch is only 250Gbps. 


\noindent\textbf{Issue 2: Arm Memory Bandwidth Bottleneck.}
\
Even if there were no switch link contention introduced by off-path architecture, \texttt{push/pull} performance would still be constrained by the SoC-based SmartNIC's Arm memory. As traffic is staged to Arm memory rather than being served by cache due to the large working set size (which matches \SystemName{}'s case), \texttt{push} traffic first goes from host interface to Arm memory, then from Arm memory to network, and \texttt{pull} traffic first goes from network to Arm memory, then from Arm memory to host interface, which incurs 2$\times$ memory access for per-direction traffic, as Figure~\ref{fig_soc_smartnic} shows.
However, an SoC-based SmartNIC lacks adequate memory bandwidth for this access pattern. For example, an off-the-shelf BlueField-2 SmartNIC requires 800Gbps memory bandwidth to serve 200Gbps packets, while it only provides 204.8Gbps theoretical memory bandwidth~\cite{bf2_spec}. As a result, we only achieve 20\% network link utilization when evaluating \texttt{push/pull} in a real BlueField-2 SmartNIC. This memory bandwidth constraint still holds for newer-generation BlueField-3, which requires 1600Gbps memory bandwidth to achieve 400Gbps line-rate throughput, while it only provides 716.8Gbps theoretical memory bandwidth~\cite{bf3_spec}. Due to the constraints of power, chip scaling, and form factor, this issue is unlikely to be resolved by near-future SoC-based SmartNICs~\cite{rambda_hpca2023, smartds_isca23}.

\subsubsection{Our Solution: FPGA-Based SmartNIC-Managed Collective Library}
\
To address the memory bandwidth issue, we propose the \textit{FPGA-based SmartNIC-managed collective library} that performs the collectives from GPU/CPU on FPGA-based SmartNICs~\cite{difference_fccm18, beyond_hpsr18, azure_nsdi18, kvdirect_sosp2017, lambdanic_icdcs2020, fcsn_fpl2022, styx_atc2023, lognic_micro2023, alkali_nsdi2025, albatross_sigcomm25, nezha_sigcomm25, enzian_asplos22, fpga_osdi20, panic_osdi2020, fcsn_fccm2022, hint_comarchletter2025, acclplus_osdi24}. The key insight is that an FPGA-based SmartNIC adopts an on-path architecture that processes packets in hardware pipelines rather than a lookaside SoC, which eliminates the internal switch bandwidth contention. Meanwhile, an FPGA-based SmartNIC allows explicitly storing packets between pipeline stages in on-chip SRAM instead of forcing packets to off-chip DRAM, which addresses the memory bandwidth constraints. Therefore, they can serve line-rate packets. To this end, we design a \textit{collective processing module} on each FPGA-based SmartNIC to process the collectives. In the following, we describe the detailed procedure for handling \texttt{push}/\texttt{pull} calls. 

\begin{figure}[!t]
    \vspace{-1ex}
    \begin{center}
        \subfigure[\label{subfig_nic_push} Pushing gradients]{
            \includegraphics[width=0.43\linewidth]{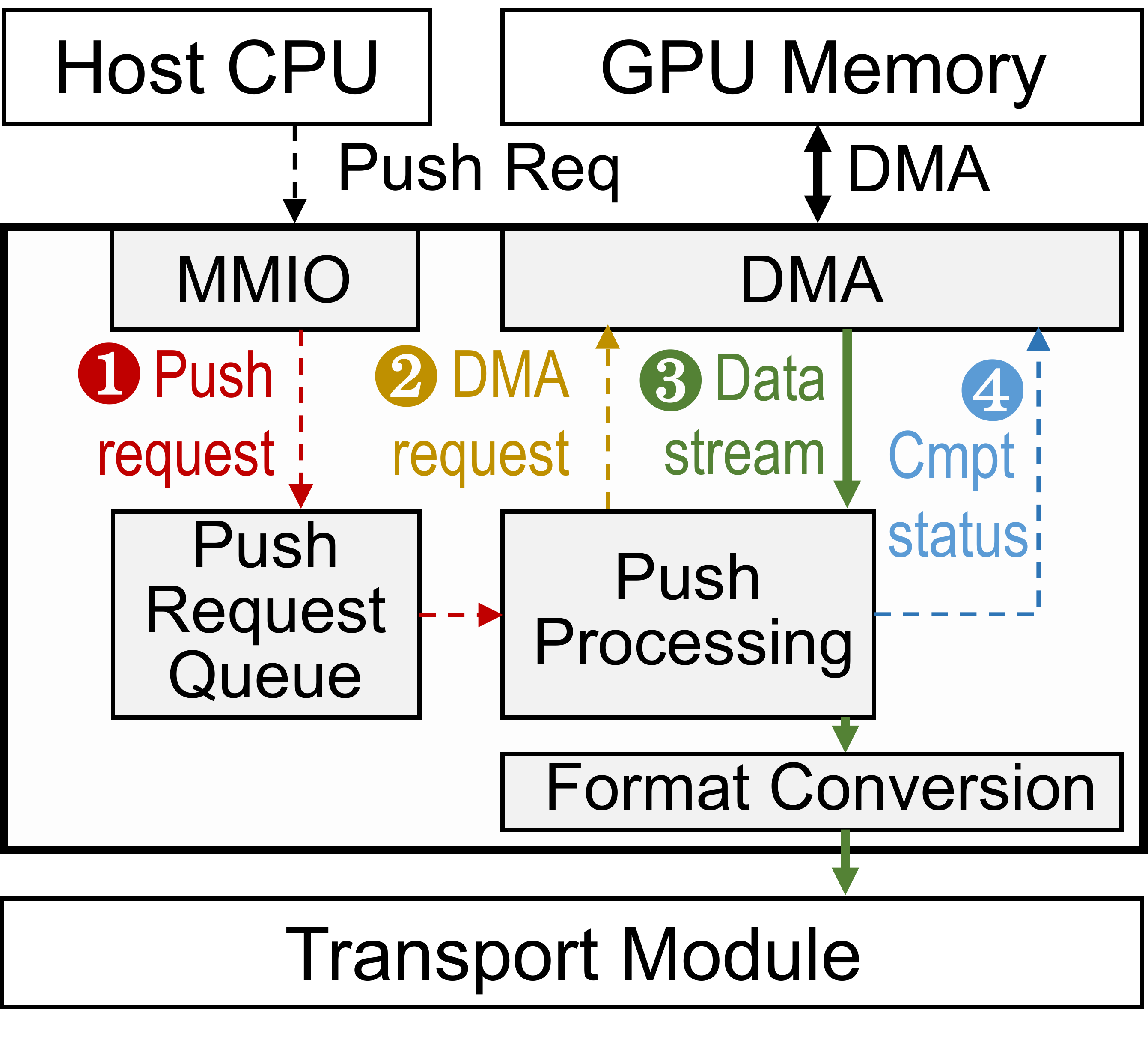}
        }
        \hfill
        \subfigure[\label{subfig_nic_pull} Pulling parameters]{
            \includegraphics[width=0.43\linewidth]{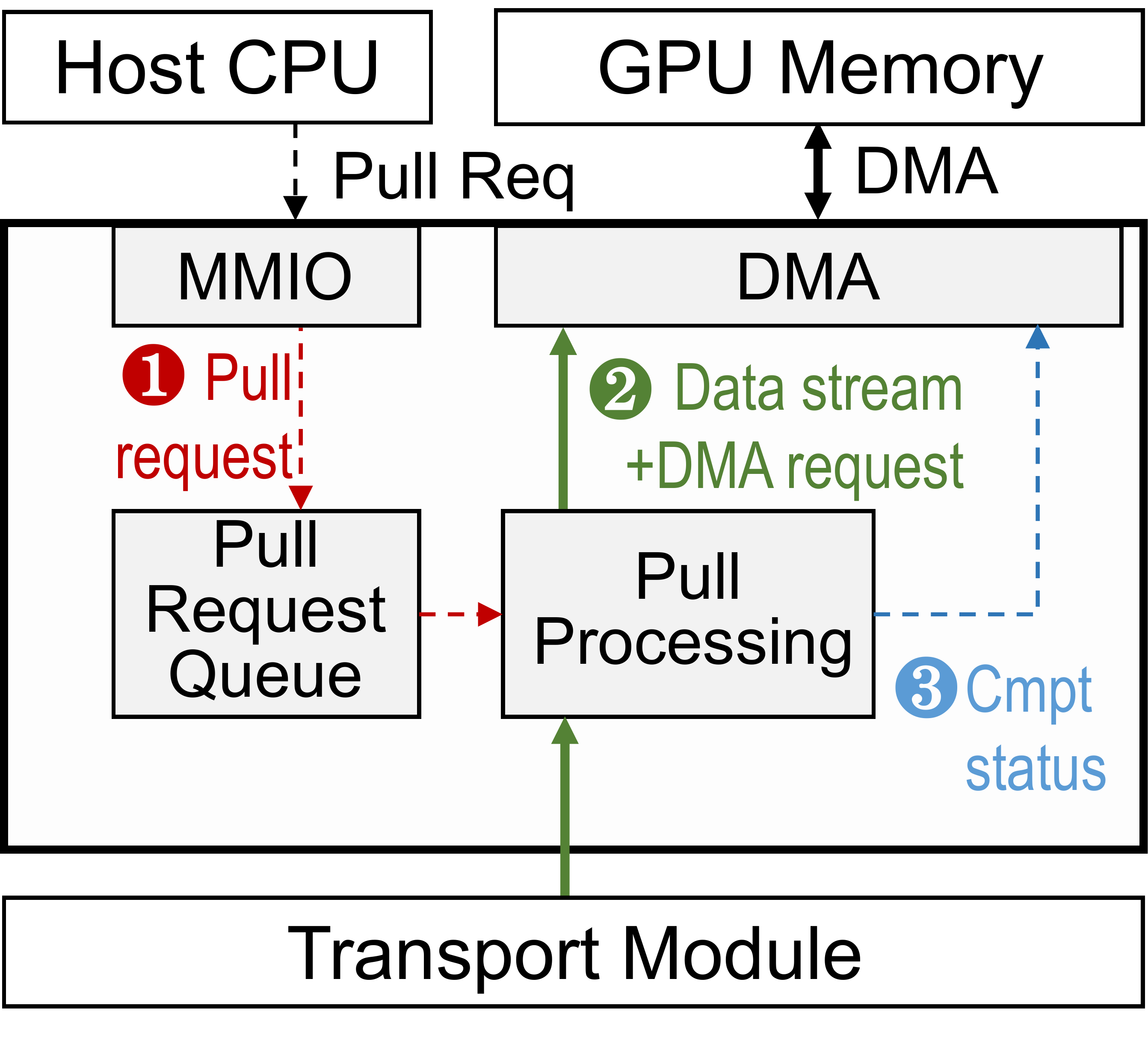}
        }
    \end{center}
    \vspace{-2ex}
    \caption{\label{fig_nic_datapath} Collective processing module of a worker.}
    \vspace{-4ex}
\end{figure}

\begin{figure*}[t]
    \vspace{-1ex}
    \centering
    \subfigure[\label{fig_reliable_param} PS broadcasts the parameters to all workers.]{
        \includegraphics[width=0.48\linewidth]{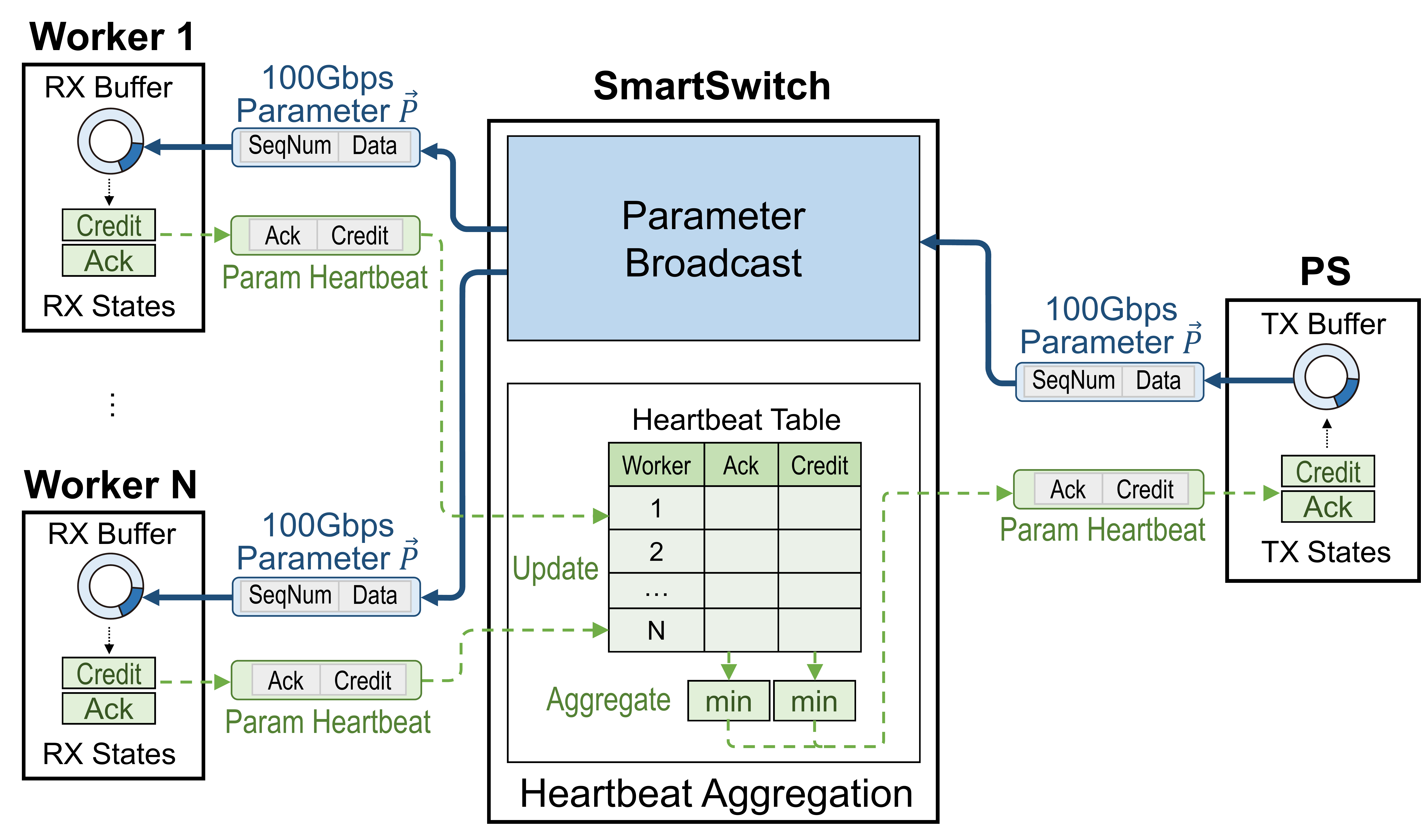}
    }
    \hfill
    \subfigure[\label{fig_reliable_grad} PS receives aggregated gradients from workers.]{
        \includegraphics[width=0.48\linewidth]{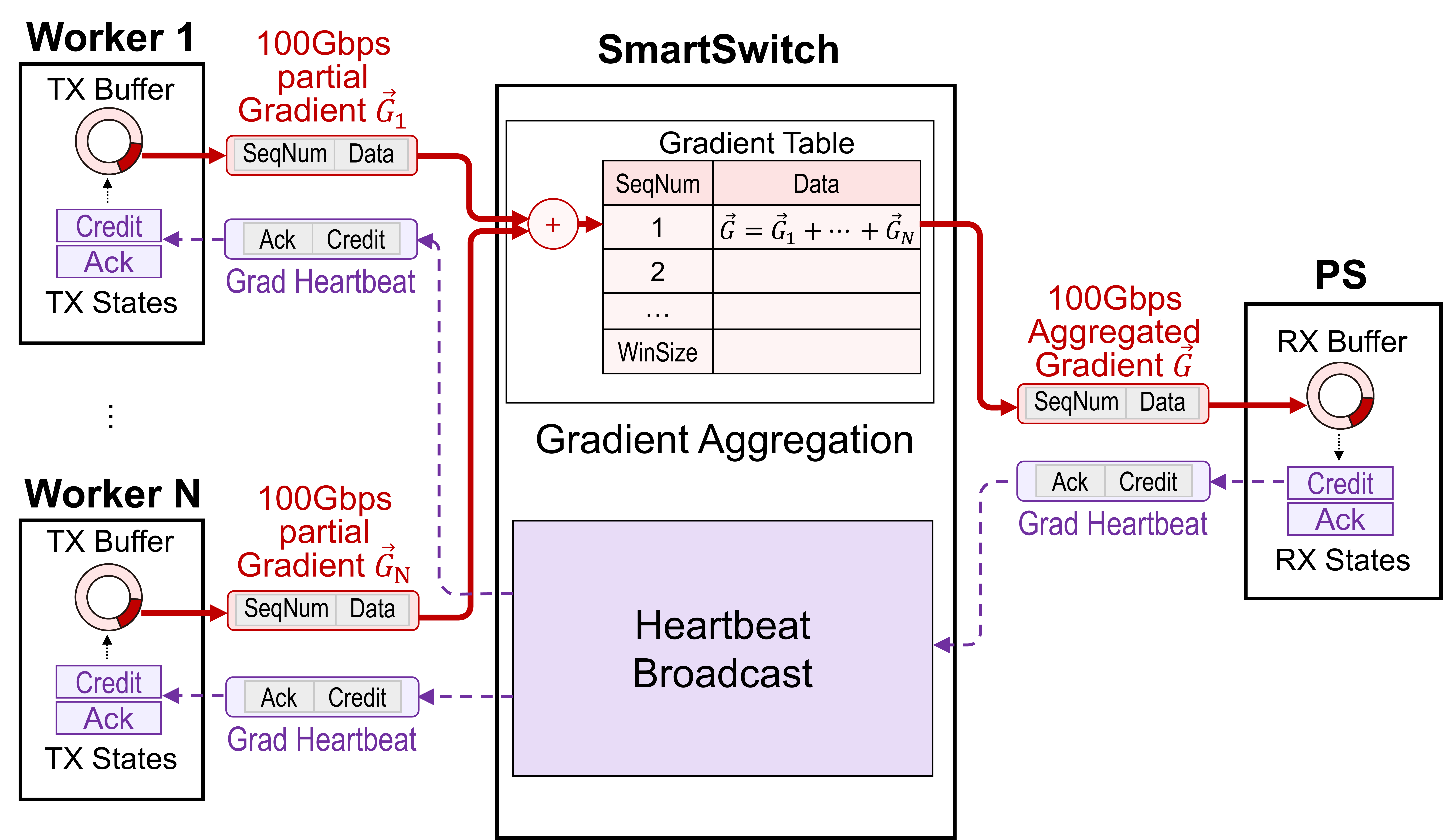}
    }
    \vspace{-1.5ex}
    \caption{\label{fig_reliable} SmartSwitch-assisted many-to-one reliable protocol that reduces the acknowledgement traffic between the SmartSwitch and the PS.}
    \vspace{-3ex}
\end{figure*}

\noindent\textbf{Handling Contention-Free Push and Pull.}
\
Figure~\ref{fig_motivation_pattern_mindp} illustrates the execution pattern of \SystemName{}. Like GPU streams, \SystemName{}'s SmartNIC provides two independent SmartNIC streams for pull and push calls, respectively, and each stream has its own hardware modules to avoid interference between push and pull requests.

To achieve this, \SystemName{} provides separate push/pull processing units for \texttt{push} and \texttt{pull} primitives.
Figure~\ref{fig_nic_datapath} shows a simplified flow of a worker's collective processing module handling requests, where \texttt{push} and \texttt{pull} are handled by different processing units to execute concurrently. 
When the host CPU calls \texttt{push} or \texttt{pull}, the software library allocates a request handle and then passes the request with the handle pointer to the SmartNIC via MMIO. Then, the request is enqueued in a \texttt{push}/\texttt{pull} request queue, such that a request called first finishes first. Next, the \texttt{push}/\texttt{pull} processing unit accepts the request from the queue and issues a DMA request to access GPU memory. The push/pull processing unit processes data from/to GPU memory in a pipelined manner to enable concurrent DMA and network data transport. Upon completion of the DMA procedure, the \texttt{push}/\texttt{pull} processing unit writes the execution status to the request handle to finalize the host \texttt{wait} function. 

\noindent\textbf{Enabling Direct Access to GPU Memory.}
\
We follow the existing works~\cite{fpganic, strom_eurosys20} to enable SmartNIC to directly access the GPU memory via GPU virtual addresses. 

\noindent\textbf{Format Conversion Unit.}
\
Large-scale model training typically produces gradients in fp16 or bf16 formats~\cite{mixedprecision}. However, current Smart\-Switches do not support floating-point arithmetic. To this end, we follow the format conversion strategy of SwitchML~\cite{switchml} and integrate the strategy into the format conversion unit that is hardcoded in SmartNICs. 

\subsection{SmartSwitch-Assisted Many-to-One Reliable Protocol}

To address \textbf{C2}, we propose a SmartSwitch-assisted reliable protocol that uses SmartSwitch to aggregate ACKs from workers to the parameter server and broadcast ACKs reversely. A reliable protocol requires 1) flow control that ensures the sender sends data at the same rate as the receiver receives data to prevent a fast sender from overwhelming the receiver, and 2) reliability assurance that the primitives can detect and recover packet loss or data error. 

\subsubsection{Na\"ive Solutions: Establishing One-to-One Reliable Connections}

A straightforward solution is to establish a one-to-one reliable connection between the SmartSwitch and each worker/PS. However, this solution requires the switch to act as an endpoint executing heavy reliable protocol, which is impractical due to insufficient hardware resources. Specifically, a Tofino SmartSwitch processes a packet in a pipeline with $\leq$20 hardware stages~\cite{tofino}, whereas a reliable RDMA or TCP packet requires more than 50 hardware stages to process. 

Another solution is to establish a one-to-one reliable connection between each worker and the PS, such that the SmartSwitch only forwards the acknowledgment packets without any processing. However, this method has two issues: First, a PS that executes reliable protocols with many (e.g., 64) workers simultaneously incurs high implementation complexity to SmartNICs; Second, existing TCP and RDMA adopt per-packet acknowledgment, so the PS must receive many acknowledgment packets from workers after sending one data packet. Consequently, massive acknowledgment packets occupy non-negligible network bandwidth, lowering the throughput. To demonstrate this, we simulate the maximum achievable data throughput with different numbers of workers under 100Gbps network. The orange line in Figure~\ref{fig_ack_overhead} shows that the throughput drops to 30Gbps with 32 workers, and only 18Gbps with 64 workers. 

\vspace{-0.5ex}
\subsubsection{Our Solution: SmartSwitch-Assisted Many-to-One Reliable Protocol}

To this end, we propose a \textit{SmartSwitch-assisted many-to-one reliable protocol} that 1)~uses periodical heartbeat packets to reduce the acknowledgment traffic between workers and the PS, and 2)~uses SmartSwitch to aggregate many heartbeat packets from workers to one heartbeat packet to the PS, and broadcast one heartbeat packet from PS to many heartbeat packets to workers. Thus, the PS only sends and receives one acknowledgment packet for each heartbeat cycle, regardless of the number of workers to maintain good scalability. The blue line in Figure~\ref{fig_ack_overhead} shows that our SmartSwitch-assisted protocol achieves a higher data throughput than one-to-one reliable connections, and the throughput does not drop as the number of workers scales. 

\noindent\textbf{Packet Transport Procedure.}
\
Figure~\ref{fig_reliable} illustrates the packet transport procedure of \SystemName{}. The SmartSwitch follows SwitchML~\cite{switchml} to broadcast parameters and aggregate gradients. Additionally, each worker and the PS periodically send heartbeat packets containing two connection states: \texttt{Ack} that indicates the next expected sequence number, and \texttt{Credit} field that indicates the maximum sequence number the heartbeat sender can accept. The SmartSwitch broadcasts the gradient heartbeat from the PS to all workers. Meanwhile, it maintains a heartbeat table to record the latest parameter heartbeat from each worker, and periodically performs minimum aggregation to Ack and Credit in the table and sends a heartbeat packet containing the aggregated states to the PS.

For flow control, a worker/PS ceases sending packets once it sends a packet with a sequence number equal to its TX \texttt{Credit} value, and resumes only after TX \texttt{Credit} value is updated by the heartbeat packets; for reliability assurance, f there are unacknowledged packets and a sender TX state's \texttt{Ack} has not increased for the Retransmission Time Out (RTO) threshold, the sender resends packets starting with the sequence number \texttt{Ack}. The SmartSwitch simply forwards any resent packet to the receiver. We follow industry practices~\cite{stellar_sigcomm25} to set the RTO threshold to \textasciitilde 250 us.

\subsection{Step-Centric Optimizer Pipelining}
\label{subsec_optimizer}

\begin{figure}[t]
    \begin{minipage}{0.47\linewidth}
        \begin{center}
            \includegraphics[width=\linewidth]{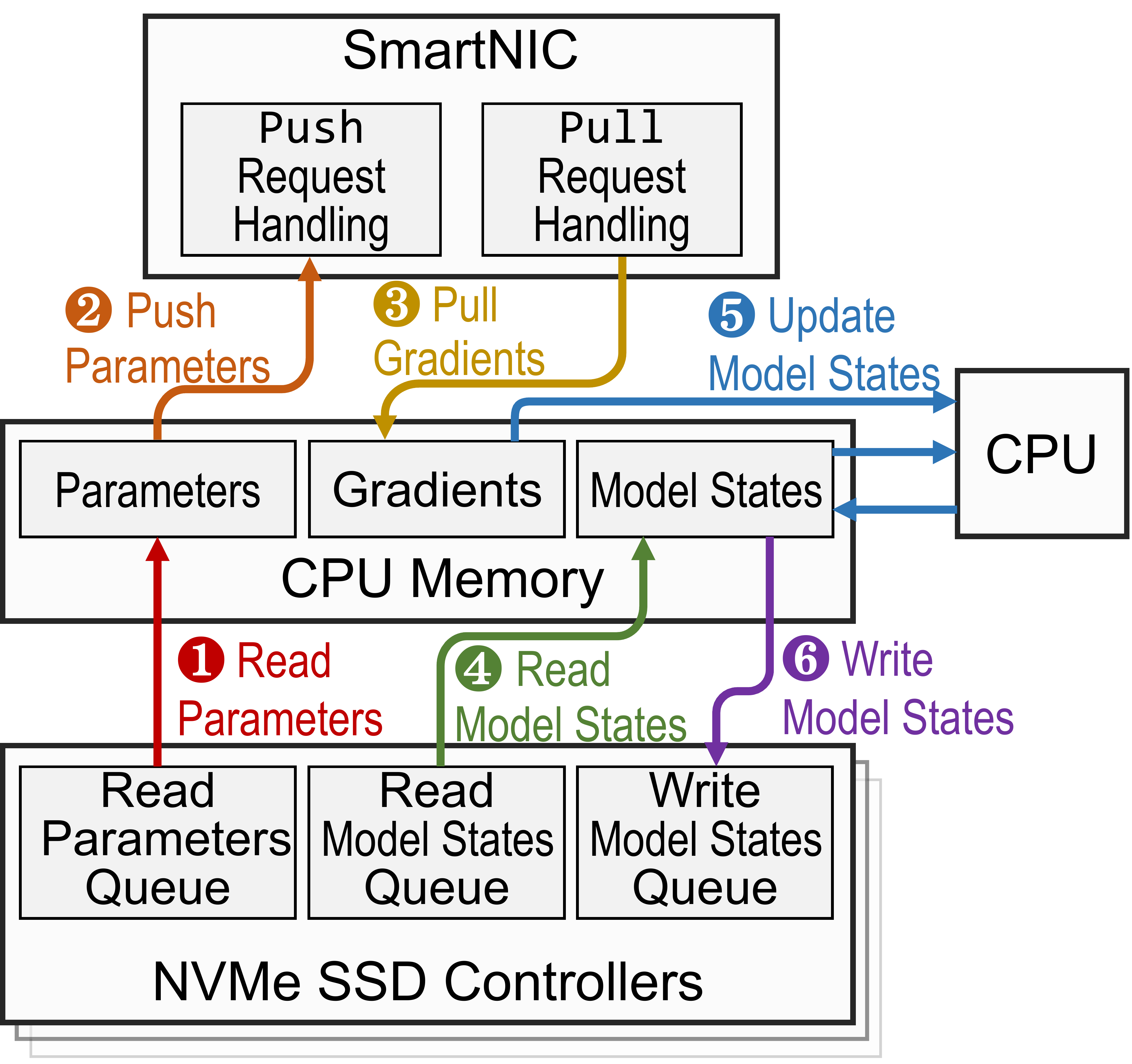}
            \vspace{-4ex}
            \caption{\label{subfig_optimizer_datapath} Data path of the in-network optimizer during the backward stage. }
            \vspace{-3ex}
        \end{center}
    \end{minipage} 
    \hfill
    \begin{minipage}{0.49\linewidth}
        \centering
        \vspace{1.3ex}
        \includegraphics[width=\linewidth]{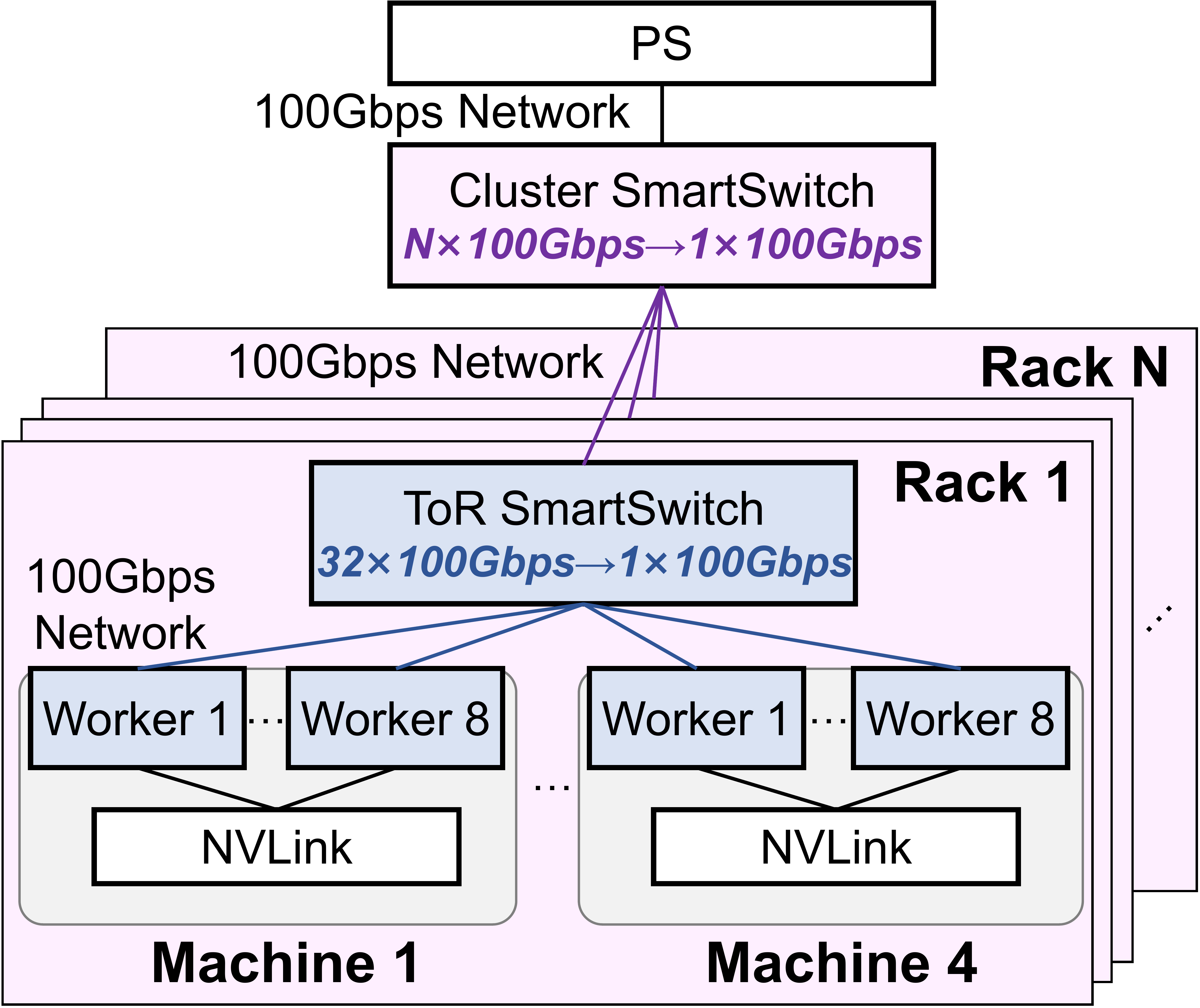}
        \vspace{-3ex}
        \caption{Multi-rack topology of \SystemName{}.}
        \label{fig_cluster_topo}
        \vspace{-2ex}
    \end{minipage} 
\end{figure}

Thanks to the SmartSwitch-assisted many-to-one reliable
protocol that provides reliable aggregated gradients, our PS needs to perform Adam optimizer on the line-rate aggregated gradients and provide on-demand parameters to GPUs. 
However, it is not trivial to achieve. 
The main challenge is that the size of model states is way larger than the host memory size, so it is generally assumed that a 100B model needs many CPU servers to implement the corresponding CPU optimizer~\cite{byteps}. To this end, to address \textbf{C3}, we propose the \textit{step-centric pipelining technology} to 1) allocate PS's CPU threads by optimizer steps, instead of by layers, and 2) deeply pipeline SSD accesses, CPU Adam, and collectives to enable an out-of-core optimizer to consume 100Gbps aggregated gradients. 
Figure~\ref{subfig_optimizer_datapath} illustrates a simplified data path of our out-of-core optimizer. In the following, we present the resource requirement of a training iteration that consists of two stages: forward and backward, followed by the detailed design of \textit{step-centric pipelining technology}. 

\noindent\textbf{Forward Stage.} 
\
During this stage, the optimizer needs to provide on-demand parameters to GPUs. It undergoes two steps: \circled{1}~Reading 2-byte parameters from the SSDs to the CPU memory, and \circled{2}~pushing 2-byte parameters from CPU memory to SmartNIC.  
Therefore, providing 100Gbps (11.6 GB/s) on-demand parameters to GPUs requires 23.3 GB/s memory bandwidth and 11.6 GB/s SSD bandwidth in Table~\ref{table_optimizer_req}.\footnote{Providing 200Gbps on-demand parameters doubles the memory and SSD bandwidths.}  

\noindent\textbf{Backward Stage. }
During this stage, the optimizer provides on-demand parameters to GPUs and performs the Adam operation on the aggregated gradients. 
It undergoes the two steps (\circled{1} and \circled{2}) and four additional steps: \circled{3}~pulling 2-byte gradients from the SmartNIC to the CPU memory, \circled{4}~reading 12-byte model states from SSDs to the CPU memory, \circled{5}~performing the \textbf{compute-intensive} CPU Adam, where the CPU reads 2-byte gradients and 12-byte model states from the memory, performs 17 floating-point operations to update model states~\cite{zero-offload}, and writes 12-byte updated model states and 2-byte parameter copy to the memory, and \circled{6}~writing 12-byte updated model states and 2-byte parameter copy back to SSDs. 
This stage requires 99 GFLOPS computation, 349 GB/s memory bandwidth, and 81.4 GB/s SSD bandwidth.

\begin{table}[t]\footnotesize
    \centering
    \caption{Resource requirements for Adam optimizer.}
    \vspace{-2ex}
    \label{table_optimizer_req}
    \resizebox{\columnwidth}{!}{
        \begin{tabular}{ccccc}
            \specialrule{.1em}{.05em}{.05em} 
             \textbf{\makecell{Network\\Thpt}} & \textbf{Stage} & \textbf{\makecell{Compute \\ FLOPS}} & \textbf{\makecell{\narrow{Mem Bandwidth} \\ \narrow{(Read+Write Total)}}} & \textbf{\makecell{\narrow{SSD Bandwidth} \\ \narrow{(Per Direction)}}} \\ \specialrule{.1em}{.05em}{.05em} 
             \multirow{2}{*}{\textbf{100Gbps}} & \textbf{Fwd} & 0 & 23.3 GB/s & 11.6 GB/s \\ \cline{2-5}
             & \textbf{Bwd} & 99.0 GFLOPS & 349 GB/s & 81.4 GB/s \\ \hline
             \multirow{2}{*}{\textbf{200Gbps}} & \textbf{Fwd} & 0 & 46.6 GB/s & 23.3 GB/s \\ \cline{2-5}
             & \textbf{Bwd} & 198 GFLOPS & 698 GB/s & 163 GB/s \\
            \specialrule{.1em}{.05em}{.05em} 
        \end{tabular}
    }
    \vspace{-4ex}
\end{table}
\begin{figure}[t]
    \centering
    \subfigure[\label{fig_pipelining_layer} Layer-centric pipelining: Limited parallelism causes pipeline bubbles.]{
        \includegraphics[width=0.95\linewidth]{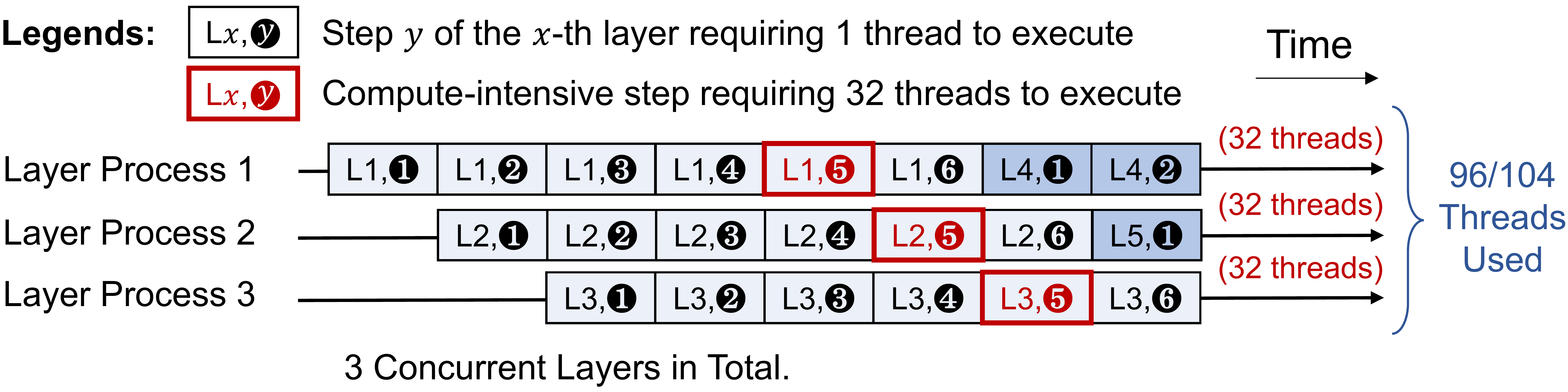}
    }
    \vspace{-1ex}
    \\
    \subfigure[\label{fig_pipelining_step} Our step-centric pipelining: Fully pipelines the optimizer steps.]{
        \includegraphics[width=0.95\linewidth]{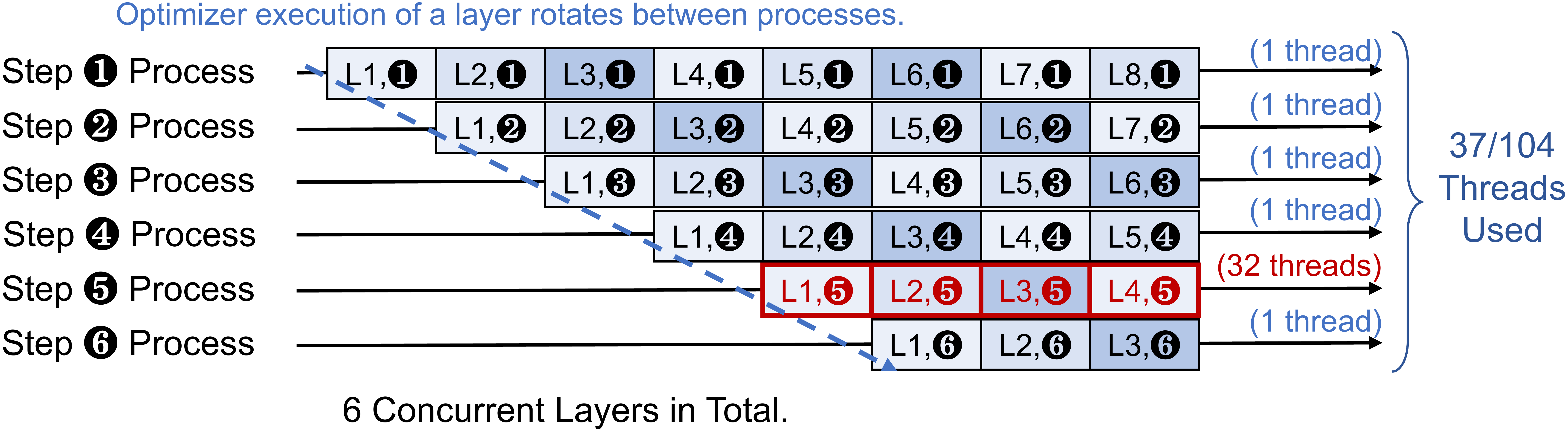}
    }
    \vspace{-1ex}
    \caption{Pipelining of collectives, SSD IO, and CPU Adam during the backward stage.}
    \vspace{-3ex}
\end{figure}

\noindent\textbf{Na\"ive Solution: Layer-Centric Pipelining.}
\
A straightforward method to pipeline the optimizer steps is \textit{layer-centric pipelining}, which assigns a fixed number of CPU threads for each model layer to sequentially execute the optimizer steps. This method naturally pipelines the steps by concurrently executing the optimizer across multiple layers. However, this approach incurs an issue: Different steps in the optimizer require varying numbers of CPU threads to saturate. In particular, CPU Adam (Step~\circled{5}) requires 32 threads on Intel Xeon 5320 CPU to consume aggregated gradients at line rate, while the remaining steps only require 1 CPU thread to saturate. Consequently, layer-centric pipelining requires 32 CPU threads to execute each model layer, thereby limiting the overall achievable parallelism in Figure~\ref{fig_pipelining_layer}.

\noindent\textbf{Our Solution: Step-Centric Pipelining.}
\
To address the limited parallelism issue, we propose \textit{step-centric pipelining} that allocates CPU threads to each optimizer step. Each step executes one layer at a time and a layer rotates from the first step to the last. This technology allows us to allocate more threads to the compute-intensive Step~\circled{5} and fewer threads to the remaining steps, as Figure~\ref{fig_pipelining_step} shows, so as to enable full pipelining of optimizer steps with a small number of CPU threads, e.g., 37 threads in Figure~\ref{fig_pipelining_step}.

\begin{table}[t]\footnotesize
    \centering
    \vspace{-1ex}
    \caption{Resources provided by different CPUs.}
    \vspace{-2ex}
    \label{table_optimizer_res}
    \resizebox{\columnwidth}{!}{
        \begin{tabular}{cccc}
            \specialrule{.1em}{.05em}{.05em} 
             \textbf{CPU} & \textbf{\narrow{Compute FLOPS}} & \textbf{\narrow{Mem Bandwidth}} & \textbf{\narrow{PCIe Bandwidth}} \\ \specialrule{.1em}{.05em}{.05em} 
             \textbf{5320 (Gen 4)} & \makecell{1.83 TFLOPS} & \makecell{375 GB/s} & \makecell{246 GB/s} \\ \hline
             \textbf{6730P (Gen 5)} & \makecell{2.56 TFLOPS} & \makecell{819 GB/s} & \makecell{678 GB/s} \\
            \specialrule{.1em}{.05em}{.05em} 
        \end{tabular}
    }
    \vspace{-4ex}
\end{table}

\noindent\textbf{100B Model Optimizer in a Scalable PS.}
\
Table~\ref{table_optimizer_res} shows the compute power, memory bandwidth, and PCIe bandwidth of two commodity machines, each with two CPUs: Intel Xeon 5320 (PCIe Gen 4) and 6730P (PCIe Gen 5). 
We observe that 6730P (or 5320) machine can satisfy more than 200Gbps (or 100Gbps) network, in case the machine provides sufficient SSD bandwidth, e.g., with 12 SSDs. We conclude that a single scalable PS is sufficient to concurrently provide on-demand parameters and perform the Adam operation on the aggregated gradients from any number of workers at line rate. 

\noindent\textbf{Supporting Multi-Rack.}
\
\SystemName{} supports training on multi-rack clusters by hierarchical switches in a topology shown in Figure~\ref{fig_cluster_topo}. Each rack employs a ToR SmartSwitch to aggregate gradients from its workers to compute partial aggregated gradients. A cluster SmartSwitch aggregates partial aggregated gradients from the ToR SmartSwitches and forwards the fully aggregated gradients to the PS. The parameters are broadcast in a reverse hierarchical manner. Supporting more workers only needs a deeper hierarchy of SmartSwitches.

\begin{table}[t]\footnotesize
    \begin{center}   
        \caption{Models for evaluation. Custom models keep the same architecture as OPT models with random parameters.}  
        \vspace{-5ex}
        \label{table_models} 
        \resizebox{\columnwidth}{!}{
            \begin{tabular}{lccc} 
                \specialrule{.1em}{.05em}{.05em} 
                \textbf{Model}&  \textbf{\#Transformer Blocks}&  \textbf{\#Head}&\textbf{Hidden Dimension}\\ \specialrule{.1em}{.05em}{.05em} 
                \textbf{OPT-1.3B}&  24&  32&2048\\ \hline 
                \textbf{OPT-2.7B}&  32&  32&2560\\ \hline 
                \textbf{OPT-6.7B}&  32&  32&4096\\ \hline 
                \textbf{OPT-13B} &  40&  40&5120\\ \hline 
                \textbf{OPT-30B} &  48&  56&7168\\ \hline
                \textbf{OPT-66B} &  64&  72&9216\\ \hline
                \textbf{Custom-175B}&  96&  96&12288\\ \hline
                \textbf{Custom-276B}& 112& 112&14336\\\hline
                \textbf{Custom-505B}& 124& 144&18432\\\hline
                \textbf{Custom-1.0T}& 172& 172&22016\\\specialrule{.1em}{.05em}{.05em} 
            \end{tabular}
        }
        \vspace{-5ex}
    \end{center}
\end{table}

\section{Evaluation}
\subsection{Experimental Setup}
\label{subsec_exp_setup}

\noindent\textbf{Workloads.}
\
We choose OPT models~\cite{opt} and our custom models of different sizes listed in Table~\ref{table_models}. Our results can be generalized to other LLMs such as Llama~\cite{llama3} since their computation/collective patterns are similar.

In each of the evaluation experiments, we select the largest model size that the baselines can train under the corresponding configuration, which is 175B for Subsection~\ref{subsec_exp_endtoend}, \ref{subsec_exp_ablation}, and end-to-end breakdown and MPS comparison in Subsection~\ref{subsec_exp_coll}, 30B for SHARP comparison in Subsection~\ref{subsec_exp_coll}, 13B for Subsection~\ref{subsec_exp_dgx}. The sequence length is set to 1024 for all experiments.

\noindent\textbf{Evaluated Cluster.}
\
We evaluate \SystemName{} and baselines on an 8-worker cluster, each worker with dual Intel Xeon Silver 4214 CPUs, 256 GB DDR4 memory, 1$\times$ NVIDIA A100 40GB GPU, and system-specific network and SSDs. All machines are connected by 100Gbps network. We will introduce the network and SSD settings of \SystemName{} and baselines below. 

\noindent\textbf{\SystemName's Configurations.}
\
We evaluate \SystemName{} on the cluster with 8 machines (each with 1 GPU and 1 SmartNIC connected to a SmartSwitch~\cite{wedge100bf}) plus a PS. The PS equips dual Intel Xeon Gold 5320 CPUs, 12 SSDs, and a SmartNIC. We implement SmartNICs on Xilinx Alveo U50 FPGAs.
Tables~\ref{table_nic_resource} and~\ref{table_switch_resource} show the resource consumption of SmartNICs and the SmartSwitch logic. 
\SystemName{} enables activation checkpointing~\cite{activationcheckpointing, checkmate} and bf16 training~\cite{mixedprecision}. 

\noindent\textbf{Baselines.}
\
We use four systems as our baselines. 

The first baseline is ZeRO-Infinity~\cite{zero-infinity}, an MSDP system that distributes the model states in workers' SSDs. We evaluate ZeRO-Infinity on the cluster with 8 machines, and 1 Mellanox SN2700 switch~\cite{sn2700}. Each machine features 2 SSDs (16 in total\footnote{ZeRO-Infinity equips 4 more SSDs than \SystemName{} because our SSDs provide different I/O bandwidth on PCIe Gen 3 worker machine of ZeRO-Infinity and Gen 4 PS machine of \SystemName{}. Our configuration ensures both systems have 26~GB/s aggregated I/O bandwidth per direction with 1:1 mixed read/write.}), 1 GPU, and 1 CX-5 NIC connected to the switch. We run ZeRO-Infinity on DeepSpeed 0.9.3~\cite{deepspeed} and NCCL 2.20.5 with activation checkpointing and bf16 training. 

The second baseline is ZeRO-Offload~\cite{zero-offload}, which is used as an alternative to ZeRO-Infinity for experiments on the NVLink machine (Subsection~\ref{subsec_exp_dgx}), because the NVLink machine we rent does not support plugging ad-hoc SSDs and thus we cannot achieve high performance from ZeRO-Infinity that relies on more SSDs to efficiently train a large model. We run ZeRO-Offload with the same configuration as ZeRO-Infinity, except that the model states are stored in CPU memory rather than SSDs.

The third baseline is ZeRO-3~\cite{zero}, which adopts MSDP but keeps the model state shard in GPU memory instead of CPU memory or SSDs. We run ZeRO-3 on the same cluster and with the same configuration as that in ZeRO-Infinity, except that model states are kept in GPU memory.

The fourth baseline is ATP~\cite{atp}, a SmartSwitch-enhanced MRDP system that provides coupled \texttt{push}-\texttt{pull} primitives (same semantics as \texttt{AllReduce}). We run ATP on the cluster with 8 machines, 8 GPUs, 8 CX-5 NICs, and 1 SmartSwitch. Each machine has 1 GPU and 1 NIC connected to the SmartSwitch. We run ATP on PyTorch 1.9.1~\cite{pytorch} with activation checkpointing and bf16 training.

\begin{table}[t]\footnotesize
    \centering
    \caption{Hardware resource consumption of SmartNIC.}
    \vspace{-2ex}
    \label{table_nic_resource}
    \begin{tabular}{cccc}
        \specialrule{.1em}{.05em}{.05em} 
        \textbf{LUT} & \textbf{FF} & \textbf{BRAM} & \textbf{URAM} \\ \specialrule{.1em}{.05em}{.05em} 
        \makecell{135K \\ (15.5\%)} & \makecell{225K \\ (12.9\%)} & \makecell{354 \\ (26.3\%)} & \makecell{128 \\ (20.0\%)} \\ 
        \specialrule{.1em}{.05em}{.05em} 
    \end{tabular}
    \vspace{-1ex}
\end{table}

\begin{table}[!t]\footnotesize
    \centering
    \vspace{-2ex}
    \caption{Hardware resource consumption of Smart\-Switch.}
    \vspace{-2ex}
    \label{table_switch_resource}
    \begin{tabular}{cccccc}
        \specialrule{.1em}{.05em}{.05em} 
        \textbf{Stage} & \textbf{MAT} & \textbf{TCAM} & \textbf{VLIW} & \textbf{Register} & \textbf{SRAM}\\ \specialrule{.1em}{.05em}{.05em} 
        \makecell{11 \\ (91.7\%)} & \makecell{99 \\ (51.6\%)} & \makecell{111B \\ (28.9\%)} & \makecell{1.64Kb \\ (12.0\%)} & \makecell{37 \\ (77.1\%)} & \makecell{48 MiB \\ (39.5\%)} \\
        \specialrule{.1em}{.05em}{.05em} 
    \end{tabular}
    \vspace{-2ex}
\end{table}

\begin{figure}[!t]
    \begin{minipage}{0.457\linewidth}
        \begin{center}
            \includegraphics[width=\linewidth]{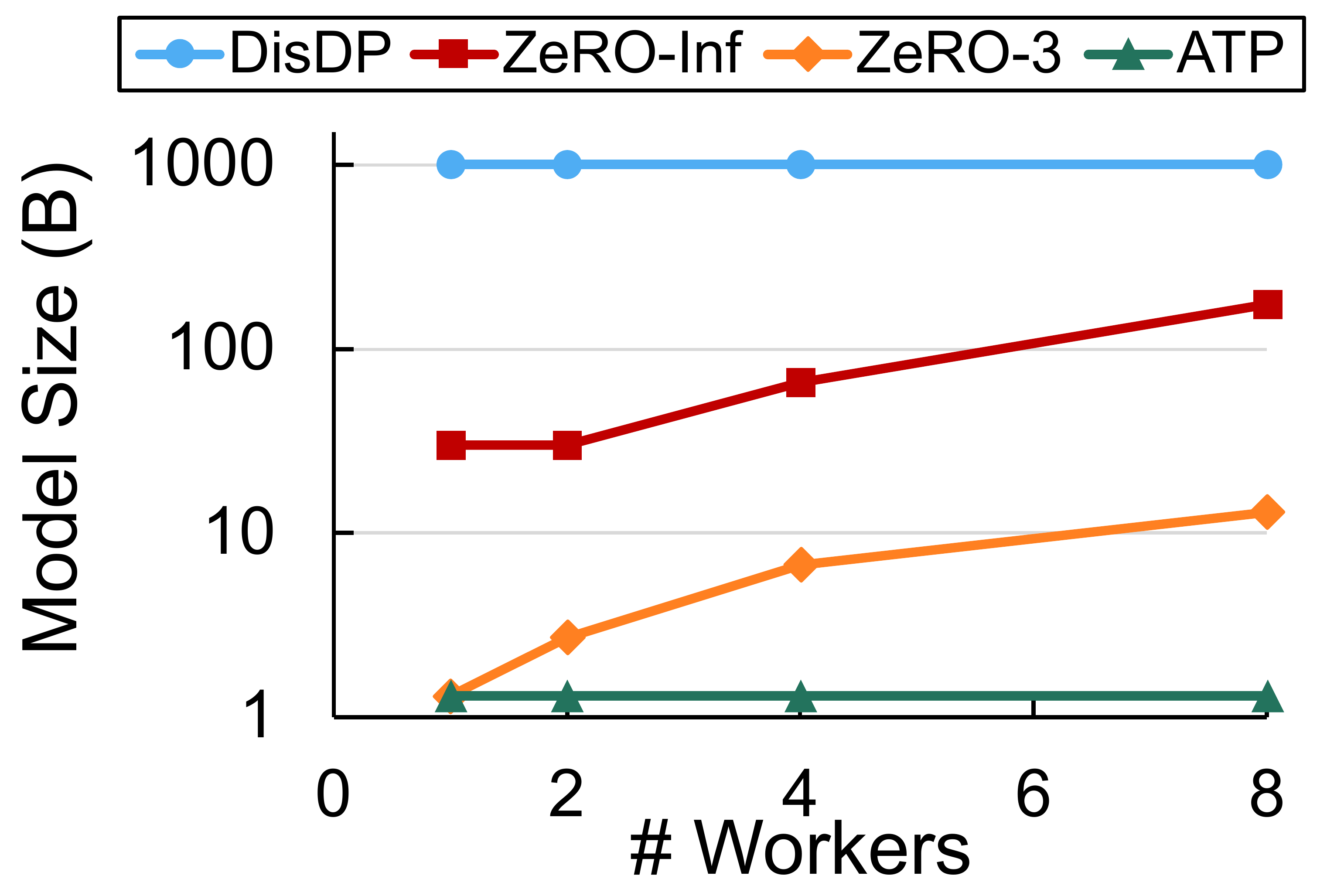}
        \end{center}
        \vspace{-3ex}
        \caption{\label{fig_exp_max_model} Maximum trainable model size.}
        \vspace{-2ex}
    \end{minipage} 
    \hfill
    \begin{minipage}{0.502\linewidth}
        \begin{center}
            \includegraphics[width=\linewidth]{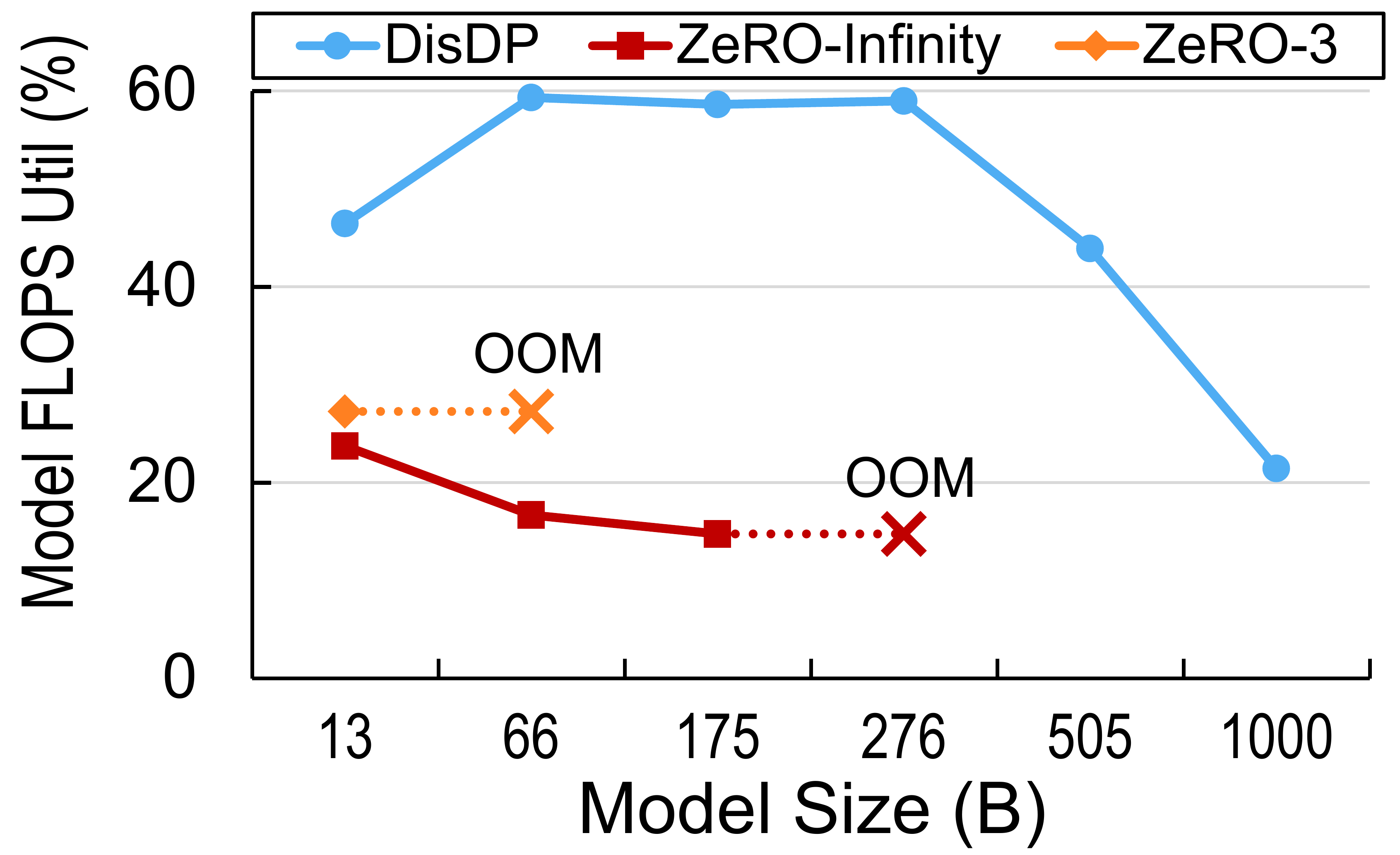}
        \end{center}
        \vspace{-3ex}
        \caption{\label{fig_exp_thpt_wrt_size} Maximum MFU on different models.}
        \vspace{-2ex}
    \end{minipage} 
\end{figure}

\subsection{Maximum Trainable Model Size}
\label{subsec_exp_model_size}

We compare the maximum trainable model sizes between \SystemName{} and baselines when varying the number of workers, as shown in Figure~\ref{fig_exp_max_model}. We make three observations. 

First, \SystemName{} allows the constant, maximum trainable model size 1T that is bounded by the size of a single layer to fit in GPU memory, because \SystemName{} disaggregates storage so that both worker CPU and GPU do not have to accommodate the entire model state shard. In contrast, ZeRO-Infinity with 8 worker machines can train only a 175B model, whose size is smaller than that of \SystemName{}, because the sharded model states require each worker to prepare auxiliary temporary buffers on the CPU memory for GPU-CPU-SSD communications, and thus limits the maximum trainable model size. 
Second, when employing an increasing number of workers, ZeRO-3 and ZeRO-Infinity increase the trainable model size, because they rely on aggregated GPU, CPU memory, and NVMe storage of worker machines to execute the distributed optimizer. However, its maximum trainable model size cannot exceed that of \SystemName{}'s to accommodate a single layer on GPU memory. 
Third, ATP is only able to train a 1.3B model, because it is based on MRDP which replicates the entire model states across all GPUs, bounding its maximum trainable model size. 

\begin{figure}[t]
    \vspace{-1ex}
    \subfigure[Training Custom-175B.]{
        \label{fig_exp_thpt_175b}
        \includegraphics[width=0.435\linewidth]{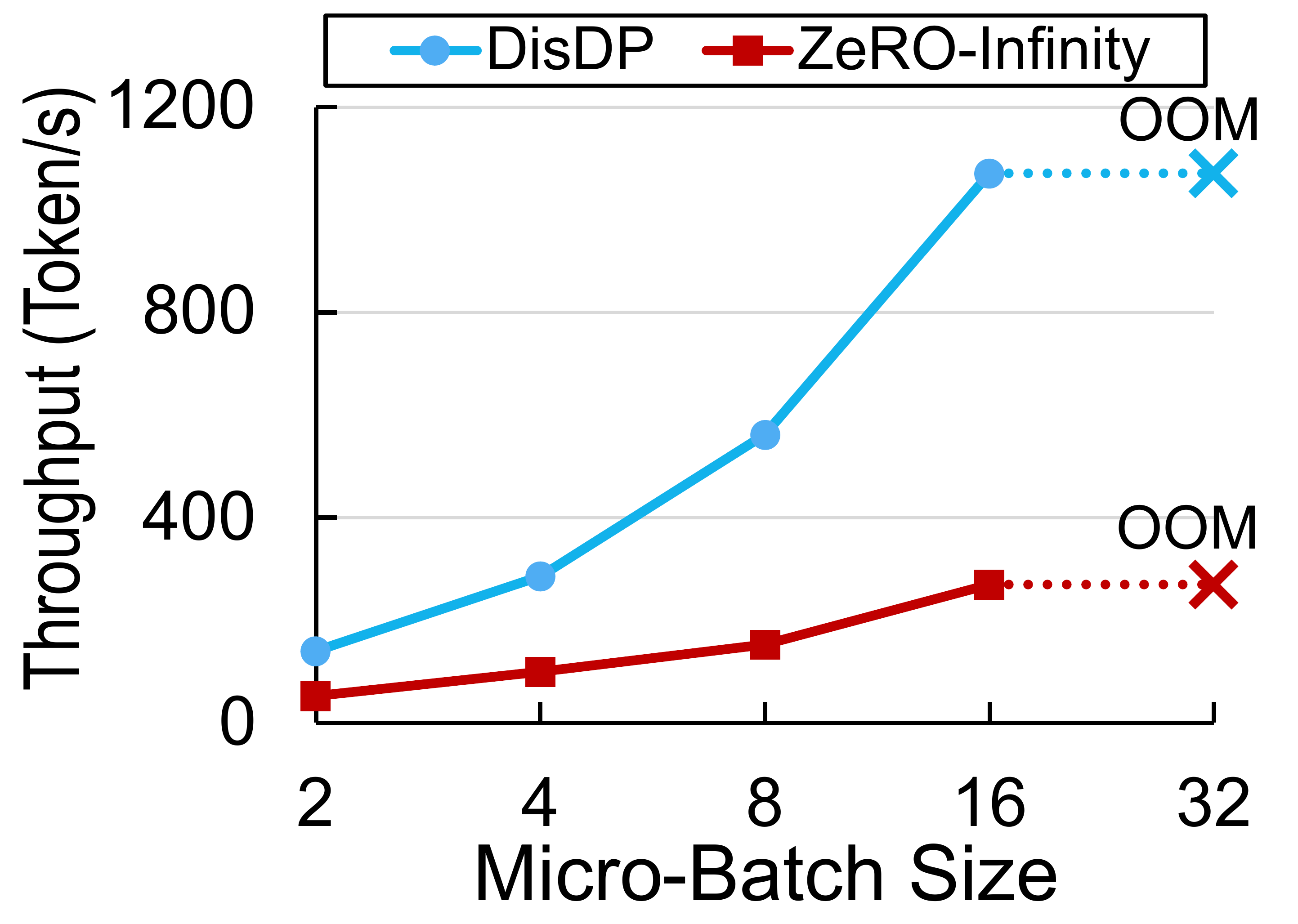}
    }
    \hfill
    \subfigure[Training OPT-1.3B.]{
        \label{fig_exp_thpt_1.3b}
        \includegraphics[width=0.505\linewidth]{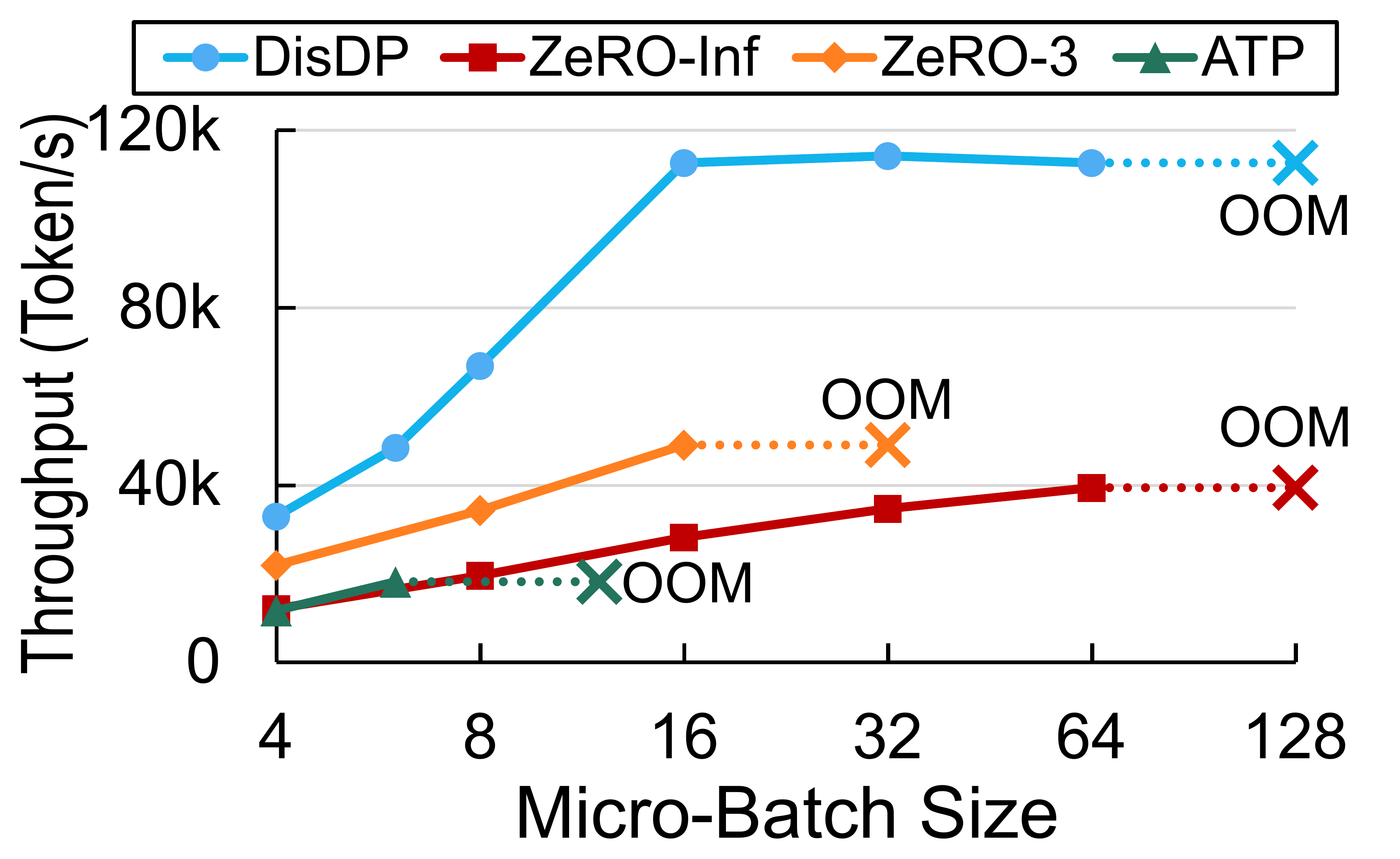}
    }
    \label{fig_exp_thpt}
    \vspace{-2ex}
    \caption{Throughput comparison when training on 8 distributed GPUs with different micro-batch sizes. }
    \vspace{-1ex}
\end{figure}

\begin{figure}[t]
    \vspace{-3ex}
    \subfigure[Training OPT-1.3B with a micro-batch size of 64.]{
        \includegraphics[width=0.302\linewidth]{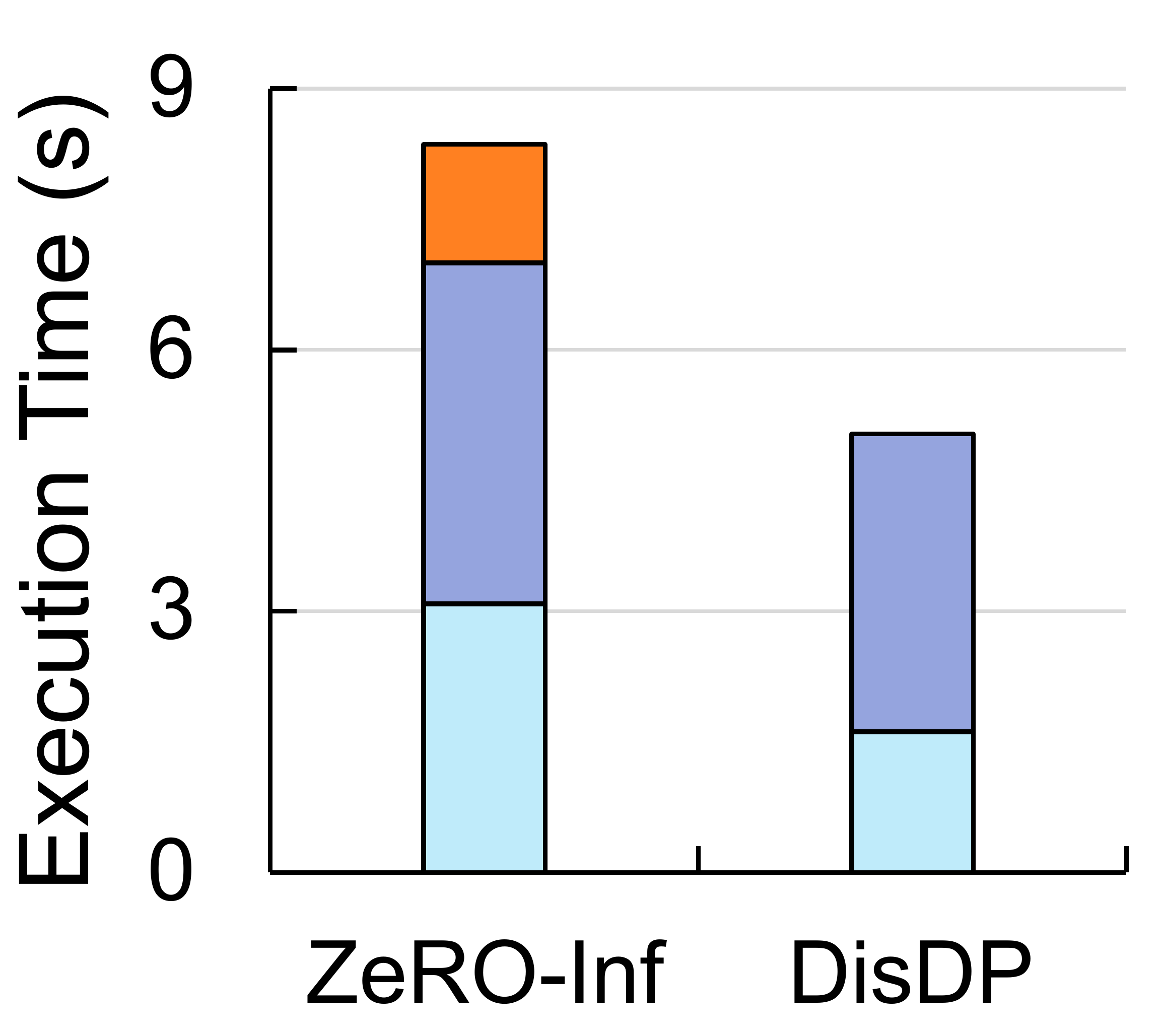}
    }
    \hfill
    \subfigure[Training OPT-30B with a micro-batch size of 32.]{
        \includegraphics[width=0.279\linewidth]{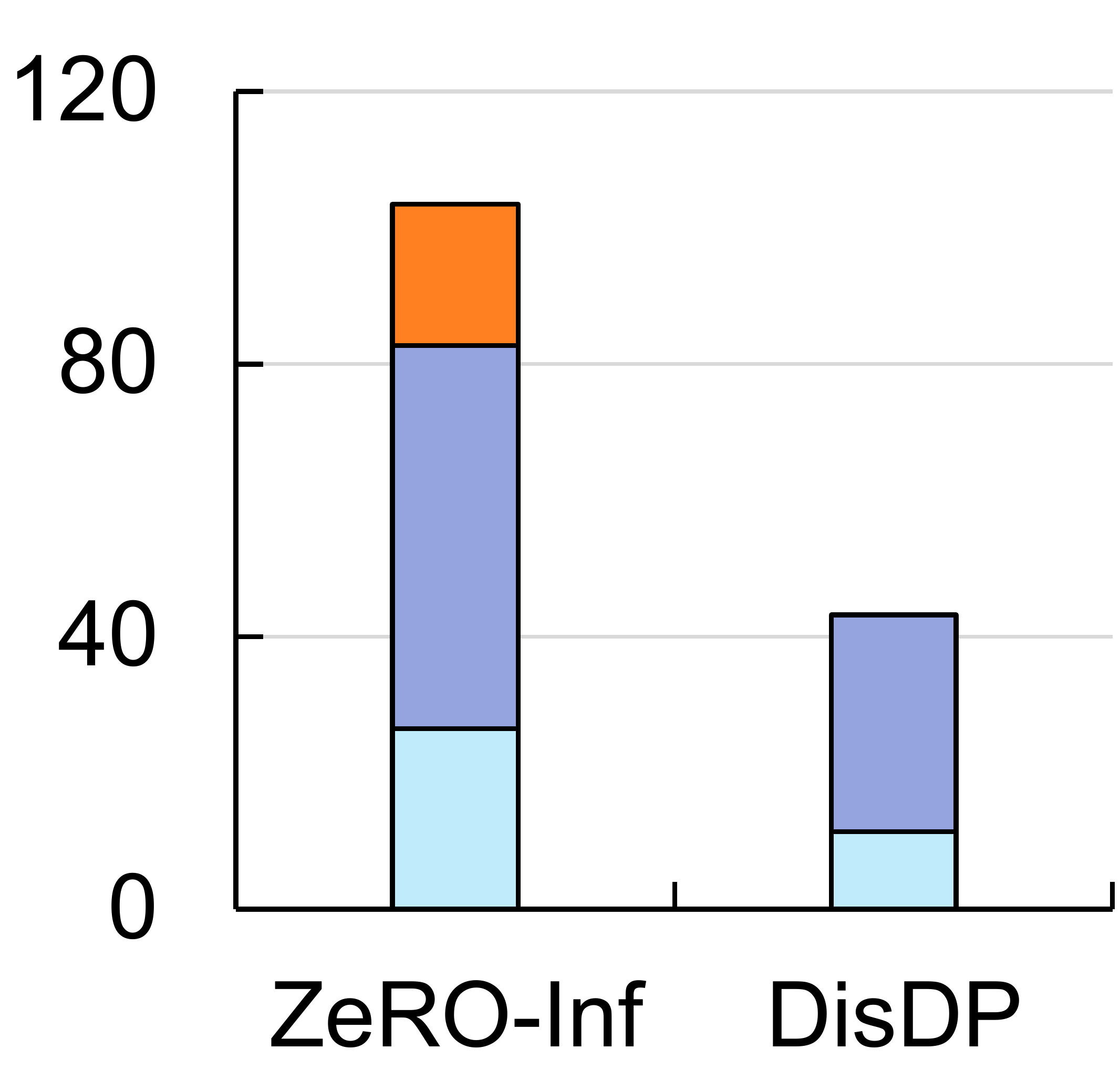}
    }
    \hfill
    \subfigure[Training Custom-175B with a micro-batch size of 16.]{
        \includegraphics[width=0.309\linewidth]{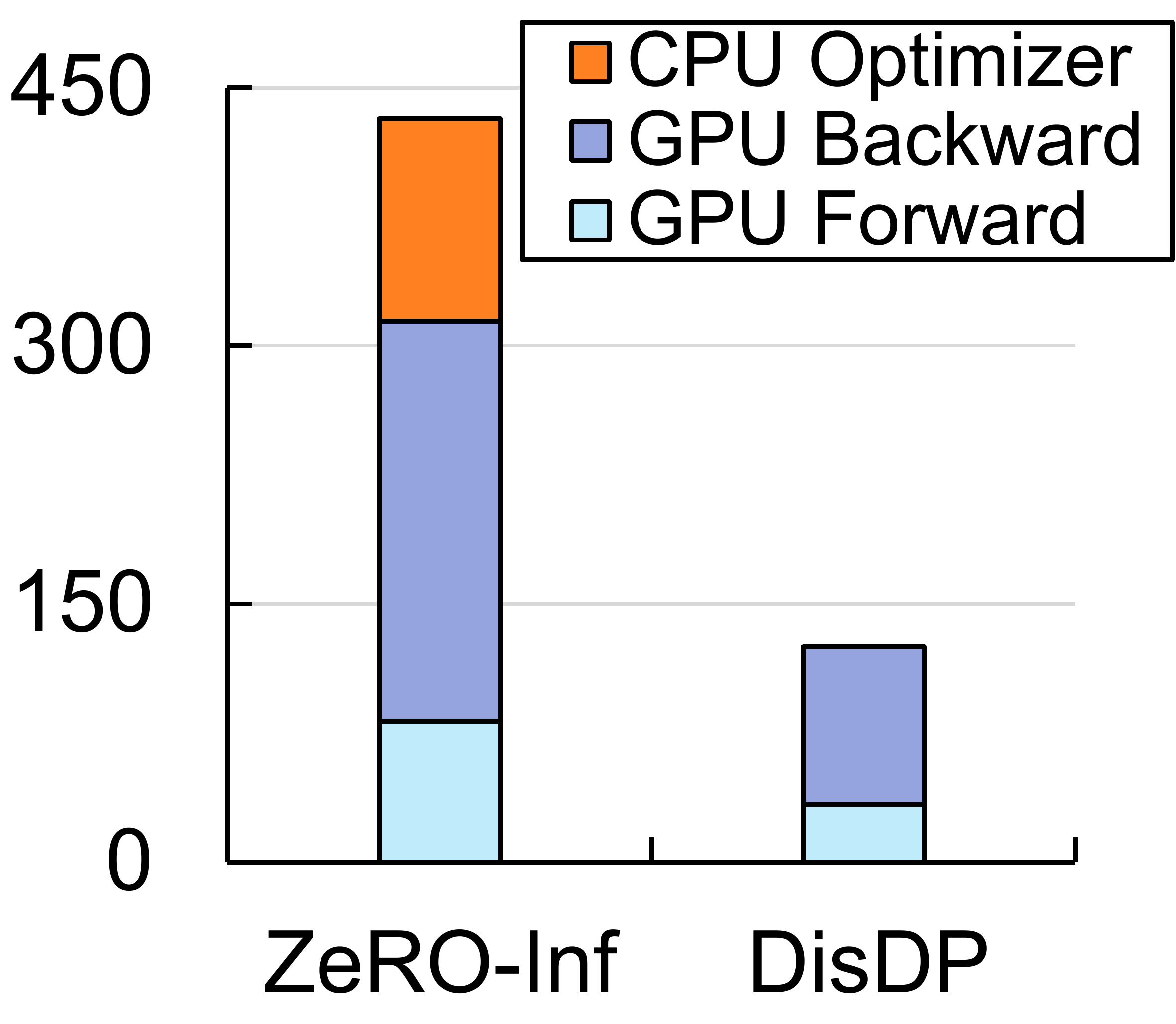}
    }
    \vspace{-2ex}
    \caption{\label{fig_exp_breakdown}Execution breakdown comparison.}
    \vspace{-2ex}
\end{figure} 

\subsection{End-to-End Throughput/MFU Comparison}
\label{subsec_exp_endtoend}

\noindent\textbf{Throughput w.r.t. Batch Size.}
\
Figure~\ref{fig_exp_thpt_175b} shows the comparison of \SystemName{} and ZeRO-Infinity on the 175B model. \SystemName{} achieves up to 134 token/s, 3.98$\times$ larger than ZeRO-Infinity.  The execution breakdown of an iteration in Figure~\ref{fig_exp_breakdown} shows the sources of performance gain: 1)~\SystemName{} achieves a significantly shorter time for the forward and the backward stages than ZeRO-Infinity because the heavy collectives bottlenecks in the two stages of ZeRO-Infinity. In contrast, \SystemName{} eliminates the interference between compute and collectives by performing collectives on SmartNICs, and uses SmartSwitch-assisted reliable protocol to further reduce the collective traffic, thus achieving high GPU utilization. 2)~The PS additionally hides the CPU optimizer behind the GPU backward stage. Notably, \SystemName{} has more benefits as model size increases. 

Figure~\ref{fig_exp_thpt_1.3b} shows a comparison of \SystemName{} and the baselines on the OPT-1.3B model. \SystemName{} achieves up to 14.1K tokens/s, which is 2.90$\times$, 2.33$\times$, and 6.28$\times$ larger than ZeRO-Infinity, ZeRO-3, and ATP at their peak throughput respectively. We make two observations.

First, \SystemName{} achieves even higher throughput than ZeRO-3 and ATP, though \SystemName{} stores the model states in slow SSDs in PS, while ZeRO-3 and ATP accommodate the model states in fast GPU memory, because \SystemName{} completely eliminates compute-network interference, while the baselines do not due to partial aggregation of compute, network, and storage.

\begin{figure}[t]
    \vspace{-1ex}
    \centering
    \subfigure[\label{fig_exp_gemm_coll_175b}Training Custom-175B.]{
        \includegraphics[width=0.379\linewidth]{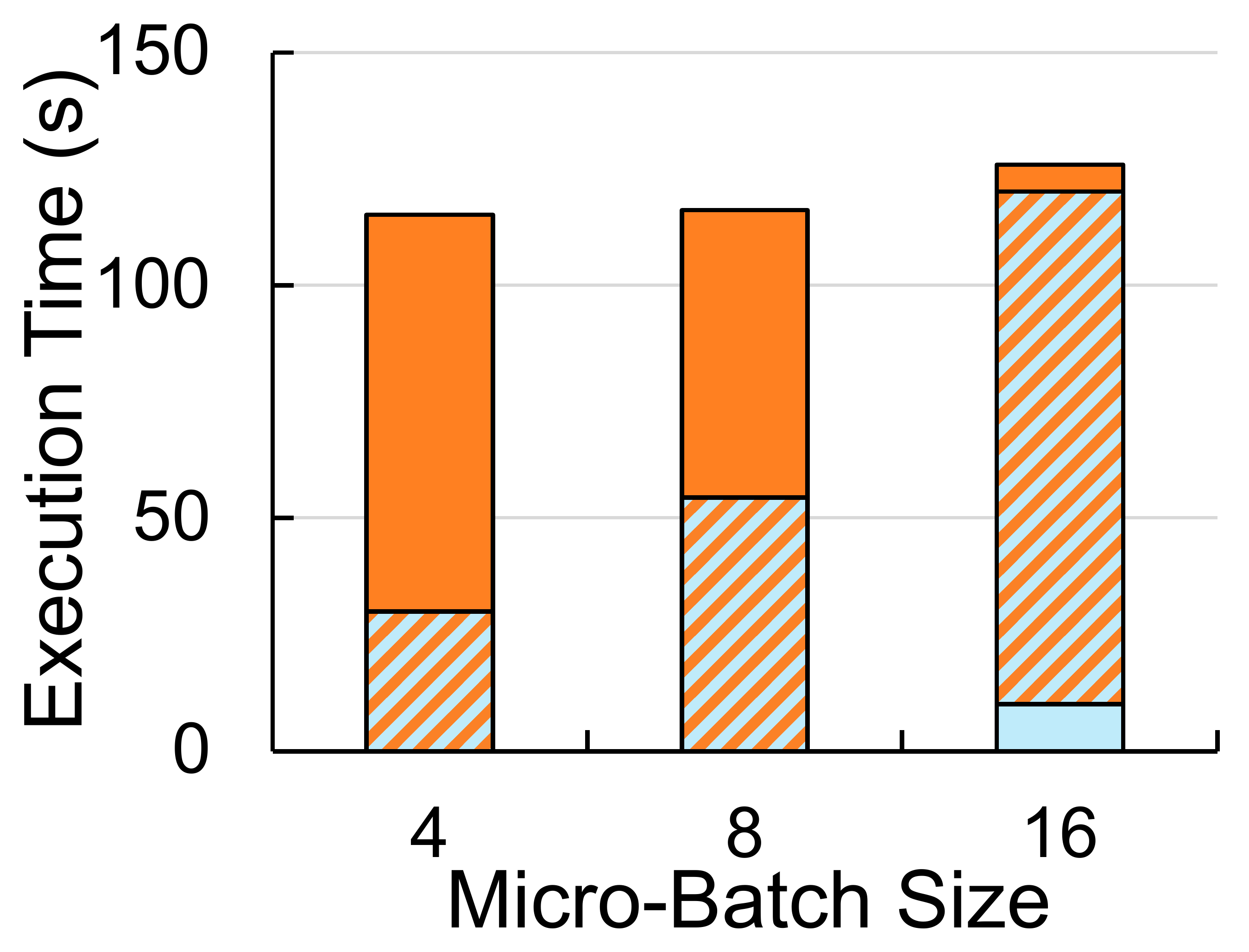}
    }
    \hfill
    \subfigure[\label{fig_exp_gemm_coll_1_3b}Training OPT-1.3B.]{
        \includegraphics[width=0.521\linewidth]{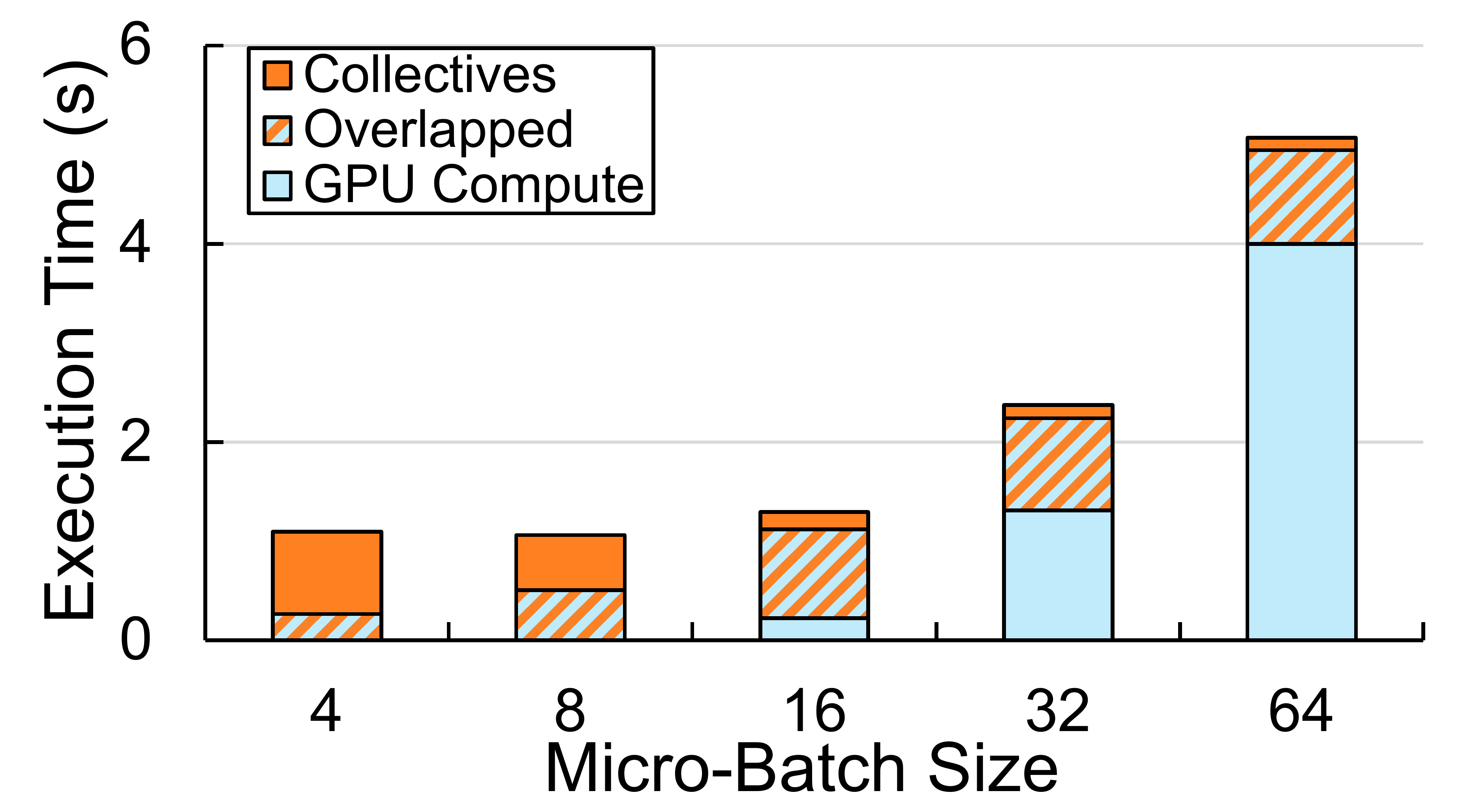}
    }
    \vspace{-1ex}
    \caption{Breakdown of compute and collectives running \SystemName{}.}
    \vspace{-1ex}
\end{figure}

Second, \SystemName{}'s throughput saturates when the micro-batch size is larger than 16. Figure~\ref{fig_exp_gemm_coll_1_3b} shows the source of the saturation: \SystemName{}'s computation time within an iteration grows linearly as micro-batch size grows, and is roughly the same as collective time at micro-batch size of 16. Since collective traffic is decided only by the model, the collective time within an iteration does not change when the micro-batch size changes. Thus, collectives are almost fully hidden behind computation at micro-batch sizes larger than 16 when there is no interference, and the training time is bounded by GPU computation instead of network bandwidth. Therefore, the saturated throughput at micro-batch sizes greater than 16 represents the maximum throughput provided by the GPUs. 

\noindent\textbf{MFU under Different Model Sizes.}
\
Figure~\ref{fig_exp_thpt_wrt_size} shows the end-to-end MFU of \SystemName{} and baselines on different models at their maximum trainable micro-batch sizes. We observe that 1)~\SystemName{} achieves at least 2.04$\times$ and 2.34$\times$ MFU over ZeRO-3 and ZeRO-Infinity when model size varies from 13B to 175B, and 2)~\SystemName{} still maintains high MFU (59\%) when training on a 276B model. Even though the MFU significantly drops on 505B and 1T models due to the small trainable micro-batch size (16 and 8, respectively) per GPU, \SystemName{} still achieves comparable MFU on a 1000B model to ZeRO-Infinity on a 13B model. 

\subsection{Effect of SmartNIC-Managed Collectives}
\label{subsec_exp_coll}

\begin{figure}[t]
    \begin{minipage}{0.559\linewidth}
        \begin{center}
        \vspace{-1ex}
        \includegraphics[width=0.9\linewidth]{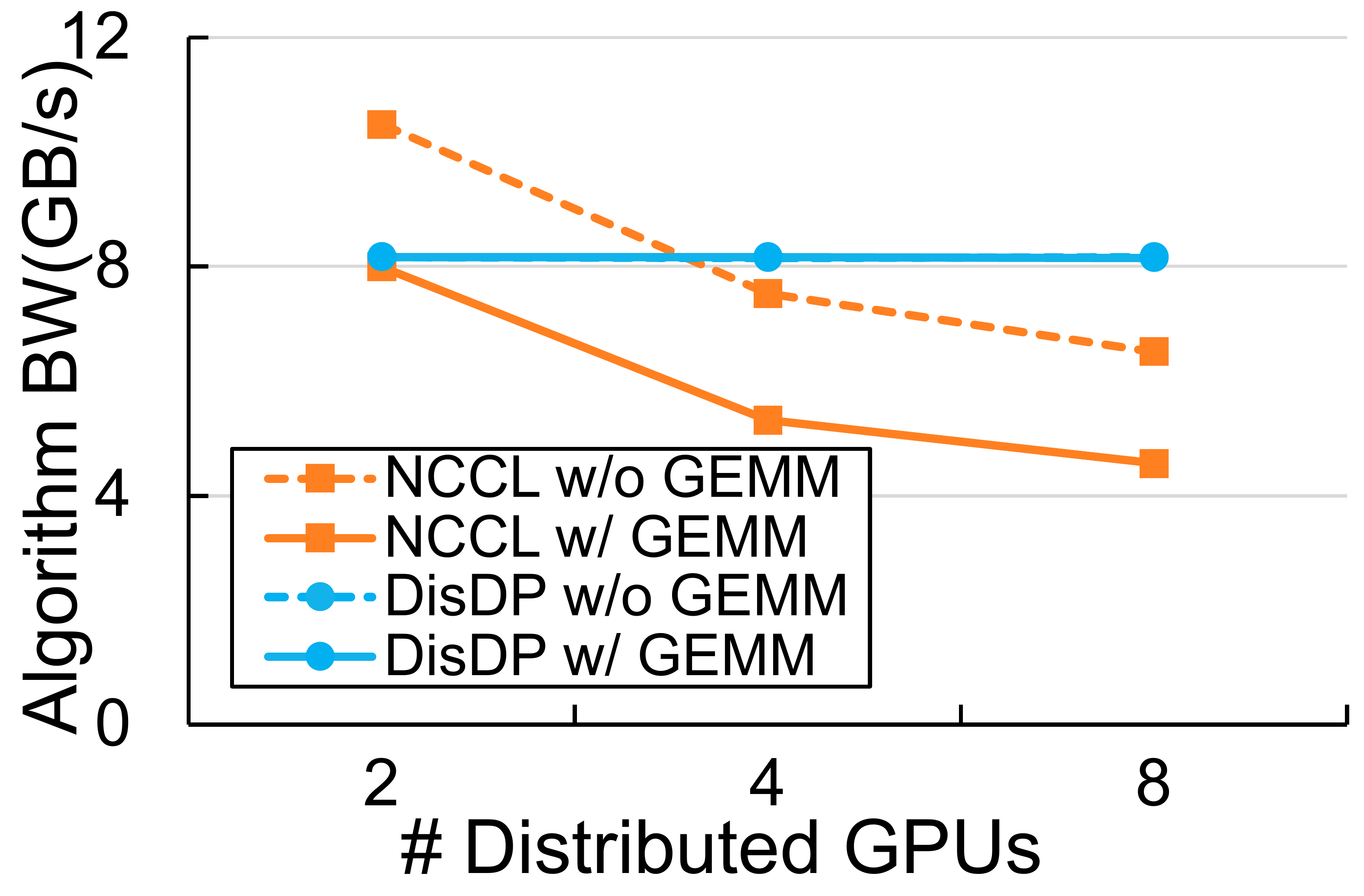}
        \end{center}
        \vspace{-3ex}
        \caption{Algorithm bandwidth of different collective primitives. }
        \label{fig_exp_interference}
        \vspace{-2ex}
    \end{minipage}
    \hfill
    \begin{minipage}{0.393\linewidth}
        \begin{center}
        \vspace{-1ex}
        \includegraphics[width=0.9\linewidth]{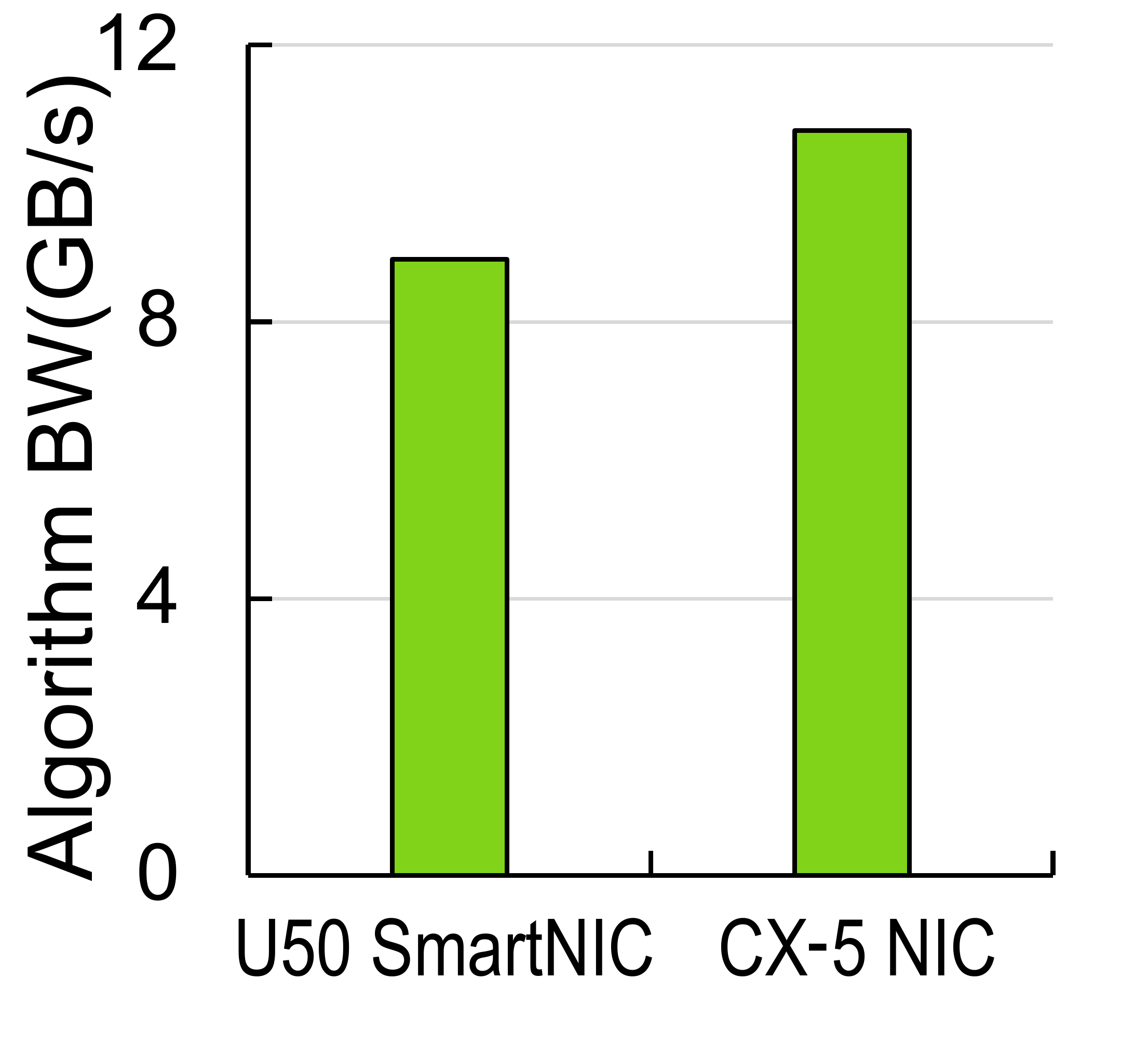}
        \end{center}
        \vspace{-3ex}
        \caption{DMA capability comparison. }
        \label{fig_exp_interference_dma}
        \vspace{-2ex}
    \end{minipage}
\end{figure}

To validate the effect of the SmartNIC-managed interference-free collectives, we run \SystemName{} \texttt{push} and \texttt{pull} and NCCL \texttt{AllReduce} to aggregate data with and without concurrent GPU GEMM kernels (experimental setup in Subsection~\ref{subsec_motivation_msdp}). 
Figure~\ref{fig_exp_interference} shows the algorithm bandwidth of \SystemName{} and NCCL. We make three observations. 

First, the concurrent GEMM does not incur overhead to \SystemName{} because \SystemName{} offloads the collectives from GPU to SmartNIC, thus minimizing the interference between collectives and GEMM. Therefore, \SystemName{} with concurrent GEMM achieves 2\%, 35\%, and 44\% higher algorithm bandwidth than that of NCCL on 2, 4, and 8 distributed GPUs, respectively. 
Second, \SystemName{} increases the algorithm bandwidth by 8\% and 20\% on 4 and 8 GPUs without concurrent GEMM compared to NCCL because the \texttt{push}-\texttt{pull} primitives reduce the collective traffic by at most half compared to NCCL \texttt{AllReduce}. 
Third, NCCL achieves 28\% higher algorithm bandwidth than \SystemName{} on 2 GPUs without concurrent GEMM. This is because the U50 SmartNIC we use has weaker DMA capability (i.e., supporting fewer outstanding PCIe transactions) than a CX-5 NIC. To illustrate this, Figure~\ref{fig_exp_interference_dma} shows the bandwidth of U50 SmartNIC and CX-5 NIC performing bidirectional DMA to GPU memory. We observe that a CX-5 NIC achieves 24\% higher bandwidth than the U50 SmartNIC. 
Therefore, we expect \SystemName{} to have a similar algorithm bandwidth to NCCL on SmartNICs with better DMA capability. 


\noindent\textbf{Effect on End-to-End Training.}
\
We break down an iteration of \SystemName{} for training Custom-175B, whose experimental setup is the same as Subsection~\ref{subsec_motivation_msdp}. The result is in Figure~\ref{fig_exp_gemm_coll_175b} and shows that 1)~\SystemName{} almost fully overlaps computation and collectives, while ZeRO-Infinity (Figure~\ref{fig_motivation_zero_interference}) does not; and 2)~\SystemName{} reduces the collective time by 20\%.

\noindent\textbf{Comparison to ZeRO-Infinity with Improved GPU Scheduling.}
\
To further break down the effect of GPU computing unit contention in compute-collective interference, we conduct a simulation comparison between \SystemName{} and ZeRO-Infinity with two hypothetical GPU scheduling optimizations: ZeRO-Infinity with MPS (ZeRO-Inf+MPS) that isolates GPU SMs, and ZeRO-Infinity with GPU SM preemption (ZeRO-Inf+Preemp). We simulate the two baselines' collectives by inserting virtual collectives with predefined bandwidth (Figure~\ref{fig_exp_interference}) to ZeRO-Infinity's computation traces, and simulate ZeRO-Inf+MPS's computation kernel by adding MPS overhead coefficient to computation traces.

\begin{figure}[t]
    \vspace{-1ex}
    \begin{minipage}{0.45\linewidth}
        \begin{center}
            \includegraphics[width=\linewidth]{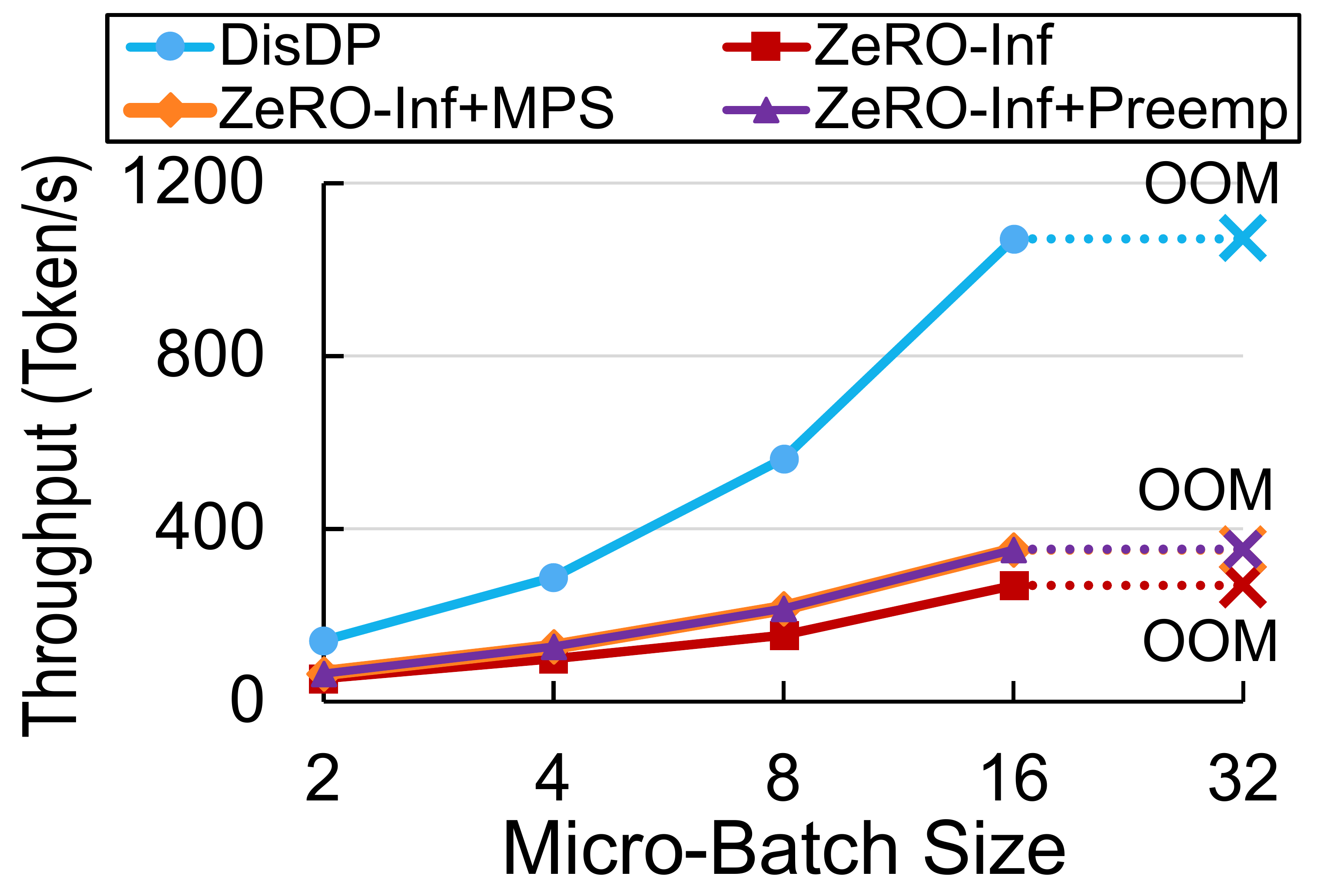}
        \end{center}
        \vspace{-3ex}
        \caption{\label{fig_exp_zero_mps}\SystemName{} vs. simulated ZeRO-Infinity w/o computing unit contention. }
        \vspace{-3ex}
    \end{minipage}
    \hfill
    \begin{minipage}{0.45\linewidth}
        \begin{center}
            \includegraphics[width=\linewidth]{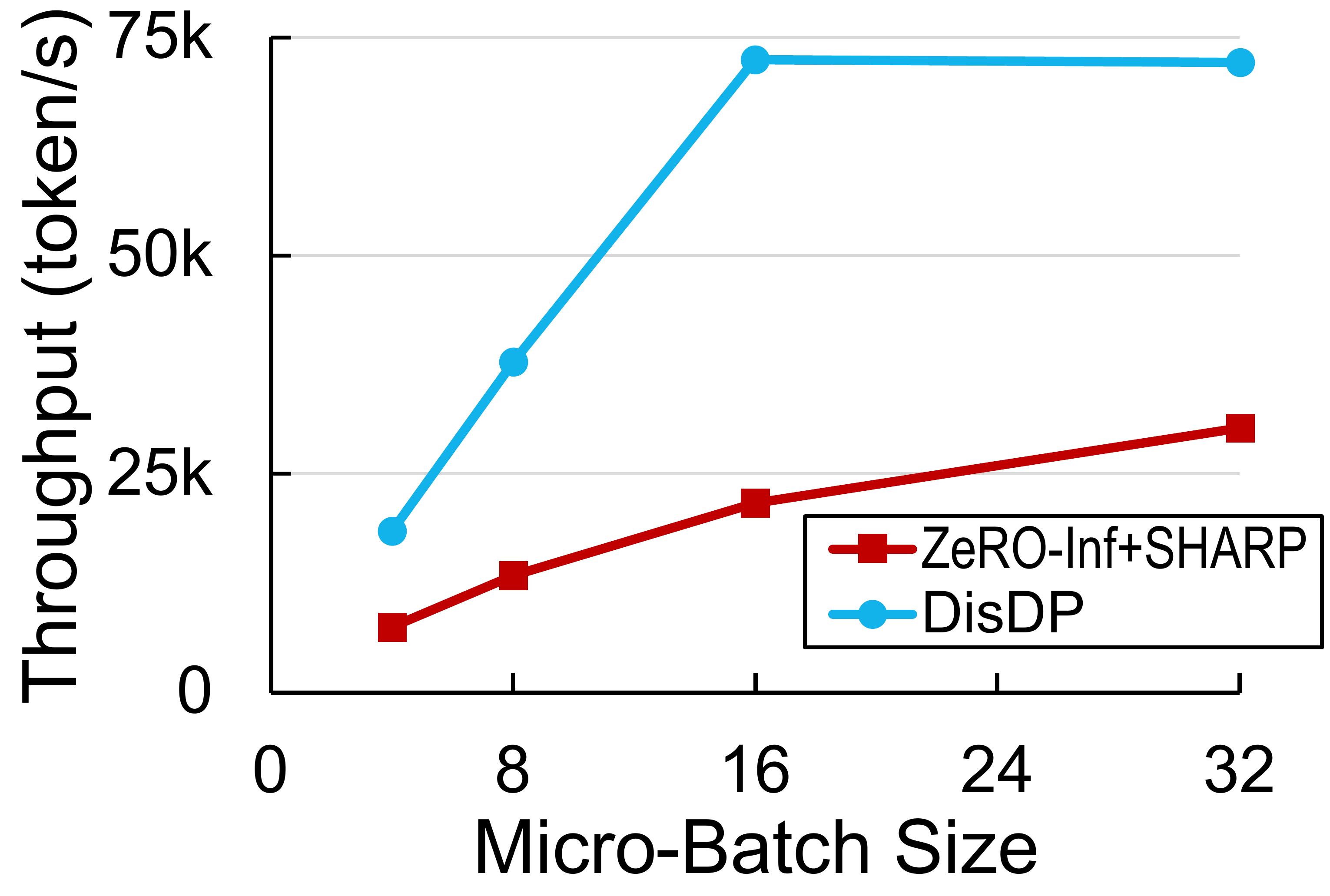}
        \end{center}
        \vspace{-3ex}
        \caption{\label{fig_exp_thpt_sharp}\SystemName{} vs. \narrow{ZeRO-Infinity w/ SHARP}.}
        \vspace{-2ex}
    \end{minipage}
\end{figure}

Figure~\ref{fig_exp_zero_mps} shows the systems' throughput for training Custom-175B. We make two observations. First, both ZeRO-Inf+MPS and ZeRO-Inf+Preemp increase ZeRO-Infinity's throughput by 30\%, and their throughput are almost overlapped (only 0.1\%\textasciitilde 0.4\% throughput differences), because both optimizations remove GPU computing unit contention, thus achieving similar effects. Second, ZeRO-Inf+MPS and ZeRO-Inf+Preemp only achieve 32.7\% and 32.8\% of \SystemName{}'s throughput, because these optimizations do not address GPU memory bandwidth contention between GEMM and collective kernels demonstrated in Subsection~\ref{subsec_motivation_msdp}, so there is still interference between computation and collectives. In contrast, \SystemName{}'s disaggregated design fully overlaps computation and collectives. In conclusion, we need full disaggregation, rather than a better GPU computing unit scheduling.

\noindent\textbf{Comparison to ZeRO-Infinity with SHARP.}
\
To eliminate the impact of in-network aggregation, we compare \SystemName{} and ZeRO-Infinity with NVIDIA SHARP~\cite{sharp, sharp-agg}, so that the main difference between the two systems is GPU-/SmartNIC-managed collectives. Because we only have 4$\times$ 100Gbps CX-6 NICs that SHARP requires, we run ZeRO-Infinity on 4 workers, each has 1$\times$ GPU, 4$\times$ SSDs (to ensure SSD bandwidth is the same as \SystemName{}), 1$\times$ CX-6 NICs connected to an Infiniband switch~\cite{qm8790}. Figure~\ref{fig_exp_thpt_sharp} shows their throughput when training on the OPT-30 B model. 

We observe that \SystemName{} achieves 2.38\textasciitilde 3.35$\times$ throughput over ZeRO-Infinity+SHARP, because ZeRO-Infinity+SHARP still suffers from GEMM-collective interference due to partial disaggregation of compute and network. In conclusion, we need full disaggregation from \SystemName{}, rather than only using in-switch aggregation. 

\subsection{Ablation Study}
\label{subsec_exp_ablation}

\subsubsection{Inteference-Free Collectives Ablation}

\begin{figure}[t]
    \vspace{-1ex}
    \begin{minipage}{0.45\linewidth}
        \begin{center}
            \includegraphics[width=\linewidth]{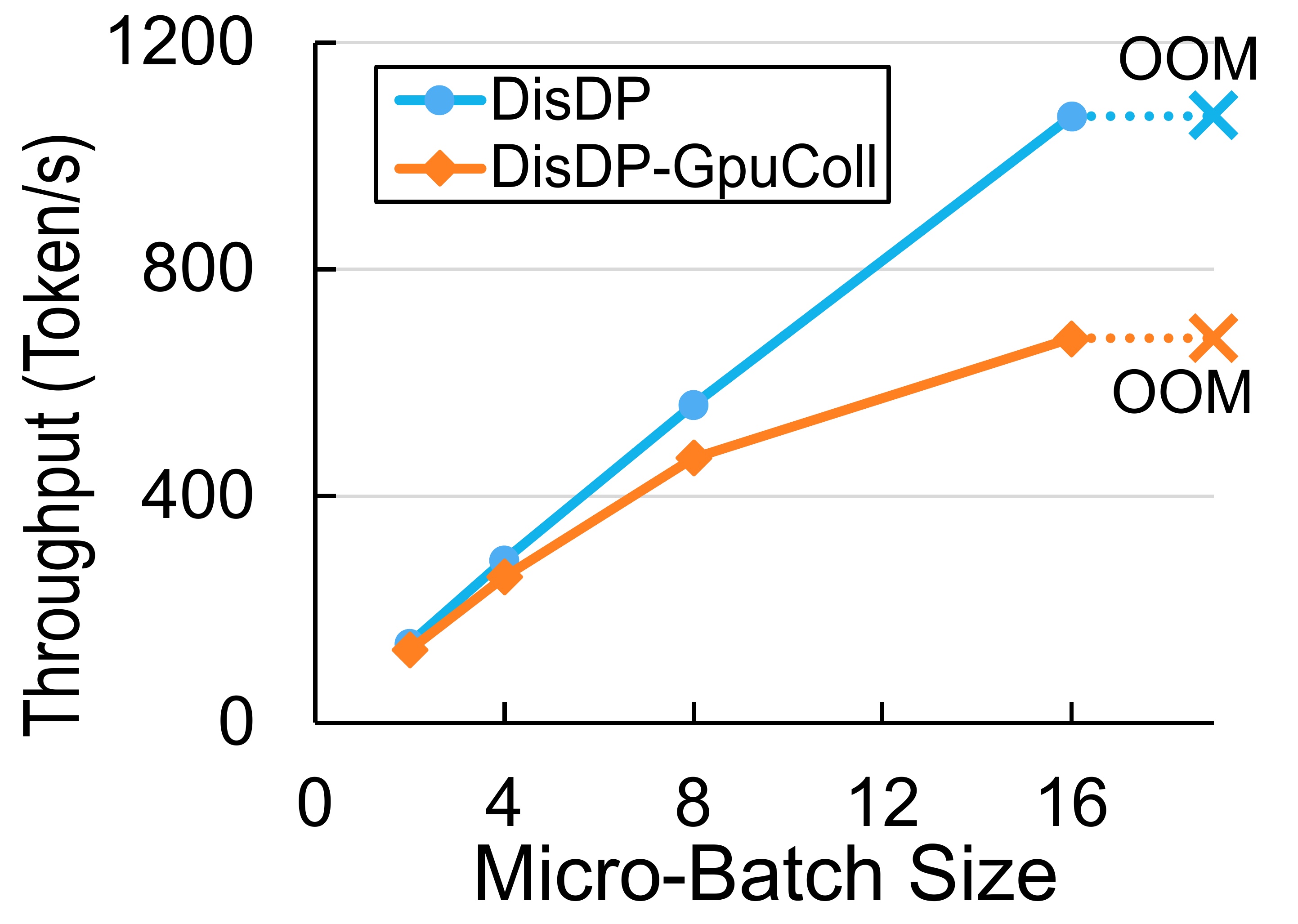}
        \end{center}
        \vspace{-3ex}
        \caption{\label{fig_exp_ablation_collectives}Effect of interference-free collectives. }
        \vspace{-2ex}
    \end{minipage}
    \hfill
    \begin{minipage}{0.45\linewidth}
        \begin{center}
            \includegraphics[width=\linewidth]{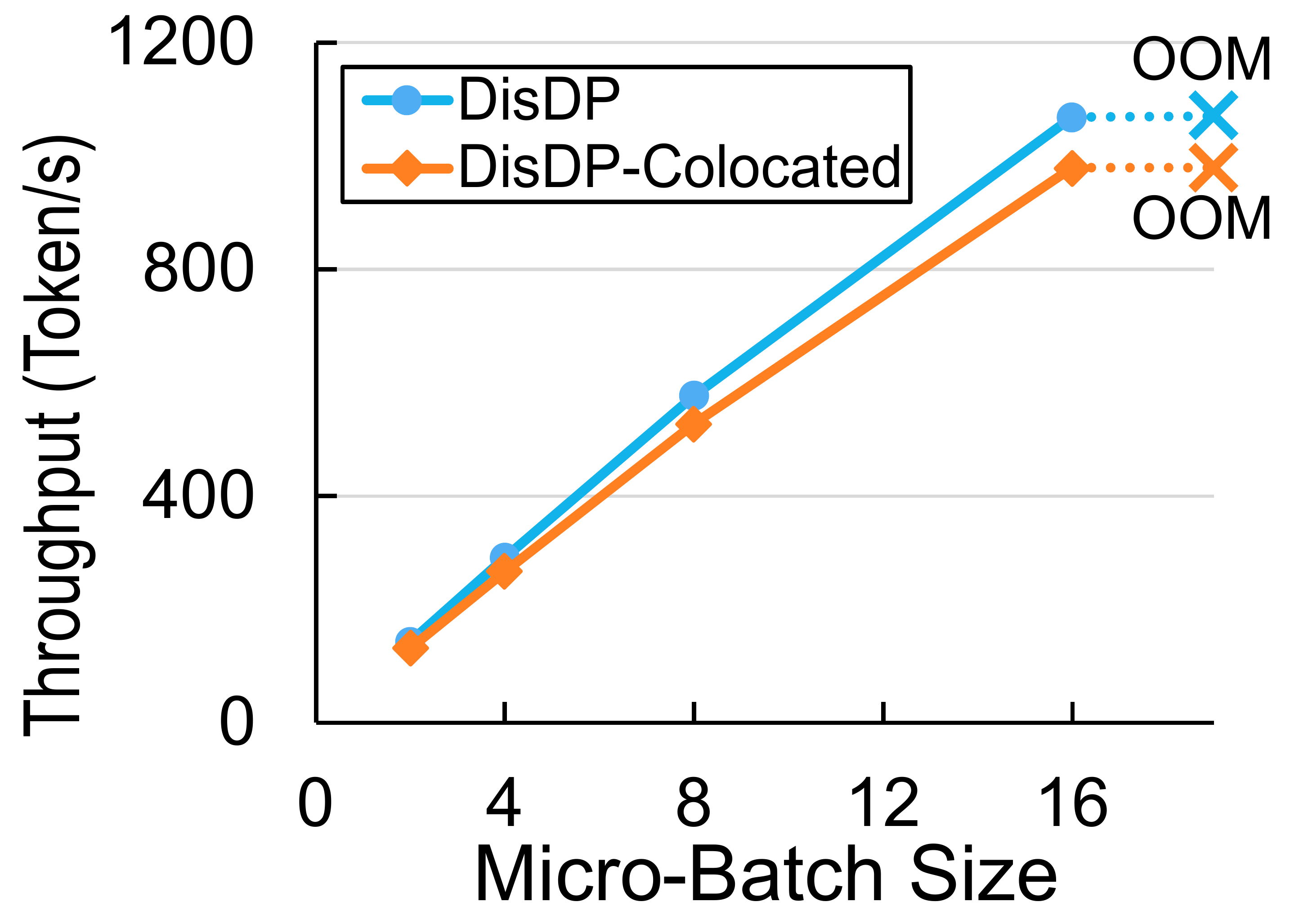}
        \end{center}
        \vspace{-3ex}
        \caption{\label{fig_exp_ablation_pushpull}Effect of many-to-one reliable protocol. }
        \vspace{-2ex}
    \end{minipage}
\end{figure}

To assess the contribution of the interference-free collectives, we compare \SystemName{} with \SystemName{}-GpuColl that uses a GPU kernel to issue \texttt{push/pull} requests to SmartNIC and poll completion, so that GEMM and \texttt{push/pull} kernels compete for GPU computing units. Since the collectives are coupled with many-to-one protocol, we cannot actually run \SystemName{}-GpuColl, so we simulate the throughput of \SystemName{}-GpuColl training Custom-175B on our 8-worker cluster, as Figure~\ref{fig_exp_ablation_collectives} shows. 

We make two observations. First, \SystemName{} achieves 1.58$\times$ maximum throughput over \SystemName{}-GpuColl, because \SystemName{}-GpuColl fails to fully overlap computation and network communication due to computing unit contention. Second, when a small micro-batch size is (2 and 4) leads to a trivial throughput gain because a small micro-batch size takes much shorter computation time compared to collectives (shown in Figure~\ref{fig_exp_gemm_coll_175b}), so the benefit of compute-network overlapping is not significant. However, the performance gain at large micro-batch size~(16) is significant, because compute and network then take roughly the same time and are fully overlapped, thus bringing the most significant benefit.

\subsubsection{Many-to-One Reliable Protocol Ablation}

To examine the effect of the many-to-one reliable protocol, we compare \SystemName{} with \SystemName{}-Colocated, a baseline that uses conventional colocated PS architecture, rather than \SystemName{}'s scalable PS to execute optimizer. Since \SystemName{}'s many-to-one protocol is coupled with SmartNIC-centric collectives, we cannot actually run \SystemName{}-Colocated, so we simulate \SystemName{}-Colocated by inserting virtual collectives to \SystemName{}'s compute traces.

Figure~\ref{fig_exp_ablation_pushpull} shows the simulated throughput of two implementations on the Custom-175B model. \SystemName{} achieves 1.10$\times$ throughput over \SystemName{}-Colocated at different batch sizes, because colocated PS incurs both worker and PS traffic on the same machine, leading to heavier collective traffic during backward propagation. During backward propagation, a SmartNIC has 1 worker push and 1 worker pulls for each layer, incurring $S$ send receive. Additionally, the SmartNIC has 1 PS push and 1 PS pull, incurring $\frac{S}{N}$ send receives on each PS. Therefore, \SystemName{}-Colocated has $(1+\frac{1}{N})$ send receives in total, which is 1.13$\times$ collective traffic than \SystemName{} on 8 workers (1 push and 1 pulls during backward stage for worker, which is $S$ send receives. Similarly, PS also has $S$ send receives) on each NIC, leading to longer collective time. At micro-batch sizes up to 16, \SystemName{}-Colocated's collectives take longer than computation, thus the overall throughput is bounded by collective time. Therefore, \SystemName{}-Colocated's heavier collective traffic causes lower throughput.

\subsubsection{Step-Centric Optimizer Pipelining Ablation}

To examine the effect of the step-centric optimizer pipelining, we compare \SystemName{} with \SystemName{}-LC that implements na\"ive layer-centric pipelining for PS. 

\begin{figure}[t]
    \centering
    \vspace{-1ex}
    \begin{minipage}{0.397\linewidth}
        \begin{center}
            \includegraphics[width=\linewidth]{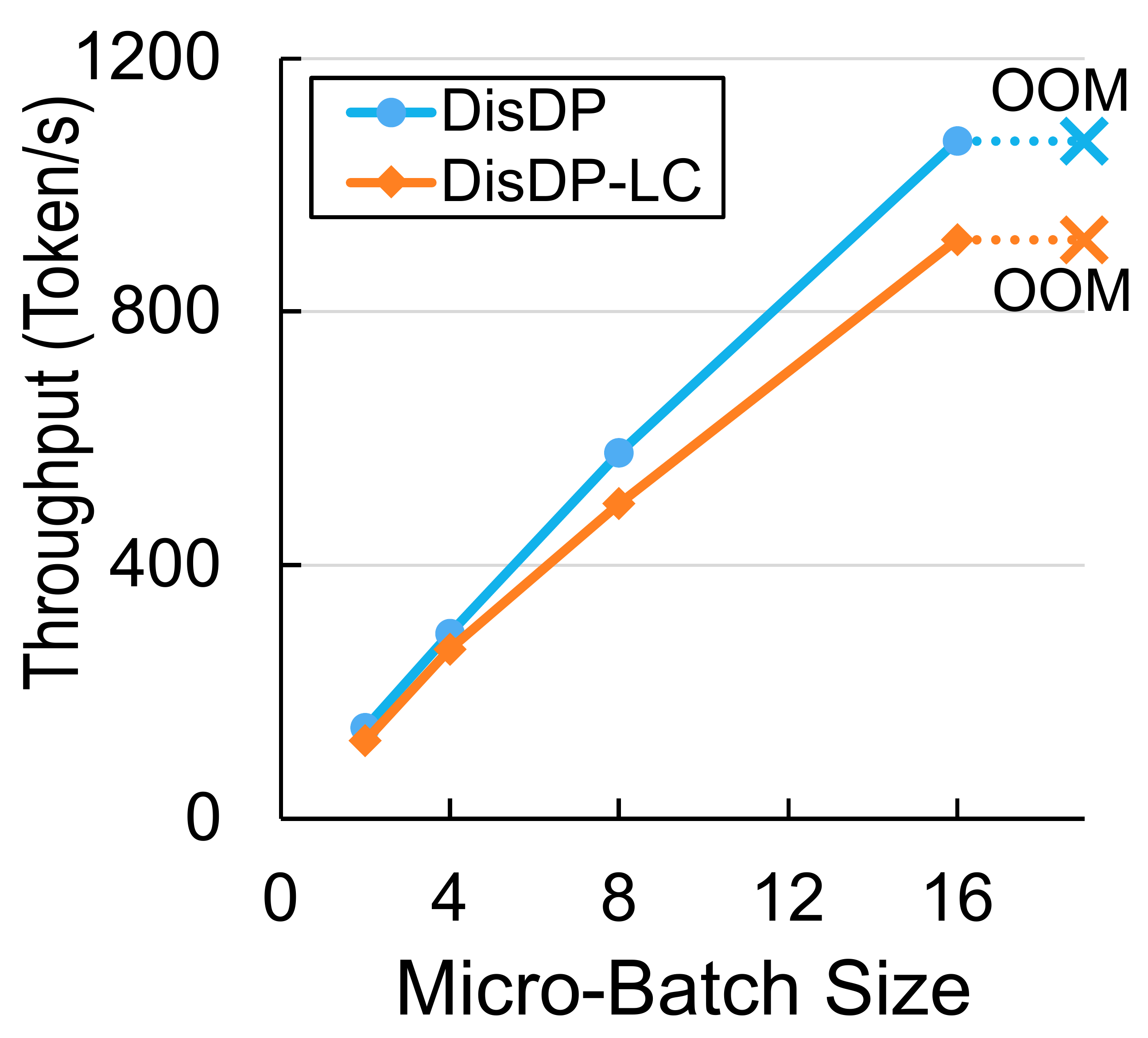}
        \end{center}
        \vspace{-3ex}
        \caption{\label{fig_exp_ablation_optimizer}Effect of step-centric pipelining.}
    \end{minipage}
    \hfill
    \begin{minipage}{0.544\linewidth}
        \begin{center}
            \includegraphics[width=\linewidth]{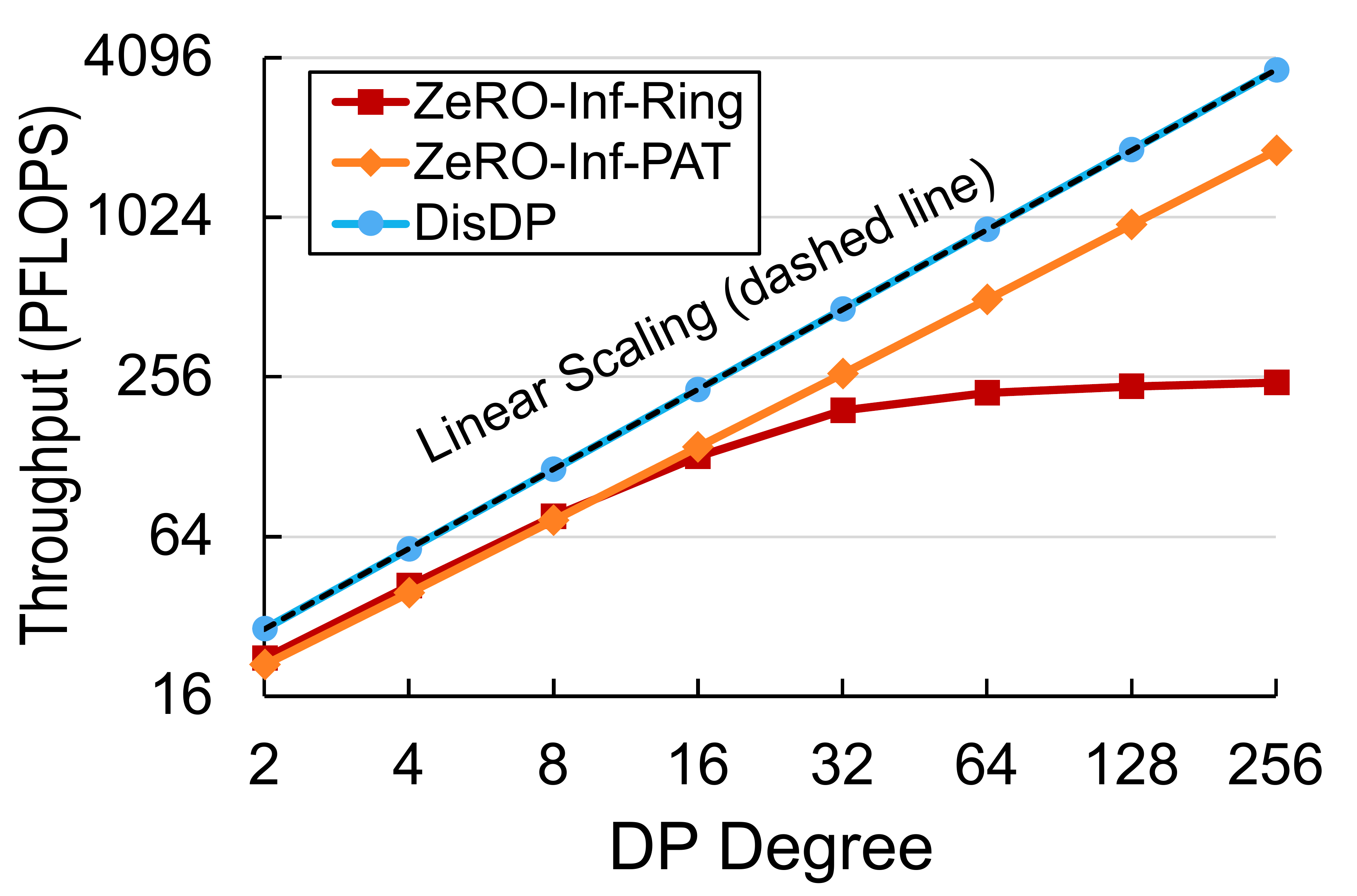}
        \end{center}
        \vspace{-3ex}
        \caption{\label{fig_exp_weak_scale}Scaling of \SystemName{}: ASTRA-sim simulation on 175B model with TP8, PP16, and varying DP degree.} 
        \vspace{-2ex}
    \end{minipage}
\end{figure}

Figure~\ref{fig_exp_ablation_optimizer} shows the throughput of two implementations on the Custom-175B model. We observe that \SystemName{} achieves 1.10\textasciitilde 1.17$\times$ higher throughput than \SystemName{}-LC, because \SystemName{}-LC's throughput at large batch sizes is bottlenecked by the layer-centric optimizer that fails to serve line-rate packets. In contrast, \SystemName{} with step-centric pipelining can serve line-rate packets, because the throughput is bottlenecked by the network rather than the optimizer execution.

\subsection{Scaling Out of \SystemName}
\label{subsec_exp_sca}

To show the horizontal scalability of \SystemName{}, we simulate \SystemName{} and ZeRO-Infinity using ASTRA-sim~\cite{astrasim_ispass23, chakra_arxiv} on industrial-scale clusters. We follow the existing industrial practices~\cite{megascale, llama3_isca2025} with 3D parallelism~\cite{deepseekv3_isca2025, fred_isca2025} to employ 8-degree TP~\cite{megatronlm, megatron2021, 2dtp, 3dtp, sarathi, meshslice_isca2025} over 600GB/s scale-up networks, and 16-degree 1F1B PP~\cite{megatron2021, gpipe, pipedream, dapple, mobius, zerobubble, narayanan2021memory, adapipe, zhao2021vpipe, hetpipe, ttgnn_micro23} and varying-degree DP over 100Gbps scale-out networks. We vary the DP degree to up to 256,
which covers the largest DP degree that industrial practices like Llama-3~\cite{llama3_isca2025} (128 DP degree) and MegaScale~\cite{megascale} (192 DP degree) adopts.
We run ZeRO-Infinity's collectives on two algorithms: The typical ring algorithms (ZeRO-Inf-Ring), and PAT~\cite{pat} (ZeRO-Inf-PAT) that performs hierarchical tree-based RS and AG.
Figure~\ref{fig_exp_weak_scale} shows the global TFLOPS with micro-batch sizes of 16 (the same setting in Llama~\cite{llama3}). 

We make two observations. First, ZeRO-Infinity with ring algorithm scales poorly with $>$16 DP degree, because ring collectives introduce a long dependency tree at a large network communication scale~\cite{rina_icnp24} and thus is vulnerable to interference~\cite{mics, fsdp}. In contrast, \SystemName{} linearly scales out because its hardware-accelerated SmartSwitch-assisted many-to-one collectives compress the dependency chain into a few hops between each worker and the PS. Second, \SystemName{} achieves higher throughput than ZeRO-Infinity with both algorithms. On 256-degree DP, \SystemName{} achieves 2.0$\times$ throughput over ZeRO-Infinity with PAT and 15.1$\times$ throughput over that with the ring algorithm, because ZeRO-Infinity with both algorithms suffers from GEMM-collective interference, while \SystemName{} eliminates the interference by SmartNIC-managed collectives. In conclusion, \SystemName{}'s disaggregation of compute, network, and storage enables efficient 3D parallelism with a large DP~degree.

\begin{figure}[t]
    \vspace{-1ex}
    \begin{minipage}{0.496\linewidth}
        \begin{center}
            \includegraphics[width=\linewidth]{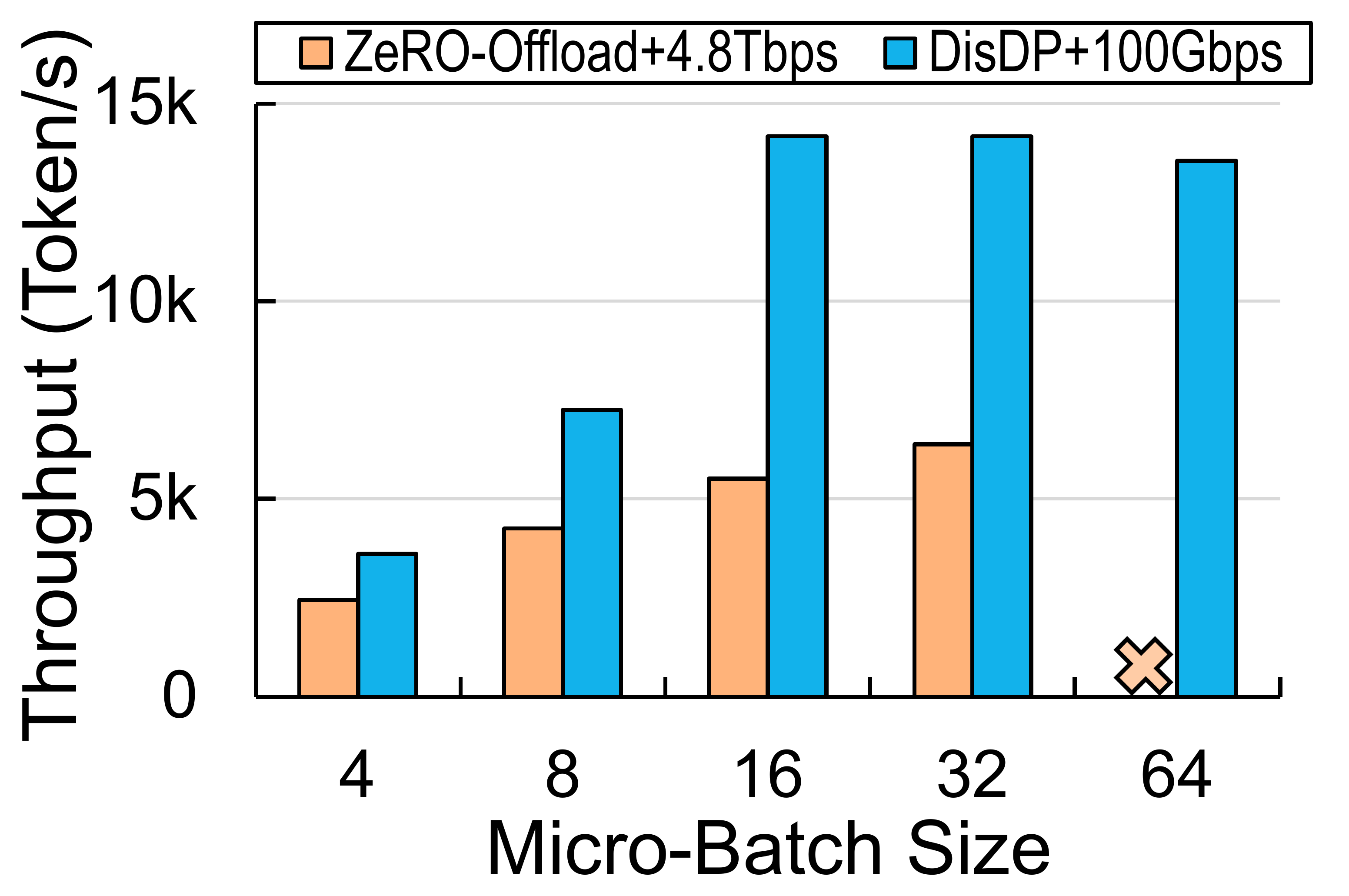}
        \end{center}
        \vspace{-3.5ex}
        \caption{\label{fig_exp_compare_dgx} \SystemName{} under 100Gbps network vs. ZeRO- \narrow{Offload under 4.8Tbps NVLink.}}
        \vspace{-2ex}
    \end{minipage} 
    \hfill
    \begin{minipage}{0.402\linewidth}
        \begin{center}
            \includegraphics[width=\linewidth]{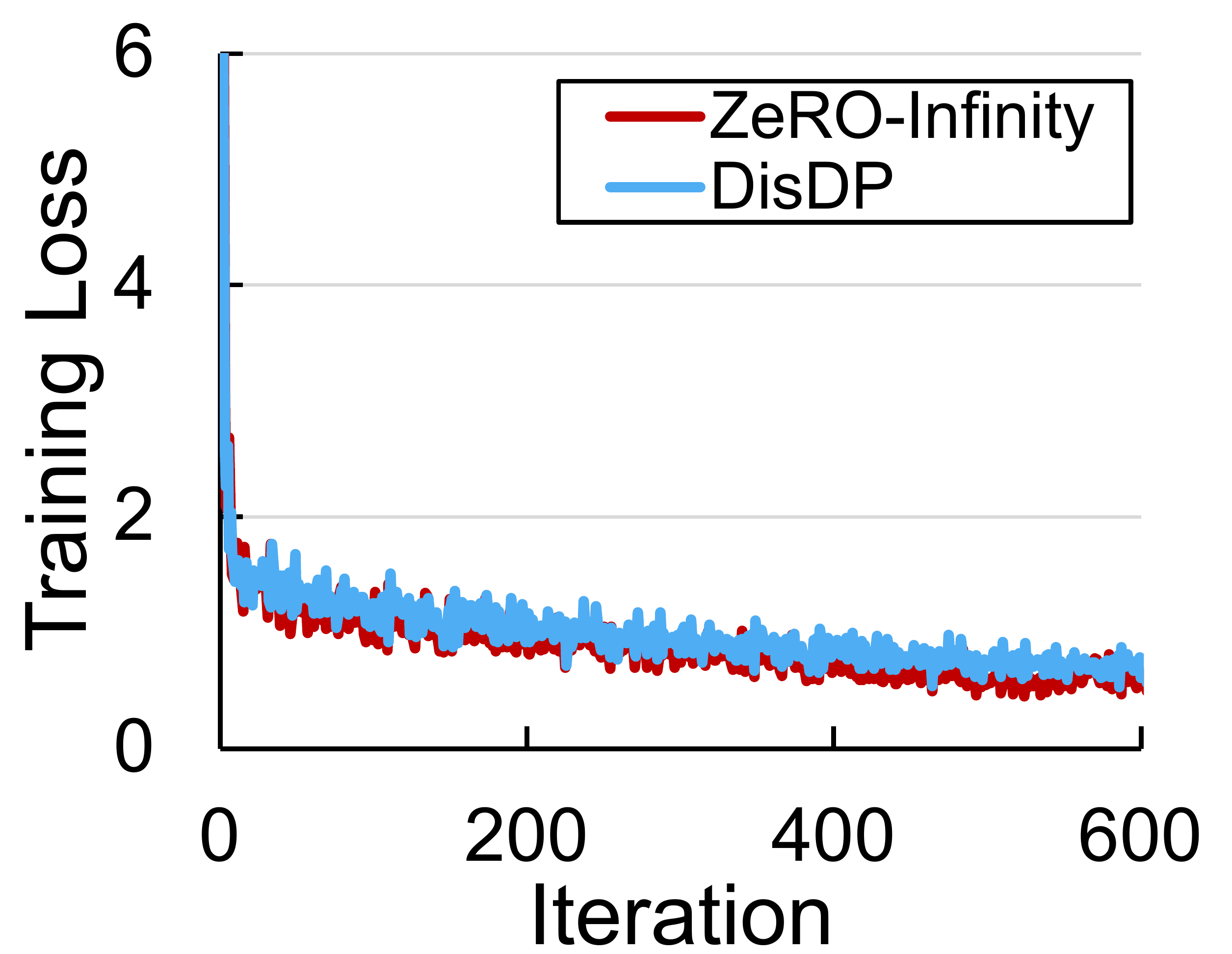}
        \end{center}
        \vspace{-4ex}
        \caption{\label{fig_exp_convergence} Convergence: \SystemName{} vs. ZeRO-Infinity}
        \vspace{-1ex}
    \end{minipage} 
\end{figure}

\subsection{Effect of Faster Network}
\label{subsec_exp_dgx}

We validate the benefit of \SystemName{} over baselines with higher network bandwidth. 
Due to a lack of 400Gbps NICs, we choose 8$\times$ A100-40GB GPUs fully connected by 600GB/s NVLink, which is an order of magnitude faster than 100Gb/s of \SystemName{}. 
We compare \SystemName{} on 100Gbps Ethernet with ZeRO-Offload on a 600GB/s (4.8Tbps) fast network when training the OPT-13B model.
Figure~\ref{fig_exp_compare_dgx} shows the results. We make two observations. 
First, \SystemName{} achieves 2.22$\times$ throughput over NVLink-enhanced ZeRO-Offload, mainly because ZeRO-Offload suffers from severe interference between GEMM and collectives, while \SystemName{} addresses this by disaggregating compute and collectives. 
Second, \SystemName{} achieves saturated throughput at micro-batch sizes greater than 16 due to GPU compute bottleneck (as demonstrated in Subsection~\ref{subsec_exp_endtoend}). Thus, a 100Gbps network rather than high-bandwidth NVLink is sufficient when 1) the micro-batch size is large so that computation takes longer than collectives, and 2) computation and collectives are fully overlapped so collective overhead are completely hidden by computation.
In conclusion, we argue for a fully disaggregated design with a relatively slow network, rather than simply upgrading to a faster network under an aggregated infrastructure, to fully address the communication bottleneck in LLM training.

\subsection{Cost-Efficiency Comparison}
\label{subsec_exp_cost_efficiency}

To show cost-efficiency benefits of \SystemName{}, we compare the price and throughput of \SystemName{} on a 32-GPU commodity cluster to ZeRO-Infinity on both the 32-GPU commodity cluster and a 32-GPU DGX cluster (intra-machine GPUs are fully connected with 600GB/s NVLink). We use throughput-per-dollar as the cost-efficiency metric, so the cost-efficiency benefit is calculated by throughput-per-dollar ratio ${\rm =\frac{Relative~Throughput}{Relative~Price}}$. We gather the price  of a DGX machine with 8$\times$ A100-40GB GPUs from~\cite{dgxprice} and the remaining components from wholesalers' public quotes,\footnote{Quote of worker and PS machines is the total price after configuring CPU, memory, system storage, and labor. } as listed in Table~\ref{table_component_price}. Due to a lack of GPUs, we use ASTRA-sim\cite{astrasim_ispass23} to simulate the throughput of \SystemName{} and ZeRO-Infinity when training the 175B model. 
Table~\ref{table_cost_efficiency} shows that \SystemName{} only costs 60\% of that of a DGX cluster, while achieving 1.17$\times$ throughput than ZeRO-Infinity on DGX.
As a result, \SystemName{} achieves a competitive 1.96$\times$ throughput-per-dollar over ZeRO-Infinity on the DGX cluster mainly due to \SystemName{}'s fully disaggregated design. In contrast, ZeRO-Infinity on the commodity cluster only achieves 0.22$\times$ throughput and 0.40$\times$ cost-efficiency over that on the DGX cluster.

\begin{table}[t]\footnotesize
    \centering
    \caption{Price of each component.}
    \vspace{-2ex}
    \label{table_component_price}
    \begin{tabular}{llll}
        \specialrule{.1em}{.05em}{.05em} 
        \textbf{Component} & \textbf{Price (\$)} & \textbf{Component} & \textbf{Price (\$)} \\ \specialrule{.1em}{.05em}{.05em} 
        \multirow{2}{*}{\textbf{\makecell[l]{Worker w/ CPU \\ and Memory}}} & \multirow{2}{*}{9,060~\cite{worker_price}} & \textbf{SSD} & 850~\cite{ssd_price} \\ \cline{3-4}
        & & \textbf{GPU} & 8,800~\cite{gpu_price} \\ \hline
        \multirow{2}{*}{\textbf{\makecell[l]{PS w/ CPU \\ and Memory}}} & \multirow{2}{*}{10,588~\cite{ps_price}} & \textbf{CX-5 NIC} & 755~\cite{nic_price} \\ \cline{3-4}
        & & \textbf{SmartNIC} & 2,965~\cite{u50} \\ \hline
        \textbf{\vnarrow{Conventional Switch}} & 18,990~\cite{switch_price} & \textbf{SmartSwitch} & 10,020~\cite{smartswitch_price} \\
        \specialrule{.1em}{.05em}{.05em} 
    \end{tabular}
    \vspace{-4ex}
\end{table}

\begin{table}[!t]\footnotesize
    \centering
    \caption{Cost-efficiency: Training 175B model on 32 GPUs.}
    \label{table_cost_efficiency}
    \vspace{-2ex}
    \resizebox{\columnwidth}{!}{
        \begin{tabular}{lccc}
            \specialrule{.1em}{.05em}{.05em} 
            & \textbf{\makecell{\SystemName{}\\on Ours}} & \textbf{\makecell{\narrow{ZeRO-Infinity}\\on Ours}} & \textbf{\makecell{ZeRO-Infinity\\on DGX}} \\ \specialrule{.1em}{.05em}{.05em} 
            \textbf{Machine Price (\$)} & \makecell[c]{10,600$\times$4 \\ + 17,000$\times$1} & 10,600$\times$4 & \multirow{2}{*}[0.5em]{\makecell[c]{154,800$\times$4 \\ (4 machines \\ + 32 GPUs \\ + 32 NICs \\ + 16 SSDs)}} \\ \cline{1-3}
            \textbf{GPU Price (\$)} & 8,000$\times$32 & 8,000$\times$32 & \\ \cline{1-3} 
            \textbf{NIC Price (\$)} & 1,600$\times$33 & 1,000$\times$32 &  \\ \cline{1-3} 
            \textbf{SSD Price (\$)} & 310$\times$12 & 310$\times$16 &  \\ \hline
            \textbf{Switch Price (\$)} & 10,000$\times$1 & 19,000$\times$1 & 19,000$\times$1 \\ \hline
            \textbf{Total Price (\$)} & 381,920 & 354,360 & 638,200 \\ \hline
            \textbf{Throughput (Token/s)} & 4,386 & 822 & 3,745 \\ \hline 
            \textbf{Relative Throughput} & 1.17$\times$ & 0.22$\times$ & 1$\times$ \\ \hline
            \textbf{Relative Price} & 0.60$\times$ & 0.56$\times$ & 1$\times$  \\ \hline
            \textbf{\narrow{Throughput/Dollar Ratio}} & 1.96$\times$ & 0.40$\times$ & 1$\times$ \\ \specialrule{.1em}{.05em}{.05em} 
        \end{tabular}
    }
    \vspace{-3ex}
\end{table}

\subsection{Training Convergence}
\label{subsec_exp_convergence}

To validate that \SystemName{} does not affect the training convergence, we fine-tune the OPT-66B model (the largest pretrained model we can access) on the 8-GPU cluster with the rm-static dataset~\cite{rm-static}. During the fine-tuning process, we set the micro-batch size to 8 per worker and the data format to bf16. The fine-tuning process takes \textasciitilde 3 days for ZeRO-Infinity and \textasciitilde 12 hours for \SystemName{}.
We compare the training loss of \SystemName{} and ZeRO-Infinity, as shown in Figure \ref{fig_exp_convergence}. We observe that \SystemName{} and ZeRO-Infinity have roughly overlapped loss curves, indicating \SystemName{} keeps the same convergence rate.

\section{Related Work}

\noindent\textbf{SmartNIC-Offloaded Collectives.}
\
Recent works~\cite{suresh2023novel, graham2024optimizing, oodsmpi_sc25} offload collective scheduling from CPU to SoC-based SmartNICs to overlap network communication and computation. However, they build on the RDMA protocol that only supports reliable one-to-one connections and cannot be applied to many-to-one \texttt{push/pull}. Khalilov et al.~\cite{khalilov2024network} enable one-to-many broadcast with SoC-based SmartNIC. However, it does not simultaneously support many-to-one aggregation because an SoC-based SmartNIC cannot process line-rate \texttt{push} and \texttt{pull} simultaneously due to its limited internal switch link and Arm memory bandwidth constraints (Subsection~\ref{subsec_design_bluefield}). In contrast, \SystemName{} uses FPGA-based SmartNICs to enable concurrent many-to-one \texttt{push} and \texttt{pull} for efficient FSDP training.

\noindent\textbf{Compute-Network Overlapping.}
\
Recent systems propose kernel decomposition~\cite{overlap_asplos22, centauri_asplos24, lancet_mlsys2024, flashoverlap_eurosys26, huang2023p4sgd, fpga_apsys25, RoPeerTo_eurosys26} for computation and collectives kernels, so as to enable fine-grained overlapping for TP and expert parallelism where compute and network cannot overlap due to data dependency. However, they do not address the compute-network interference in MSDP, so they still suffer from GPU computing unit contention that prevents compute and network from fully overlapping. In contrast, \SystemName{} enables full compute-network disaggregation with SmartNICs.

Some works enable compute-network overlapping by GPU kernel fusion~\cite{coconet_asplos22, tilelink_mlsys2025, optimizing_sc2024, mirage_arxiv2025}, which addresses GPU computing unit contention by manually allocating appropriate GPU threads for GEMM and collectives. However, kernel fusion does not address GPU memory bandwidth and L2 contention, so they still suffer from collective bandwidth drop, as shown in Subsection~\ref{subsec_motivation_msdp}. In contrast, \SystemName{} eliminates the interference with SmartNIC-centric collectives. 

\noindent\textbf{In-Switch Aggregation.}
\
Recent works~\cite{netagg, incbricks, switchml, atp, esa, programmable_switch_nsdi23, a2tp, drl_isca19, rackblox_sosp23, taurus_asplos2022, network_isca2020, netreduce_asplos2023} leverage SmartSwitches to reduce the network data volume during DNN training. However, these systems only support \texttt{AllReduce}-like semantics, thus not compatible with LLM training. SHARP~\cite{sharp-agg, sharp, mina} integrates collective logics on dedicated InfiniBand switches, However, SHARP relies on GPUs for data chunk management and consequently suffers from computation-collective interference. In contrast, \SystemName{} proposes novel smart network-managed collectives to improve GPU utilization.



\section{Conclusion}

In this paper, we propose \SystemName, a fully disaggregated data-parallel architecture for 100B-scale distributed training. 
The key idea is 1) offloading collectives to SmartNICs to avoid interference between GEMM kernels and collective kernels, and 2) offloading the optimizer to a scalable PS that supports serving arbitrary numbers of workers.
The experimental results show that \SystemName{} achieves 1.96$\times$ throughput-per-dollar over ZeRO-Offload on DGX when training on a 175B model.
\section*{Acknowledgement.}

We thank the anonymous ISCA reviewers for improving this paper. 
The work is supported by the following grants: the Major Project of the Zhejiang Provincial Natural Science Foundation under Grant No. LD26F020002, 
the National Natural Science Foundation of China under the grant numbers (62472384, 62441236, U24A20326). Zeke Wang is the corresponding author.

\bibliographystyle{IEEEtran}
\bibliography{bigmodelcite}

@inproceedings{deepseek_infra_sc24,
author = {An, Wei and Bi, Xiao and Chen, Guanting and Chen, Shanhuang and Deng, Chengqi and Ding, Honghui and Dong, Kai and Du, Qiushi and Gao, Wenjun and Guan, Kang and Guo, Jianzhong and Guo, Yongqiang and Fu, Zhe and He, Ying and Huang, Panpan and Li, Jiashi and Liang, Wenfeng and Liu, Xiaodong and Liu, Xin and Liu, Yiyuan and Liu, Yuxuan and Lu, Shanghao and Lu, Xuan and Nie, Xiaotao and Pei, Tian and Qiu, Junjie and Qu, Hui and Ren, Zehui and Sha, Zhangli and Su, Xuecheng and Sun, Xiaowen and Tan, Yixuan and Tang, Minghui and Wang, Shiyu and Wang, Yaohui and Wang, Yongji and Xie, Ziwei and Xiong, Yiliang and Xu, Yanhong and Ye, Shengfeng and Yu, Shuiping and Zha, Yukun and Zhang, Liyue and Zhang, Haowei and Zhang, Mingchuan and Zhang, Wentao and Zhang, Yichao and Zhao, Chenggang and Zhao, Yao and Zhou, Shangyan and Zhou, Shunfeng and Zou, Yuheng},
title = {Fire-Flyer AI-HPC: A Cost-Effective Software-Hardware Co-Design for Deep Learning},
year = {2024},
booktitle = {SC}
}

@inproceedings{RoPeerTo_eurosys26,
author = {Venere, Marco and Sorrentino, Giuseppe and Ramhorst, Benjamin and Heer, Maximilian Jakob and Petrica, Lucian and Korolija, Dario and Santambrogio, Marco D. and Conficconi, Davide and Alonso, Gustavo and O'Brien, Ken},
title = {RoPeerTo: A Datacenter-Scale Architecture for Peer-To-Peer DMA between GPUs and FPGAs},
year = {2026},
booktitle = {EuroSys}
}

@inproceedings{fpga_apsys25,
author = {Fahmy, Suhaib A. and Yang, Ziyi and Chen, Yixi and Alonso, Gustavo and Istv\'{a}n, Zsolt and Canini, Marco},
title = {FPGAs Are the Hero In-Network Computing Needs},
year = {2025},
booktitle = {APSys}
}

@article{huang2023p4sgd,
  title={P4sgd: Programmable switch enhanced model-parallel training on generalized linear models on distributed fpgas},
  author={Huang, Hongjing and Li, Yingtao and Sun, Jie and Zhu, Xueying and Zhang, Jie and Luo, Liang and Li, Jialin and Wang, Zeke},
  journal={IEEE Transactions on Parallel and Distributed Systems},
  volume={34},
  number={8},
  pages={2311--2324},
  year={2023},
  publisher={IEEE}
}

@inproceedings{in_network_computing_eurosys19,
    author = {Tokusashi, Yuta and Dang, Huynh Tu and Pedone, Fernando and Soul\'{e}, Robert and Zilberman, Noa},
    title = {The Case For In-Network Computing On Demand},
    year = {2019},
    booktitle = {EuroSys}
}

@inproceedings{inc_hotnets19,
    author = {Xiong, Zhaoqi and Zilberman, Noa},
    title = {Do Switches Dream of Machine Learning? Toward In-Network Classification},
    year = {2019},
    booktitle = {HotNets}
}

@article{bert,
  title={Bert: Pre-training of deep bidirectional transformers for language understanding},
  author={Devlin, Jacob and Chang, Ming-Wei and Lee, Kenton and Toutanova, Kristina},
  journal={arXiv preprint},
  year={2018}
}

@article{t5,
  title={Exploring the limits of transfer learning with a unified text-to-text transformer.},
  author={Colin Raffel and Noam Shazeer and Adam Roberts and Katherine Lee and Sharan Narang and Michael Matena and Yanqi Zhou and Wei Li and Peter J. Liu},
  journal={J. Mach. Learn. Res.},
  year={2020}
}

@inproceedings{gpt-3,
  title={Language models are few-shot learners},
  author={Tom Brown and Benjamin Mann and Nick Ryder and Melanie Subbiah and Jared D Kaplan and Prafulla Dhariwal and Arvind Neelakantan and Pranav Shyam and Girish Sastry and Amanda Askell and Sandhini Agarwal and Ariel Herbert-Voss and Gretchen Krueger and Tom Henighan and Rewon Child and Aditya Ramesh and Daniel Ziegler and Jeffrey Wu and Clemens Winter and Chris Hesse and Mark Chen and Eric Sigler and Mateusz Litwin and Scott Gray and Benjamin Chess and Jack Clark and Christopher Berner and Sam McCandlish and Alec Radford and Ilya Sutskever and Dario Amodei},
  booktitle={NeurIPS},
  year={2020}
}

@inproceedings{swin-transformer,
  title={Swin transformer: Hierarchical vision transformer using shifted windows},
  author={Liu, Ze and Lin, Yutong and Cao, Yue and Hu, Han and Wei, Yixuan and Zhang, Zheng and Lin, Stephen and Guo, Baining},
  booktitle={ICCV},
  year={2021}
}

@article{vit,
  title={An image is worth 16x16 words: Transformers for image recognition at scale},
  author={Alexey Dosovitskiy and Lucas Beyer and Alexander Kolesnikov and Dirk Weissenborn and Xiaohua Zhai and Thomas Unterthiner and Mostafa Dehghani and Matthias Minderer and Georg Heigold and Sylvain Gelly and Jakob Uszkoreit and Neil Houlsby},
  journal={arXiv preprint},
  year={2020}
}

@article{dalle,
  title={Dall-e: Creating images from text},
  author={Reddy, Mr D Murahari and Basha, Mr Sk Masthan and Hari, Mr M Chinnaiahgari and Penchalaiah, Mr N},
  journal={UGC Care Group I Journal},
  year={2021}
}

@article{palm,
  title={Palm: Scaling language modeling with pathways},
  author={Aakanksha Chowdhery and Sharan Narang and Jacob Devlin and Maarten Bosma and Gaurav Mishra and Adam Roberts and Paul Barham and Hyung Won Chung and Charles Sutton and Sebastian Gehrmann and Parker Schuh and Kensen Shi and Sasha Tsvyashchenko and Joshua Maynez and Abhishek Rao and Parker Barnes and Yi Tay and Noam Shazeer and Vinodkumar Prabhakaran and Emily Reif and Nan Du and Ben Hutchinson and Reiner Pope and James Bradbury and Jacob Austin and Michael Isard and Guy Gur-Ari and Pengcheng Yin and Toju Duke and Anselm Levskaya and Sanjay Ghemawat and Sunipa Dev and Henryk Michalewski and Xavier Garcia and Vedant Misra and Kevin Robinson and Liam Fedus and Denny Zhou and Daphne Ippolito and David Luan and Hyeontaek Lim and Barret Zoph and Alexander Spiridonov and Ryan Sepassi and David Dohan and Shivani Agrawal and Mark Omernick and Andrew M. Dai and Thanumalayan Sankaranarayana Pillai and Marie Pellat and Aitor Lewkowycz and Erica Moreira and Rewon Child and Oleksandr Polozov and Katherine Lee and Zongwei Zhou and Xuezhi Wang and Brennan Saeta and Mark Diaz and Orhan Firat and Michele Catasta and Jason Wei and Kathy Meier-Hellstern and Douglas Eck and Jeff Dean and Slav Petrov and Noah Fiedel},
  journal={JMLR},
  year={2023}
}

@article{mt-nlg,
  title={Using deepspeed and megatron to train megatron-turing nlg 530b, a large-scale generative language model},
  author={Shaden Smith and Mostofa Patwary and Brandon Norick and Patrick LeGresley and Samyam Rajbhandari and Jared Casper and Zhun Liu and Shrimai Prabhumoye and George Zerveas and Vijay Korthikanti and Elton Zhang and Rewon Child and Reza Yazdani Aminabadi and Julie Bernauer and Xia Song and Mohammad Shoeybi and Yuxiong He and Michael Houston and Saurabh Tiwary and Bryan Catanzaro},
  journal={arXiv preprint},
  year={2022}
}

@inproceedings {programmable_switch_nsdi23,
    author = {Mariano Scazzariello and Tommaso Caiazzi and Hamid Ghasemirahni and Tom Barbette and Dejan Kosti{\'c} and Marco Chiesa},
    title = {A High-Speed Stateful Packet Processing Approach for Tbps Programmable Switches},
    booktitle = {NSDI},
    year = {2023},
}

@article{patrickstar,
  title={PatrickStar: Parallel Training of Pre-trained Models via Chunk-based Memory Management},
  author={Fang, Jiarui and Yu, Yang and Zhu, Zilin and Li, Shenggui and You, Yang and Zhou, Jie},
  journal={arXiv preprint},
  year={2021}
}

@misc{qm8790,
  title={QM8790 Mellanox Quantum™ HDR Edge Switch},
  author={NVIDIA},
  howpublished={\url{https://network.nvidia.com/files/doc-2020/pb-qm8790.pdf}},
  year={2022}
}

@inproceedings{zero,
  title={Zero: Memory optimizations toward training trillion parameter models},
  author={Rajbhandari, Samyam and Rasley, Jeff and Ruwase, Olatunji and He, Yuxiong},
  booktitle={SC},
  year={2020}
}

@inproceedings{strom_eurosys20,
author = {Sidler, David and Wang, Zeke and Chiosa, Monica and Kulkarni, Amit and Alonso, Gustavo},
title = {StRoM: Smart Remote Memory},
year = {2020},
booktitle = {EuroSys}
}

@inproceedings{byteps,
  title={A unified architecture for accelerating distributed DNN training in heterogeneous GPU/CPU clusters},
  author={Jiang, Yimin and Zhu, Yibo and Lan, Chang and Yi, Bairen and Cui, Yong and Guo, Chuanxiong},
  booktitle={OSDI},
  year={2020}
}

@inproceedings{zero-offload,
  title={ZeRO-Offload: Democratizing Billion-Scale Model Training},
  author={Ren, Jie and Rajbhandari, Samyam and Aminabadi, Reza Yazdani and Ruwase, Olatunji and Yang, Shuangyan and Zhang, Minjia and Li, Dong and He, Yuxiong},
  booktitle={ATC},
  year={2021}
}

@inproceedings{zero-infinity,
  title={Zero-infinity: Breaking the gpu memory wall for extreme scale deep learning},
  author={Rajbhandari, Samyam and Ruwase, Olatunji and Rasley, Jeff and Smith, Shaden and He, Yuxiong},
  booktitle={SC},
  year={2021}
}

@inproceedings {ps,
  author = {Mu Li and David G. Andersen and Jun Woo Park and Alexander J. Smola and Amr Ahmed and Vanja Josifovski and James Long and Eugene J. Shekita and Bor-Yiing Su},
  title = {Scaling Distributed Machine Learning with the Parameter Server},
  booktitle = {OSDI},
  year = {2014},
}

@inproceedings{psnips,
  title={Communication efficient distributed machine learning with the parameter server},
  author={Li, Mu and Andersen, David G and Smola, Alexander J and Yu, Kai},
  booktitle={NIPS},
  year={2014}
}

@inproceedings{geeps,
  title={Geeps: Scalable deep learning on distributed gpus with a gpu-specialized parameter server},
  author={Cui, Henggang and Zhang, Hao and Ganger, Gregory R and Gibbons, Phillip B and Xing, Eric P},
  booktitle={EuroSys},
  year={2016}
}

@article{allreduce,
  title={Bandwidth optimal all-reduce algorithms for clusters of workstations},
  author={Patarasuk, Pitch and Yuan, Xin},
  journal={Journal of Parallel and Distributed Computing},
  year={2009}
}

@inproceedings{switchml,
  title={Scaling distributed machine learning with In-Network aggregation},
  author={Sapio, Amedeo and Canini, Marco and Ho, Chen-Yu and Nelson, Jacob and Kalnis, Panos and Kim, Changhoon and Krishnamurthy, Arvind and Moshref, Masoud and Ports, Dan and Richt{\'a}rik, Peter},
  booktitle={NSDI},
  year={2021}
}

@inproceedings{atp,
  title={ATP: In-network Aggregation for Multi-tenant Learning},
  author={Lao, ChonLam and Le, Yanfang and Mahajan, Kshiteej and Chen, Yixi and Wu, Wenfei and Akella, Aditya and Swift, Michael},
  booktitle={NSDI},
  year={2021}
}

@misc{tofino,
  title={Intel Tofino Programmable Ethernet Switch ASIC},
  author={Intel},
  howpublished={\url{https://www.intel.com/content/www/us/en/products/network-io/programmable-ethernet-switch/tofino-series.html}},
  year={2020}
}

@inproceedings{fpganic,
  title={FpgaNIC: An FPGA-based Versatile 100Gb SmartNIC for GPUs},
  author={Wang, Zeke and Huang, Hongjing and Zhang, Jie and Wu, Fei and Alonso, Gustavo},
  booktitle={ATC},
  year={2022}
}

@article{mixedprecision,
  title={Mixed precision training},
  author={Paulius Micikevicius and Sharan Narang and Jonah Alben and Gregory Diamos and Erich Elsen and David Garcia and Boris Ginsburg and Michael Houston and Oleksii Kuchaiev and Ganesh Venkatesh and Hao Wu},
  journal={arXiv preprint},
  year={2017}
}

@article{activationcheckpointing,
  title={Training deep nets with sublinear memory cost},
  author={Chen, Tianqi and Xu, Bing and Zhang, Chiyuan and Guestrin, Carlos},
  journal={arXiv preprint},
  year={2016}
}

@inproceedings{gpipe,
  title={Gpipe: Efficient training of giant neural networks using pipeline parallelism},
  author={Yanping Huang and Youlong Cheng and Ankur Bapna and Orhan Firat and Dehao Chen and Mia Chen and HyoukJoong Lee and Jiquan Ngiam and Quoc V. Le and Yonghui Wu and zhifeng Chen},
  booktitle={NeurIPS},
  year={2019}
}

@article{3dtp,
  title={Maximizing parallelism in distributed training for huge neural networks},
  author={Bian, Zhengda and Xu, Qifan and Wang, Boxiang and You, Yang},
  journal={arXiv preprint},
  year={2021}
}

@article{colossal-ai,
  title={Colossal-AI: A Unified Deep Learning System For Large-Scale Parallel Training},
  author={Bian, Zhengda and Liu, Hongxin and Wang, Boxiang and Huang, Haichen and Li, Yongbin and Wang, Chuanrui and Cui, Fan and You, Yang},
  journal={arXiv preprint},
  year={2021}
}

@article{dp,
  title={A bridging model for parallel computation},
  author={Valiant, Leslie G},
  journal={Communications of the ACM},
  year={1990},
}

@inproceedings{incbricks,
  title={Incbricks: Toward in-network computation with an in-network cache},
  author={Liu, Ming and Luo, Liang and Nelson, Jacob and Ceze, Luis and Krishnamurthy, Arvind and Atreya, Kishore},
  booktitle={ASPLOS},
  year={2017}
}

@article{programmableswitch,
  title={Forwarding metamorphosis: Fast programmable match-action processing in hardware for SDN},
  author={Bosshart, Pat and Gibb, Glen and Kim, Hun-Seok and Varghese, George and McKeown, Nick and Izzard, Martin and Mujica, Fernando and Horowitz, Mark},
  journal={ACM SIGCOMM Computer Communication Review},
  year={2013},
}

@article{checkmate,
  title={Checkmate: Breaking the memory wall with optimal tensor rematerialization},
  author={Jain, Paras and Jain, Ajay and Nrusimha, Aniruddha and Gholami, Amir and Abbeel, Pieter and Gonzalez, Joseph and Keutzer, Kurt and Stoica, Ion},
  journal={MLSys},
  year={2020}
}

@inproceedings{phub,
  title={Parameter hub: a rack-scale parameter server for distributed deep neural network training},
  author={Luo, Liang and Nelson, Jacob and Ceze, Luis and Phanishayee, Amar and Krishnamurthy, Arvind},
  booktitle={SoCC},
  year={2018}
}

@article{mxnet,
  title={Mxnet: A flexible and efficient machine learning library for heterogeneous distributed systems},
  author={Chen, Tianqi and Li, Mu and Li, Yutian and Lin, Min and Wang, Naiyan and Wang, Minjie and Xiao, Tianjun and Xu, Bing and Zhang, Chiyuan and Zhang, Zheng},
  journal={arXiv preprint},
  year={2015}
}

@inproceedings{pytorch,
  title={Automatic differentiation in pytorch},
  author={Paszke, Adam and Gross, Sam and Chintala, Soumith and Chanan, Gregory and Yang, Edward and DeVito, Zachary and Lin, Zeming and Desmaison, Alban and Antiga, Luca and Lerer, Adam},
  booktitle={Autodiff@NIPS},
  year={2017}
}

@inproceedings{pytorch-ddp,
  title={PyTorch Distributed: Experiences on Accelerating Data Parallel Training},
  author={Shen Li and Yanli Zhao and Rohan Varma and Omkar Salpekar and Pieter Noordhuis and Teng Li and Adam Paszke and Jeff Smith and Brian Vaughan and Pritam Damania and Soumith Chintala},
  booktitle={VLDB},
  year={2020}
}

@inproceedings{tensorflow,
  title={TensorFlow: a system for Large-Scale machine learning},
  author={Mart{\'\i}n Abadi and Paul Barham and Jianmin Chen and Zhifeng Chen and Andy Davis and Jeffrey Dean and Matthieu Devin and Sanjay Ghemawat and Geoffrey Irving and Michael Isard and Manjunath Kudlur and Josh Levenberg and Rajat Monga and Sherry Moore and Derek G. Murray and Benoit Steiner and Paul Tucker and Vijay Vasudevan and Pete Warden and Martin Wicke and Yuan Yu and Xiaoqiang Zheng},
  booktitle={OSDI},
  year={2016}
}

@inproceedings{deepspeed,
  title={Deepspeed: System optimizations enable training deep learning models with over 100 billion parameters},
  author={Rasley, Jeff and Rajbhandari, Samyam and Ruwase, Olatunji and He, Yuxiong},
  booktitle={SIGKDD},
  year={2020}
}

@inproceedings{pipedream,
  title={Memory-efficient pipeline-parallel dnn training},
  author={Narayanan, Deepak and Phanishayee, Amar and Shi, Kaiyu and Chen, Xie}, 
  booktitle={ICML},
  year={2021},
}

@inproceedings{dapple,
  title={DAPPLE: A pipelined data parallel approach for training large models},
  author={Shiqing Fan and Yi Rong and Chen Meng and Zongyan Cao and Siyu Wang and Zhen Zheng and Chuan Wu and Guoping Long and Jun Yang and Lixue Xia and Lansong Diao and Xiaoyong Liu and Wei Li},
  booktitle={PPoPP},
  year={2021}
}

@article{2dtp,
  title={An efficient 2d method for training super-large deep learning models},
  author={Xu, Qifan and Li, Shenggui and Gong, Chaoyu and You, Yang},
  journal={arXiv preprint},
  year={2021}
}

@article{megatronlm,
  title={Megatron-lm: Training multi-billion parameter language models using model parallelism},
  author={Shoeybi, Mohammad and Patwary, Mostofa and Puri, Raul and LeGresley, Patrick and Casper, Jared and Catanzaro, Bryan},
  journal={arXiv preprint},
  year={2019}
}

@inproceedings{megatron2021,
  title={Efficient large-scale language model training on gpu clusters using megatron-lm},
  author={Narayanan, Deepak and Shoeybi, Mohammad and Casper, Jared and LeGresley, Patrick and Patwary, Mostofa and Korthikanti, Vijay and Vainbrand, Dmitri and Kashinkunti, Prethvi and Bernauer, Julie and Catanzaro, Bryan and Phanishayee, Amar},
  booktitle={SC},
  year={2021}
}

@inproceedings{herring,
  title={Herring: Rethinking the parameter server at scale for the cloud},
  author={Thangakrishnan, Indu and Cavdar, Derya and Karakus, Can and Ghai, Piyush and Selivonchyk, Yauheni and Pruce, Cory},
  booktitle={SC},
  year={2020}
}

@misc{sn2700,
  title={SN2700 Open Ethernet Switch},
  author={NVIDIA},
  howpublished={\url{https://network.nvidia.com/files/doc-2020/pb-sn2700.pdf}},
  year={2022}
}

@misc{wedge100bf,
  title={Edgecore Wedge100BF-32X Product Info},
  author={Edgecore},
  howpublished={\url{https://www.edge-core.com/productsInfo.php?cls=1&cls2=5&cls3=181&id=335}},
  year={2020}
}

@inproceedings{projectadam,
  title={Project adam: Building an efficient and scalable deep learning training system},
  author={Chilimbi, Trishul and Suzue, Yutaka and Apacible, Johnson and Kalyanaraman, Karthik},
  booktitle={OSDI},
  year={2014}
}

@article{psplus,
  title={PS+: A Simple yet Effective Framework for Fast Training on Parameter Server},
  author={Jin, A-Long and Xu, Wenchao and Guo, Song and Hu, Bing and Yeung, Kwan},
  journal={TPDS},
  year={2022},
}

@inproceedings{netagg,
  title={Netagg: Using middleboxes for application-specific on-path aggregation in data centres},
  author={Mai, Luo and Rupprecht, Lukas and Alim, Abdul and Costa, Paolo and Migliavacca, Matteo and Pietzuch, Peter and Wolf, Alexander L},
  booktitle={CoNEXT},
  year={2014}
}

@inproceedings{sharp,
  title={Scalable hierarchical aggregation protocol (SHArP): A hardware architecture for efficient data reduction},
  author={Richard L. Graham and Devendar Bureddy and Pak Lui and Hal Rosenstock and Gilad Shainer and Gil Bloch and Dror Goldenerg and Mike Dubman and Sasha Kotchubievsky and Vladimir Koushnir and Lion Levi and Alex Margolin and Tamir Ronen and Alexander Shpiner and Oded Wertheim and Eitan Zahavi},
  booktitle={COMHPC@SC},
  pages={1--10},
  year={2016},
  organization={IEEE}
}

@inproceedings{sharp-agg,
  title={Scalable hierarchical aggregation and reduction protocol (sharp) streaming-aggregation hardware design and evaluation},
  author={Richard L. Graham and Lion Levi and Devendar Burredy and Gil Bloch and Gilad Shainer, David Cho and George Elias and Daniel Klein and Joshua Ladd and Ophir Maor and Ami Marelli, Valentin Petrov and Evyatar Romlet and Yong Qin and Ido Zemah },
  booktitle={ISC High Performance},
  year={2020},
}

@article{esa,
  title={Efficient data-plane memory scheduling for in-network aggregation},
  author={Wang, Hao and Qin, Yuxuan and Lao, ChonLam and Le, Yanfang and Wu, Wenfei and Chen, Kai},
  journal={arXiv preprint},
  year={2022}
}

@article{opt,
  title={Opt: Open pre-trained transformer language models},
  author={Susan Zhang and Stephen Roller and Naman Goyal and Mikel Artetxe and Moya Chen and Shuohui Chen and Christopher Dewan and Mona Diab and Xian Li and Xi Victoria Lin and Todor Mihaylov and Myle Ott and Sam Shleifer and Kurt Shuster and Daniel Simig and Punit Singh Koura and Anjali Sridhar and Tianlu Wang and Luke Zettlemoyer},
  journal={arXiv preprint},
  year={2022}
}

@article{colossal-auto,
  title={Colossal-Auto: Unified Automation of Parallelization and Activation Checkpoint for Large-scale Models},
  author={Liu, Yuliang and Li, Shenggui and Fang, Jiarui and Shao, Yanjun and Yao, Boyuan and You, Yang},
  journal={arXiv preprint},
  year={2023}
}

@misc{rm-static,
  commentkey={1rmstaticdatasetathiggingface},
  title={rm-static dataset at HuggingFace},
  howpublished={\url{https://huggingface.co/datasets/Dahoas/rm-static}}
}

@inproceedings{mobius,
  title={Mobius: Fine tuning large-scale models on commodity gpu servers},
  author={Feng, Yangyang and Xie, Minhui and Tian, Zijie and Wang, Shuo and Lu, Youyou and Shu, Jiwu},
  booktitle={ASPLOS},
  year={2023}
}

@misc{nccl,
  title={NVIDIA Collective Communications Library},
  author={NVIDIA},
  howpublished={\url{https://developer.nvidia.com/nccl}},
  year={2017}
}

@inproceedings{kim2019parallax,
  title={Parallax: Sparsity-aware data parallel training of deep neural networks},
  author={Kim, Soojeong and Yu, Gyeong-In and Park, Hojin and Cho, Sungwoo and Jeong, Eunji and Ha, Hyeonmin and Lee, Sanha and Jeong, Joo Seong and Chun, Byung-Gon},
  booktitle={EuroSys},
  year={2019}
}

@article{zhao2021vpipe,
  title={vPipe: A virtualized acceleration system for achieving efficient and scalable pipeline parallel dnn training},
  author={Shixiong Zhao and Fanxin Li and Xusheng Chen and Xiuxian Guan and Jianyu Jiang and Dong Huang and Yuhao Qing and Sen Wang and Peng Wang and Gong Zhang and Cheng Li and Ping Luo and Heming Cui},
  journal={TPDS},
  year={2021}
}

@article{hashemi2016performance,
  title={Performance modeling of distributed deep neural networks},
  author={Hashemi, Sayed Hadi and Noghabi, Shadi A and Gropp, William and Campbell, Roy H},
  journal={arXiv preprint},
  year={2016}
}

@inproceedings{watcharapichat2016ako,
  title={Ako: Decentralised deep learning with partial gradient exchange},
  author={Watcharapichat, Pijika and Morales, Victoria Lopez and Fernandez, Raul Castro and Pietzuch, Peter},
  booktitle={SoCC},
  year={2016}
}

@article{amsp,
  title={AMSP: Super-Scaling LLM Training via Advanced Model States Partitioning},
  author={Chen, Qiaoling and Hu, Qinghao and Ye, Zhisheng and Wang, Guoteng and Sun, Peng and Wen, Yonggang and Zhang, Tianwei},
  journal={arXiv preprint},
  year={2023}
}

@inproceedings{mics,
  title={MiCS: near-linear scaling for training gigantic model on public cloud},
  author={Zhang, Zhen and Zheng, Shuai and Wang, Yida and Chiu, Justin and Karypis, George and Chilimbi, Trishul and Li, Mu and Jin, Xin},
  booktitle={VLDB},
  year={2022}
}

@inproceedings{zero++,
  title={Zero++: Extremely efficient collective communication for giant model training},
  author={Wang, Guanhua and Qin, Heyang and Jacobs, Sam Ade and Holmes, Connor and Rajbhandari, Samyam and Ruwase, Olatunji and Yan, Feng and Yang, Lei and He, Yuxiong},
  booktitle={arXiv preprint},
  year={2023}
}

@inproceedings{fsdp,
  title={PyTorch FSDP: Experiences on Scaling Fully Sharded Data Parallel},
  author={Yanli Zhao and Andrew Gu and Rohan Varma and Liang Luo and Chien-Chin Huang and Min Xu and Less Wright and Hamid Shojanazeri and Myle Ott and Sam Shleifer and Alban Desmaison and Can Balioglu and Pritam Damania and Bernard Nguyen and Geeta Chauhan and Yuchen Hao and Ajit Mathews and Shen Li},
  booktitle={VLDB},
  year={2023},
}

@article{xu2020automatic,
  title={Automatic cross-replica sharding of weight update in data-parallel training},
  author={Xu, Yuanzhong and Lee, HyoukJoong and Chen, Dehao and Choi, Hongjun and Hechtman, Blake and Wang, Shibo},
  journal={arXiv preprint},
  year={2020}
}

@article{zerobubble,
  title={Zero Bubble Pipeline Parallelism},
  author={Qi, Penghui and Wan, Xinyi and Huang, Guangxing and Lin, Min},
  journal={arXiv preprint},
  year={2023}
}

@inproceedings{narayanan2021memory,
  title={Memory-efficient pipeline-parallel dnn training},
  author={Narayanan, Deepak and Phanishayee, Amar and Shi, Kaiyu and Chen, Xie and Zaharia, Matei},
  booktitle={ICML},
  year={2021}
}

@article{sarathi,
  title={Sarathi: Efficient llm inference by piggybacking decodes with chunked prefills},
  author={Agrawal, Amey and Panwar, Ashish and Mohan, Jayashree and Kwatra, Nipun and Gulavani, Bhargav S and Ramjee, Ramachandran},
  journal={arXiv preprint},
  year={2023}
}

@inproceedings{a2tp,
  title={A2TP: Aggregator-aware In-network Aggregation for Multi-tenant Learning},
  author={Li, Zhaoyi and Huang, Jiawei and Li, Yijun and Xu, Aikun and Zhou, Shengwen and Liu, Jingling and Wang, Jianxin},
  booktitle={EuroSys},
  year={2023}
}

@inproceedings{adapipe,
  title={AdaPipe: Optimizing Pipeline Parallelism with Adaptive Recomputation and Partitioning},
  author={Sun, Zhenbo and Cao, Huanqi and Wang, Yuanwei and Feng, Guanyu and Chen, Shengqi and Wang, Haojie and Chen, Wenguang},
  booktitle={ASPLOS},
  year={2024}
}

@inproceedings{enzian_asplos22,
    author = {Cock, David and Ramdas, Abishek and Schwyn, Daniel and Giardino, Michael and Turowski, Adam and He, Zhenhao and Hossle, Nora and Korolija, Dario and Licciardello, Melissa and Martsenko, Kristina and Achermann, Reto and Alonso, Gustavo and Roscoe, Timothy},
    title = {Enzian: an open, general, CPU/FPGA platform for systems software research},
    year = {2022},
    booktitle = {ASPLOS}
}

@inproceedings {fpga_osdi20,
    author = {Dario Korolija and Timothy Roscoe and Gustavo Alonso},
    title = {Do {OS} abstractions make sense on {FPGAs}?},
    booktitle = {OSDI},
    year = {2020}
}

@inproceedings{plink,
  title={PLink: Efficient cloud-based training with topology-aware dynamic hierarchical aggregation},
  author={Luo, Liang and West, Peter and Nelson, Jacob and Krishnamurthy, Arvind and Ceze, Luis},
  booktitle={MLSys},
  year={2020}
}

@article{ultima,
  title={Ultima: Robust and Tail-Optimal AllReduce for Distributed Deep Learning in the Cloud},
  author={Warraich, Ertza and Shabtai, Omer and Manaa, Khalid and Vargaftik, Shay and Piasetzky, Yonatan and Kadosh, Matty and Suresh, Lalith and Shahbaz, Muhammad},
  journal={arXiv preprint},
  year={2023}
}

@article{zenops,
  title={ZenoPS: A Distributed Learning System Integrating Communication Efficiency and Security},
  author={Xie, Cong and Koyejo, Oluwasanmi and Gupta, Indranil},
  journal={Algorithms},
  year={2022}
}

@inproceedings{optimus,
  title={Optimus: an efficient dynamic resource scheduler for deep learning clusters},
  author={Peng, Yanghua and Bao, Yixin and Chen, Yangrui and Wu, Chuan and Guo, Chuanxiong},
  booktitle={EuroSys},
  year={2018}
}

@inproceedings{mina,
  title={MINA: Auto-scale In-network Aggregation for Machine Learning Service},
  author={Shichen Dong and Zhixiong Niu and Mingchao Zhang and Zhiying Xu and Chuntao Hu and Wei Wang and Pengzhi Zhu and Qingchun Song and Lei Qu and Peng Cheng and Yongqiang Xiong and Chen Tian and Camtu Nguyen and Xiaoliang Wang},
  booktitle={APNet},
  year={2023}
}

@inproceedings{bml,
  title={Bml: A high-performance, low-cost gradient synchronization algorithm for dml training},
  author={Wang, Songtao and Li, Dan and Cheng, Yang and Geng, Jinkun and Wang, Yanshu and Wang, Shuai and Xia, Shu-Tao and Wu, Jianping},
  booktitle={NIPS},
  year={2018}
}

@article{megascale,
  title={MegaScale: Scaling Large Language Model Training to More Than 10,000 GPUs},
  author={Ziheng Jiang and Haibin Lin and Yinmin Zhong and Qi Huang and Yangrui Chen and Zhi Zhang and Yanghua Peng and Xiang Li and Cong Xie and Shibiao Nong and Yulu Jia and Sun He and Hongmin Chen and Zhihao Bai and Qi Hou and Shipeng Yan and Ding Zhou and Yiyao Sheng and Zhuo Jiang and Haohan Xu and Haoran Wei and Zhang Zhang and Pengfei Nie and Leqi Zou and Sida Zhao and Liang Xiang and Zherui Liu and Zhe Li and Xiaoying Jia and Jianxi Ye and Xin Jin and Xin Liu},
  journal={arXiv preprint},
  year={2024}
}

@article{fold3d,
  title={Fold3d: Rethinking and parallelizing computational and communicational tasks in the training of large dnn models},
  author={Li, Fanxin and Zhao, Shixiong and Qing, Yuhao and Chen, Xusheng and Guan, Xiuxian and Wang, Sen and Zhang, Gong and Cui, Heming},
  journal={TPDS},
  year={2023}
}

@inproceedings{topoopt,
  title={TopoOpt: Co-optimizing Network Topology and Parallelization Strategy for Distributed Training Jobs},
  author={Wang, Weiyang and Khazraee, Moein and Zhong, Zhizhen and Ghobadi, Manya and Jia, Zhihao and Mudigere, Dheevatsa and Zhang, Ying and Kewitsch, Anthony},
  booktitle={NSDI},
  year={2023}
}

@inproceedings{olmedo2020dissecting,
  title={Dissecting the CUDA scheduling hierarchy: a performance and predictability perspective},
  author={Olmedo, Ignacio Sanudo and Capodieci, Nicola and Martinez, Jorge Luis and Marongiu, Andrea and Bertogna, Marko},
  booktitle={RTAS},
  year={2020}
}

@misc{mps,
  title={Multi-Process Service Documentation},
  author={NVIDIA},
  howpublished={\url{https://docs.nvidia.com/deploy/mps/}},
  year={2017}
}

@inproceedings{ace,
  title={Enabling compute-communication overlap in distributed deep learning training platforms},
  author={Rashidi, Saeed and Denton, Matthew and Sridharan, Srinivas and Srinivasan, Sudarshan and Suresh, Amoghavarsha and Nie, Jade and Krishna, Tushar},
  booktitle={ISCA},
  year={2021}
}

@article{llama3,
  title={The llama 3 herd of models},
  author={LlamaTeam},
  journal={arXiv preprint},
  year={2024}
}

@inproceedings{sentinel,
  title={Sentinel: Efficient tensor migration and allocation on heterogeneous memory systems for deep learning},
  author={Ren, Jie and Luo, Jiaolin and Wu, Kai and Zhang, Minjia and Jeon, Hyeran and Li, Dong},
  booktitle={HPCA},
  year={2021}
}

@inproceedings{dynn-offload,
  title={Enabling Large Dynamic Neural Network Training with Learning-based Memory Management},
  author={Ren, Jie and Xu, Dong and Yang, Shuangyan and Zhao, Jiacheng and Li, Zhicheng and Navasca, Christian and Wang, Chenxi and Xu, Harry and Li, Dong},
  booktitle={HPCA},
  year={2024}
}

@inproceedings{hetpipe,
  title={HetPipe: Enabling large DNN training on (whimpy) heterogeneous GPU clusters through integration of pipelined model parallelism and data parallelism},
  author={Park, Jay H and Yun, Gyeongchan and Chang, M Yi and Nguyen, Nguyen T and Lee, Seungmin and Choi, Jaesik and Noh, Sam H and Choi, Young-ri},
  booktitle={ATC},
  year={2020}
}

@INPROCEEDINGS{difference_fccm18,
  author={Cong, Jason and Fang, Zhenman and Lo, Michael and Wang, Hanrui and Xu, Jingxian and Zhang, Shaochong},
  booktitle={FCCM}, 
  title={Understanding Performance Differences of FPGAs and GPUs}, 
  year={2018}
}

@inproceedings{ttgnn_micro23,
author = {Qu, Zheng and Niu, Dimin and Li, Shuangchen and Zheng, Hongzhong and Xie, Yuan},
title = {TT-GNN: Efficient On-Chip Graph Neural Network Training via Embedding Reformation and Hardware Optimization},
year = {2023},
booktitle = {MICRO}
}

@inproceedings{drl_isca19,
author = {Li, Youjie and Liu, Iou-Jen and Yuan, Yifan and Chen, Deming and Schwing, Alexander and Huang, Jian},
title = {Accelerating distributed reinforcement learning with in-switch computing},
year = {2019},
booktitle = {ISCA}
}

@inproceedings{beyond_hpsr18,
  title={Beyond SmartNICs: Towards a fully programmable cloud},
  author={Caulfield, Adrian and Costa, Paolo and Ghobadi, Monia},
  booktitle={HPSR},
  year={2018}
}

@inproceedings{azure_nsdi18,
  title={Azure accelerated networking: SmartNICs in the public cloud},
  author={Daniel Firestone and Andrew Putnam and Sambhrama Mundkur and Derek Chiou and Alireza Dabagh and Mike Andrewartha and Hari Angepat and Vivek Bhanu and Adrian Caulfield and Eric Chung and Harish Kumar Chandrappa and Somesh Chaturmohta and Matt Humphrey and Jack Lavier and Norman Lam and Fengfen Liu and Kalin Ovtcharov and Jitu Padhye and Gautham Popuri and Shachar Raindel and Tejas Sapre and Mark Shaw and Gabriel Silva and Madhan Sivakumar and Nisheeth Srivastava and Anshuman Verma and Qasim Zuhair and Deepak Bansal and Doug Burger and Kushagra Vaid and David A. Maltz and Albert Greenberg},
  booktitle={NSDI},
  year={2018}
}

@inproceedings{submodel_ccgrid2024,
  title={Sub-model Parallelism: A Scale-out Deployment Method for Large Multi-modal DNNs},
  author={Huang, Tianhao and Sun, Lingyu and Hou, Xiaofeng and Zhu, Xiaozhi and Xia, Xinfeng and Wang, Yutong and Chen, Mingxi and Li, Chao},
  booktitle={CCGrid},
  year={2024},
}

@inproceedings{multitree_allreduce_isca2021,
  title={Communication algorithm-architecture co-design for distributed deep learning},
  author={Huang, Jiayi and Majumder, Pritam and Kim, Sungkeun and Muzahid, Abdullah and Yum, Ki Hwan and Kim, Eun Jung},
  booktitle={ISCA},
  year={2021}
}

@article{comet_arxiv2022,
  title={COMET: A comprehensive cluster design methodology for distributed deep learning training},
  author={Kadiyala, Divya Kiran and Rashidi, Saeed and Heo, Taekyung and Bambhaniya, Abhimanyu Rajeshkumar and Krishna, Tushar and Daglis, Alexandros},
  journal={arXiv preprint},
  year={2022}
}

@inproceedings{rackblox_sosp23,
  title={RackBlox: A Software-Defined Rack-Scale Storage System with Network-Storage Co-Design},
  author={Reidys, Benjamin and Xue, Yuqi and Li, Daixuan and Sukhwani, Bharat and Hwu, Wen-Mei and Chen, Deming and Asaad, Sameh and Huang, Jian},
  booktitle={SOSP},
  year={2023}
}

@article{lsmgnn_arxiv2024,
  title={LSM-GNN: Large-scale Storage-based Multi-GPU GNN Training by Optimizing Data Transfer Scheme},
  author={Park, Jeongmin Brian and Wu, Kun and Mailthody, Vikram Sharma and Quresh, Zaid and Mahlke, Scott and Hwu, Wen-mei},
  journal={arXiv preprint},
  year={2024}
}

@inproceedings{taurus_asplos2022,
  title={Taurus: a data plane architecture for per-packet ML},
  author={Swamy, Tushar and Rucker, Alexander and Shahbaz, Muhammad and Gaur, Ishan and Olukotun, Kunle},
  booktitle={ASPLOS},
  year={2022}
}

@inproceedings{network_isca2020,
  title={An in-network architecture for accelerating shared-memory multiprocessor collectives},
  author={Klenk, Benjamin and Jiang, Nan and Thorson, Greg and Dennison, Larry},
  booktitle={ISCA},
  year={2020}
}

@inproceedings{fcsn_fpl2022,
  title={A framework for neural network inference on fpga-centric smartnics},
  author={Guo, Anqi and Geng, Tong and Zhang, Yongan and Haghi, Pouya and Wu, Chunshu and Tan, Cheng and Lin, Yingyan and Li, Ang and Herbordt, Martin},
  booktitle={FPL},
  year={2022}
}

@article{deepseek-v3,
  title={DeepSeek-V3 Technical Report},
  author={DeepSeek-AI},
  journal={arXiv preprint},
  year={2024}
}

@article{mist,
  title={Mist: Efficient Distributed Training of Large Language Models via Memory-Parallelism Co-Optimization},
  author={Zhu, Zhanda and Giannoula, Christina and Andoorveedu, Muralidhar and Su, Qidong and Mangalam, Karttikeya and Zheng, Bojian and Pekhimenko, Gennady},
  journal={arXiv preprint},
  year={2025}
}

@article{wlbllm,
  title={WLB-LLM: Workload-Balanced 4D Parallelism for Large Language Model Training},
  author={Zheng Wang and Anna Cai and Xinfeng Xie and Zaifeng Pan and Yue Guan and Weiwei Chu and Jie Wang and Shikai Li and Jianyu Huang and Chris Cai and Yuchen Hao and Yufei Ding},
  journal={arXiv preprint},
  year={2025}
}

@inproceedings{deepseekv3_isca2025,
  title={Insights into deepseek-v3: Scaling challenges and reflections on hardware for ai architectures},
  author={Chenggang Zhao and Chengqi Deng and Chong Ruan and Damai Dai and Huazuo Gao and Jiashi Li and Liyue Zhang and Panpan Huang and Shangyan Zhou and Shirong Ma and Wenfeng Liang and Ying He and Yuqing Wang and Yuxuan Liu and Y.X. Wei},
  booktitle={ISCA},
  year={2025}
}

@inproceedings{llama3_isca2025,
  title={Scaling Llama 3 Training with Efficient Parallelism Strategies},
  author={Weiwei Chu and Xinfeng Xie and Jiecao Yu and Jie Wang and Amar Phanishayee and Chunqiang Tang and Yuchen Hao and Jianyu Huang and Mustafa Ozdal and Jun Wang and Vedanuj Goswami and Naman Goyal and Abhishek Kadian and Andrew Gu and Chris Cai and Feng Tian and Xiaodong Wang and Min Si and Pavan Balaji and Ching-Hsiang Chu and Jongsoo Park },
  booktitle={ISCA},
  year={2025}
}

@inproceedings{longcontext_naaclhlt2024,
  author={Wenhan Xiong and Jingyu Liu and Igor Molybog and Hejia Zhang and Prajjwal Bhargava and Rui Hou and Louis Martin and Rashi Rungta and Karthik Abinav Sankararaman and Barlas Oguz and Madian Khabsa and Han Fang and Yashar Mehdad and Sharan Narang and Kshitiz Malik and Angela Fan and Shruti Bhosale and Sergey Edunov and Mike Lewis and Sinong Wang and Hao Ma},
  title={Effective Long-Context Scaling of Foundation Models},
  year={2024},
  booktitle={NAACL-HLT}
}

@inproceedings{kvdirect_sosp2017,
  title={Kv-direct: High-performance in-memory key-value store with programmable nic},
  author={Li, Bojie and Ruan, Zhenyuan and Xiao, Wencong and Lu, Yuanwei and Xiong, Yongqiang and Putnam, Andrew and Chen, Enhong and Zhang, Lintao},
  booktitle={SOSP},
  year={2017}
}

@inproceedings{vela_asplos2025,
  title={Vela: A Virtualized LLM Training System with GPU Direct RoCE},
  author={Apoorve Mohan and Robert Walkup and Bengi Karacali and Ming-hung Chen and Abdullah Kayi and Liran Schour and Shweta Salaria and Sophia Wen and I-hsin Chung and Abdul Alim and Constantinos Evangelinos and Lixiang Luo and Marc Dombrowa and Laurent Schares and Ali Sydney and Pavlos Maniotis and Sandhya Koteshwara and Brent Tang and Joel Belog and Rei Odaira and Vasily Tarasov and Eran Gampel and Drew Thorstensen and Talia Gershon and Seetharami Seelam },
  booktitle={ASPLOS},
  year={2025}
}

@inproceedings{p4v_sigcomm2018,
  title={P4v: Practical verification for programmable data planes},
  author={Liu, Jed and Hallahan, William and Schlesinger, Cole and Sharif, Milad and Lee, Jeongkeun and Soul{\'e}, Robert and Wang, Han and Ca{\c{s}}caval, C{\u{a}}lin and McKeown, Nick and Foster, Nate},
  booktitle={SIGCOMM},
  year={2018}
}

@inproceedings{panic_osdi2020,
  title={PANIC: A High-Performance programmable NIC for multi-tenant networks},
  author={Lin, Jiaxin and Patel, Kiran and Stephens, Brent E and Sivaraman, Anirudh and Akella, Aditya},
  booktitle={OSDI},
  year={2020}
}

@inproceedings{alkali_nsdi2025,
  title={Enabling Portable and High-Performance SmartNIC Programs with Alkali},
  author={Lin, Jiaxin and Guo, Zhiyuan and Shah, Mihir and Ji, Tao and Zhang, Yiying and Kim, Daehyeok and Akella, Aditya},
  booktitle={NSDI},
  year={2025}
}

@inproceedings{fcsn_fccm2022,
  title={FCsN: A FPGA-Centric SmartNIC Framework for Neural Networks},
  author={Guo, Anqi and Geng, Tong and Zhang, Yongan and Haghi, Pouya and Wu, Chunshu and Tan, Cheng and Lin, Yingyan and Li, Ang and Herbordt, Martin},
  booktitle={FCCM},
  year={2022},
}

@inproceedings{mist_eurosys2025,
  title={Mist: Efficient Distributed Training of Large Language Models via Memory-Parallelism Co-Optimization},
  author={Zhu, Zhanda and Giannoula, Christina and Andoorveedu, Muralidhar and Su, Qidong and Mangalam, Karttikeya and Zheng, Bojian and Pekhimenko, Gennady},
  booktitle={EuroSys},
  year={2025}
}

@inproceedings{pccl_iccd2024,
  title={PCCL: Energy-efficient LLM Training with Power-aware Collective Communication},
  author={Jia, Ziyang and Bhuyan, Laxmi N and Wong, Daniel},
  booktitle={ICCD},
  year={2024},
}

@inproceedings{chimera_isca2025,
  title={Chimera: Communication fusion for hybrid parallelism in large language models},
  author={Qin, Le and Cui, Junwei and Cai, Weilin and Huang, Jiayi},
  booktitle={ISCA},
  year={2025}
}

@inproceedings{meshslice_isca2025,
  title={MeshSlice: Efficient 2D Tensor Parallelism for Distributed DNN Training},
  author={Nam, Hyoungwook and Gerogiannis, Gerasimos and Torrellas, Josep},
  booktitle={ISCA},
  year={2025}
}

@inproceedings{lognic_micro2023,
  title={Lognic: A high-level performance model for smartnics},
  author={Guo, Zerui and Lin, Jiaxin and Bai, Yuebin and Kim, Daehyeok and Swift, Michael and Akella, Aditya and Liu, Ming},
  booktitle={MICRO},
  year={2023}
}

@article{baft_fcs2025,
  title={BAFT: bubble-aware fault-tolerant framework for distributed DNN training with hybrid parallelism},
  author={Chen, Runzhe and Lu, Guandong and Wang, Yakai and Zhang, Rui and Hu, Zheng and Miao, Yanming and Cai, Zhifang and Leng, Jingwen and Guo, Minyi},
  journal={Frontiers of Computer Science},
  year={2025},
}

@inproceedings{lambdanic_icdcs2020,
  title={$\lambda$-nic: Interactive serverless compute on programmable smartnics},
  author={Choi, Sean and Shahbaz, Muhammad and Prabhakar, Balaji and Rosenblum, Mendel},
  booktitle={ICDCS},
  year={2020}
}

@inproceedings{fred_isca2025,
  title={FRED: A Wafer-scale Fabric for 3D Parallel DNN Training},
  author={Rashidi, Saeed and Won, William and Srinivasan, Sudarshan and Gupta, Puneet and Krishna, Tushar},
  booktitle={ISCA},
  year={2025}
}

@inproceedings{vega_cgo2025,
  title={Vega: Automatically generating compiler backends using a pre-trained transformer model},
  author={Zhong, Ming and Lv, Fang and Wang, Lulin and Qiu, Lei and Wang, Yingying and Liu, Ying and Cui, Huimin and Feng, Xiaobing and Xue, Jingling},
  booktitle={CGO},
  year={2025}
}

@inproceedings{traci_isca2025,
  title={TRACI: Network Acceleration of Input-Dynamic Communication for Large-Scale Deep Learning Recommendation Model},
  author={Huang, Guyue and Li, Hao and Qin, Le and Huang, Jiayi and Kang, Yangwook and Ding, Yufei and Xie, Yuan},
  booktitle={ISCA},
  year={2025}
}

@inproceedings{styx_atc2023,
  title={STYX: Exploiting SmartNIC capability to reduce datacenter memory tax},
  author={Ji, Houxiang and Mansi, Mark and Sun, Yan and Yuan, Yifan and Huang, Jinghan and Kuper, Reese and Swift, Michael M and Kim, Nam Sung},
  booktitle={ATC},
  year={2023}
}

@article{hint_comarchletter2025,
  title={HINT: A Hardware Platform for Intra-host NIC Traffic and SmartNIC Emulation},
  author={Li, Yu and Lou, Jiaqi and Vanavasam, Srikar and Kim, Nam Sung},
  journal={IEEE Computer Architecture Letters},
  year={2025},
}

@inproceedings{yue_isscc2024,
  title={15.1 A 0.795 fJ/bit Physically-Unclonable Function-Protected TCAM for a Software-Defined Networking Switch},
  author={Yue, Zhiheng and Xiang, Xujiang and Tu, Fengbin and Wang, Yang and Wang, Yiming and Wei, Shaojun and Hu, Yang and Yin, Shouyi},
  booktitle={ISSCC},
  year={2024}
}

@inproceedings{inceptionn_micro2018,
  title={A network-centric hardware/algorithm co-design to accelerate distributed training of deep neural networks},
  author={Li, Youjie and Park, Jongse and Alian, Mohammad and Yuan, Yifan and Qu, Zheng and Pan, Peitian and Wang, Ren and Schwing, Alexander and Esmaeilzadeh, Hadi and Kim, Nam Sung},
  booktitle={MICRO},
  year={2018}
}

@misc{worker_price,
  title={Supermicro SuperServer 4029GP-TRT 4U Dual socket},
  author={{Acme Micro Systems}},
  howpublished={\url{https://www.acmemicro.com/Product/16479/Supermicro-SuperServer-4029GP-TRT-4U-Dual-socket-P-(LGA-3647)-24xDDR4-8xGPU-2x10GbE-11xPCIe-R2000W-SYS-4029GP-TRT} (accessed Nov. 13, 2025)}
}

@misc{ps_price,
  title={SuperWorksta SYS-740A-T Tower Xeon Dual Socket P+},
  author={{Acme Micro Systems}},
  howpublished={\url{https://www.acmemicro.com/Product/18368/SuperWorksta-SYS-740A-T-Tower-Xeon-Dual-Socket-P+-(LGA-4189)-DDR4-8x-3-5-SATA-2x-1GbE-6-PCI-E-Dual-1200W} (accessed Nov. 13, 2025)}
}

@misc{ssd_price,
  title={Intel SSDPF2KX038T11Z - 3.84TB SSD NVMe U.2 15mm},
  author={{Acme Micro Systems}},
  howpublished={\url{https://www.acmemicro.com/Product/18775/Intel-SSDPF2KX038T11Z---3-84TB-SSD-NVMe-U-2-15mm-PCIe-4-0-D7-P5520-Series-6500-MB-s-Read-3D4-TLC-NAND} (accessed Nov. 13, 2025)}
}

@misc{gpu_price,
  title={NVIDIA A100 40GB PCIe GPU},
  author={{Network Outlet}},
  howpublished={\url{https://networkoutlet.com/products/nvidia-a100-40gb-pcie-gpu-ampere-architecture-with-nvlink-mig} (accessed Nov. 13, 2025)}
}

@misc{nic_price,
  title={Mellanox MCX515A-CCAT ConnectX-5 EN Network Interface Card 100GbE Single-Port QSFP28 PCIe3.0 x16 Tall Bracket},
  author={{Acme Micro Systems}},
  howpublished={\url{https://www.acmemicro.com/Product/16314/Mellanox-MCX515A-CCAT-ConnectX-5-EN-Network-Interface-Card-100GbE-Single-Port-QSFP28-PCIe3-0-x16-Tall-Bracket} (accessed Nov. 13, 2025)}
}

@misc{u50,
  title={AMD Alveo U50 Data Center Accelerator Card},
  author={AMD},
  howpublished={\url{https://www.amd.com/en/products/accelerators/alveo/u50/a-u50-p00g-pq-g.html} (accessed Nov. 13, 2025)}
}

@misc{switch_price,
  title={Mellanox MSN2700-CS2F 100GbE, 1U Open Ethernet Switch w/MLNX-OS,32 QSFP28 ports},
  author={{Acme Micro Systems}},
  howpublished={\url{https://www.acmemicro.com/Product/15874/Mellanox-MSN2700-CS2F-100GbE-1U-Open-Ethernet-Switch-w-MLNX-OS-32-QSFP28-ports-2-PS(AC)-x86-CPU-Standardepth-P2C-airflow} (accessed Nov. 13, 2025)}
}

@misc{smartswitch_price,
  title={Edgecore Wedge 100BF-32X 32-Port 100GbE Bare Metal Switch with ONIE},
  author={{Colfax Direct}},
  howpublished={\url{https://www.colfaxdirect.com/store/pc/viewPrd.asp?idproduct=3485} (accessed Nov. 13, 2025)}
}

@misc{dgxprice,
  title={Dell DGX A100 Price List},
  author={{Router-switch}},
  howpublished={\url{https://itprice.com/dell-price-list/a100%20p3687%20320gb.html} (accessed Nov. 13, 2025)}
}

@inproceedings{astrasim_ispass23,
  title={Astra-sim2.0: Modeling hierarchical networks and disaggregated systems for large-model training at scale},
  author={Won, William and Heo, Taekyung and Rashidi, Saeed and Sridharan, Srinivas and Srinivasan, Sudarshan and Krishna, Tushar},
  booktitle={ISPASS},
  year={2023},
}

@inproceedings{chakra_arxiv,
  title={Chakra: Advancing performance benchmarking and co-design using standardized execution traces},
  author={Sridharan, Srinivas and Heo, Taekyung and Feng, Louis and Wang, Zhaodong and Bergeron, Matt and Fu, Wenyin and Zheng, Shengbao and Coutinho, Brian and Rashidi, Saeed and Man, Changhai and Tushar Krishna},
  booktitle={Benchmarking Machine Learning Workloads on Emerging Hardware@MLSys},
  year={2023}
}

@inproceedings{disttrain_sigcomm25,
  title={Disttrain: Addressing model and data heterogeneity with disaggregated training for multimodal large language models},
  author={Zhang, Zili and Zhong, Yinmin and Jiang, Yimin and Hu, Hanpeng and Sun, Jianjian and Ge, Zheng and Zhu, Yibo and Jiang, Daxin and Jin, Xin},
  booktitle={SIGCOMM},
  year={2025}
}

@inproceedings{bytescale_sigcomm25,
  title={ByteScale: Communication-Efficient Scaling of LLM Training with a 2048K Context Length on 16384 GPUs},
  author={Ge, Hao and Feng, Junda and Huang, Qi and Fu, Fangcheng and Nie, Xiaonan and Zuo, Lei and Lin, Haibin and Cui, Bin and Liu, Xin},
  booktitle={SIGCOMM},
  year={2025}
}

@inproceedings{resccl_sigcomm25,
  title={ResCCL: Resource-Efficient Scheduling for Collective Communication},
  author={Liu, Tongrui and Hei, Chenyang and Li, Fuliang and Gao, Chengxi and Cao, Jiamin and Wang, Tianshu and Zhai, Ennan and Wang, Xingwei},
  booktitle={SIGCOMM},
  year={2025}
}

@inproceedings{parserhawk_sigcomm25,
  title={ParserHawk: Hardware-aware parser generator using program synthesis},
  author={Gao, Xiangyu and Gao, Jiaqi and G, Karan Kumar and Haseeb, Muhammad and Zhai, Ennan and Dong, Bili and Tassarotti, Joseph and Narayana, Srinivas and Sivaraman, Anirudh},
  booktitle={SIGCOMM},
  year={2025}
}

@inproceedings{albatross_sigcomm25,
  title={Albatross: A Containerized Cloud Gateway Platform with FPGA-accelerated Packet-level Load Balancing},
  author={Jianyuan Lu and Shunmin Zhu and Jun Liang and Yuxiang Lin and Tian Pan and Yisong Qiao and Yang Song and Wenqiang Su and Yixin Xie and Yanqiang Li and Enge Song and Shize Zhang and Xiaoqing Sun and Rong Wen and Xionglie Wei and Biao Lyu and Xing Li},
  booktitle={SIGCOMM},
  year={2025}
}

@inproceedings{nezha_sigcomm25,
  title={Nezha: SmartNIC-based Virtual Switch Load Sharing},
  author={Xing Li and Enge Song and Bowen Yang and Tian Pan and Ye Yang and Qiang Fu and Yang Song and Yilong Lv and Zikang Chen and Jianyuan Lu and Shize Zhang and Xiaoqing Sun and Rong Wen and Xionglie Wei and Biao Lyu and Zhigang Zong and Qinming He and Shunmin Zhu },
  booktitle={SIGCOMM},
  year={2025}
}

@inproceedings{rina_icnp24,
  title={Rina: Enhancing ring-AllReduce with in-network aggregation in distributed model training},
  author={Chen, Zixuan and Liu, Xuandong and Li, Minglin and Hu, Yinfan and Mei, Hao and Xing, Huifeng and Wang, Hao and Shi, Wanxin and Liu, Sen and Xu, Yang},
  booktitle={ICNP},
  year={2024},
}

@misc{bf2_spec,
  title={NVIDIA BlueField-2 InfiniBand/Ethernet DPU Specifications},
  author={NVIDIA},
  howpublished={\url{https://docs.nvidia.com/networking/display/bluefield2dpuvpi/specifications} (accessed Mar. 6, 2026)}
}

@misc{bf3_spec,
  title={NVIDIA BlueField-3 Networking Platform Specifications},
  author={NVIDIA},
  howpublished={\url{https://docs.nvidia.com/networking/display/bf3dpu/specifications} (accessed Feb. 21, 2026)}
}

@misc{bluefield,
  title={NVIDIA BlueField Platform},
  author={NVIDIA},
  howpublished={\url{https://www.nvidia.com/en-us/networking/products/data-processing-unit/} (accessed Feb. 21, 2026)}
}

@misc{nccl_sharp,
    title={Using NVIDIA SHARP with NVIDIA NCCL},
    author={NVIDIA},
    howpublished={\url{https://docs.nvidia.com/networking/display/sharpv300/using+nvidia+sharp+with+nvidia+nccl}}
}

@inproceedings{intel_ipu_hcs2025,
  title={Intel{\textregistered} IPU E2200: Second Generation Infrastructure Processing Unit (IPU)},
  author={Fleming, Pat and Chang, Chihjen and Collier, Derek and Singhai, Anjali and Doyle, Stephen and Louzoun, Eliel and Lee, David and Ayyavu, Vetrivel and Livne, Sarig and Hathaway, Robert and Tony Hurson and Jackson Ellis and Tamar Bar-Kanarik and Jonathan Kenny and Cristine Dumitrescu and Yaron Wolberger},
  booktitle={HCS},
  year={2025},
}

@article{pat,
  title={PAT: a new algorithm for all-gather and reduce-scatter operations at scale},
  author={Jeaugey, Sylvain},
  journal={arXiv preprint},
  year={2025}
}

@inproceedings{overlap_asplos22,
  title={Overlap communication with dependent computation via decomposition in large deep learning models},
  author={Shibo Wang and Jinliang Wei and Amit Sabne and Andy Davis and Berkin Ilbeyi and Blake Hechtman and Dehao Chen and Karthik Srinivasa Murthy and Marcello Maggioni and Qiao Zhang and Sameer Kumar and Tongfei Guo and Yuanzhong Xu and Zongwei Zhou},
  booktitle={ASPLOS},
  year={2022}
}

@inproceedings{centauri_asplos24,
  title={Centauri: Enabling efficient scheduling for communication-computation overlap in large model training via communication partitioning},
  author={Chen, Chang and Li, Xiuhong and Zhu, Qianchao and Duan, Jiangfei and Sun, Peng and Zhang, Xingcheng and Yang, Chao},
  booktitle={ASPLOS},
  year={2024}
}

@inproceedings{lancet_mlsys2024,
  title={Lancet: Accelerating mixture-of-experts training by overlapping weight gradient computation and all-to-all communication},
  author={Jiang, Chenyu and Tian, Ye and Jia, Zhen and Wu, Chuan and Wang, Yida and Zheng, Shuai},
  booktitle={MLSys},
  year={2024}
}

@inproceedings{flashoverlap_eurosys26,
  title={Efficient and Adaptable Overlapping for Computation and Communication via Signaling and Reordering},
  author={Ke Hong and Xiuhong Li and Minxu Liu and Qiuli Mao and Tianqi Wu and Zixiao Huang and Lufang Chen and Zhong Wang and Yichong Zhang and Zhenhua Zhu and Guohao Dai and Yu Wang},
  booktitle={EuroSys},
  year={2026}
}

@inproceedings{coconet_asplos22,
  title={Breaking the computation and communication abstraction barrier in distributed machine learning workloads},
  author={Jangda, Abhinav and Huang, Jun and Liu, Guodong and Sabet, Amir Hossein Nodehi and Maleki, Saeed and Miao, Youshan and Musuvathi, Madanlal and Mytkowicz, Todd and Saarikivi, Olli},
  booktitle={ASPLOS},
  year={2022}
}

@inproceedings{optimizing_sc2024,
  title={Optimizing distributed ml communication with fused computation-collective operations},
  author={Punniyamurthy, Kishore and Hamidouche, Khaled and Beckmann, Bradford M},
  booktitle={SC},
  year={2024},
}

@inproceedings{tilelink_mlsys2025,
  title={Tilelink: Generating efficient compute-communication overlapping kernels using tile-centric primitives},
  author={Size Zheng and Jin Fang and Xuegui Zheng and Qi Hou and Wenlei Bao and Ningxin Zheng and Ziheng Jiang and Dongyang Wang and Jianxi Ye and Haibin Lin and Li-Wen Chang and Xin Liu},
  booktitle={MLSys},
  year={2025}
}

@article{mirage_arxiv2025,
  title={Mirage Persistent Kernel: A Compiler and Runtime for Mega-Kernelizing Tensor Programs},
  author={Xinhao Cheng and Zhihao Zhang and Yu Zhou and Jianan Ji and Jinchen Jiang and Zepeng Zhao and Ziruo Xiao and Zihao Ye and Yingyi Huang and Ruihang Lai and Hongyi Jin and Bohan Hou and Mengdi Wu and Yixin Dong and Anthony Yip and Zihao Ye and Songting Wang and Wenqin Yang and Xupeng Miao and Tianqi Chen and Zhihao Jia},
  journal={arXiv preprint},
  year={2025}
}

@inproceedings{suresh2023novel,
  title={A novel framework for efficient offloading of communication operations to bluefield smartnics},
  author={Suresh, Kaushik Kandadi and Michalowicz, Benjamin and Ramesh, Bharath and Contini, Nick and Yao, Jinghan and Xu, Shulei and Shafi, Aamir and Subramoni, Hari and Panda, Dhabaleswar},
  booktitle={IPDPS},
  year={2023},
}

@inproceedings{acclplus_osdi24,
  title={ACCL+: an FPGA-Based collective engine for distributed applications},
  author={He, Zhenhao and Korolija, Dario and Zhu, Yu and Ramhorst, Benjamin and Laan, Tristan and Petrica, Lucian and Blott, Michaela and Alonso, Gustavo},
  booktitle={OSDI},
  year={2024}
}

@inproceedings{graham2024optimizing,
  title={Optimizing application performance with bluefield: accelerating large-message blocking and nonblocking collective operations},
  author={Graham, Richard and Bosilca, George and Qin, Yong and Settlemyer, Bradley and Shainer, Gilad and Stunkel, Craig and Vallee, Geoffroy and Williams, Brody and Cisneros-Stoianowski, Gerardo and Ohlmann, Sebastian and Rampp, Markus},
  booktitle={ISC},
  year={2024},
}

@inproceedings{oodsmpi_sc25,
  title={ODOS-MPI: HPC-Friendly SmartNIC Offloading of Computation/Communication Kernels},
  author={Usman, Muhammad and Benito, Mariano and Iserte, Sergio and Pe{\~n}a, Antonio J},
  booktitle={SC},
  year={2025}
}

@inproceedings{netreduce_asplos2023,
  title={In-network aggregation with transport transparency for distributed training},
  author={Liu, Shuo and Wang, Qiaoling and Zhang, Junyi and Wu, Wenfei and Lin, Qinliang and Liu, Yao and Xu, Meng and Canini, Marco and Cheung, Ray CC and He, Jianfei},
  booktitle={ASPLOS},
  year={2023}
}

@inproceedings{khalilov2024network,
  title={Network-offloaded bandwidth-optimal broadcast and Allgather for distributed AI},
  author={Khalilov, Mikhail and Di Girolamo, Salvatore and Chrapek, Marcin and Nudelman, Rami and Bloch, Gil and Hoefler, Torsten},
  booktitle={SC},
  year={2024},
}

@inproceedings{smartds_isca23,
  title={Smartds: Middle-tier-centric smartnic enabling application-aware message split for disaggregated block storage},
  author={Zhang, Jie and Huang, Hongjing and Zhu, Lingjun and Ma, Shu and Rong, Dazhong and Hou, Yijun and Sun, Mo and Gu, Chaojie and Cheng, Peng and Shi, Chao and Wang, Zeke},
  booktitle={ISCA},
  year={2023}
}

@inproceedings{rambda_hpca2023,
  title={Rambda: Rdma-driven acceleration framework for memory-intensive $\mu$s-scale datacenter applications},
  author={Yuan, Yifan and Huang, Jinghan and Sun, Yan and Wang, Tianchen and Nelson, Jacob and Ports, Dan RK and Wang, Yipeng and Wang, Ren and Tai, Charlie and Kim, Nam Sung},
  booktitle={HPCA},
  year={2023}
}

@inproceedings{smartns_eurosys26,
  title={FlexiNS: A SmartNIC-Centric, Line-Rate and Flexible Network Stack},
  author={Chen, Xuzheng and Zhang, Jie and Zhu, Baolin and Zhu, Xueying and Chen, Zhongqing and Ting Fu and Ma, Shu and Zhu, Lingjun and Shi, Chao and Zhang, Yin and Shu, Yuanchao and Cheng, Peng and Wang, Zeke},
  booktitle={EuroSys},
  year={2026}
}



\end{document}